\documentclass[%
 aip,
 amsmath,amssymb,
 reprint,%
]{revtex4-1}

\usepackage{graphicx}% Include figure files
\usepackage{dcolumn}% Align table columns on decimal point
\usepackage{bm}% bold math
\usepackage[utf8]{inputenc}
\usepackage{bbold}
\usepackage{color}
\begin{document}

%\preprint{}

\title{Ferromagnetism-induced phase separation in a two-dimensional spin fluid}


\author{Mathias Casiulis}
\affiliation{Sorbonne Universit\'e, Laboratoire de Physique Th\'eorique de la Mati\`ere Condens\'ee,  CNRS UMR 7600, 4 Place Jussieu, F-75005 Paris, France}
\email{casiulis@lptmc.jussieu.fr}
\author{Marco Tarzia}
\affiliation{Sorbonne Universit\'e, Laboratoire de Physique Th\'eorique de la Mati\`ere Condens\'ee,  CNRS UMR 7600, 4 Place Jussieu, F-75005 Paris, France}
\author{Leticia F. Cugliandolo}
\affiliation{Sorbonne Universit\'e, Laboratoire de Physique Th\'eorique et Hautes
\'Energies,  CNRS UMR 7589, 4 Place Jussieu, F-75005 Paris, France}
\author{Olivier Dauchot}
\affiliation{PSL Research University, Laboratoire Gulliver, CNRS UMR 7083,
ESPCI Paris, 10 rue Vauquelin, 75005 Paris, France}
%\thanks{}


\date{\today}

\begin{abstract}
We study the liquid-gas phase separation observed in a system of repulsive particles dressed with ferromagnetically aligning spins, a so-called ``spin fluid''. 
Microcanonical ensemble numerical simulations of finite-size systems reveal that magnetization sets in and induces a liquid-gas phase separation between a disordered gas and a ferromagnetic dense phase at low enough energies and large enough densities. 
The dynamics after a quench into the coexistence region show that the order parameter associated to the liquid-vapour phase separation follows an algebraic law with an unusual exponent, as it is forced to synchronize with the growth of the magnetization: this suggests that for finite size systems the magnetization sets in along a Curie line, which is also the gas-side spinodal line, and that
the coexistence region ends at a tricritical point. 
This picture is confirmed at the mean-field level with different approximation schemes, namely a Bethe lattice resolution and a virial expansion complemented by the introduction of a self-consistent Weiss-like molecular field.
However, a detailed finite-size scaling analysis shows that in two dimensions the ferromagnetic phase escapes the Berezinskii-Kosterlitz-Thouless scenario, and that the long-range order is not destroyed by the unbinding of topological defects. The Curie line becomes thus a magnetic crossover in the thermodynamic limit.
Finally, the effects of the magnetic interaction range and those of the interaction softness are characterized within a mean-field semi-analytic low-density approach.
\end{abstract}

\pacs{}
\keywords{Theory of liquids, Spin magnetism, Phase separation}

\maketitle

% References should be done using the~\cite, \ref, and \label commands
\section{Introduction \label{sec:Introduction}}
Spin fluids are a family of models introduced by Weis and coworkers~\cite{Tavares1995,Lomba1998} as simplified models of magnetic colloids or alloys, and more recently used to study binary fluids~\cite{Wilding2003,Das2006,Puosi2016}.
They are, essentially, off-lattice versions of interacting spin systems, in which (anti)ferromagnetically coupled spins are carried by particles and (usually short-range) distance-dependent interactions. 
An {\it ideal spin fluid} is one such system with no explicit attractive interaction, but only kinetic energy, core repulsion, and magnetic spin coupling. 
Up to now, continuous-spin spin fluids were only studied in $3d$, first with Heisenberg spins~\cite{Tavares1995,Lomba1998}, and later with planar spins~\cite{Omelyan2009}.
The aim of these studies was to determine the phase diagram of these unusual systems, that can display both isotropic-(anti)ferromagnetic and gas-liquid-solid phase transitions.
The phase diagrams, with the magnetic and gas-liquid transitions, were found by solving integral equations under various closure hypotheses that used both Born-Green-like and Hypernetted-chain-like relations. 
The crystal, when predicted~\cite{Lomba1998}, was found through density-functional theory, following standard textbook methods,~\cite{Hansen2006} and well-established literature on the matter.~\cite{Barrat1987}

In two-dimensional space, the magnetization transition of the on-lattice XY model is known to be highly peculiar. 
The Mermin-Wagner theorem~\cite{Mermin1966,Mermin1967} prohibits spontaneous symmetry breaking at any finite temperature.
However, a static phase transition occurs at a finite critical temperature, $T_{KT}$ between a high-temperature disordered paramagnet and a low-temperature ferromagnetic phase with quasi-long-range order.
This phase transition obeys the Berezinskii-Kosterlitz-Thouless (BKT) scenario,~\cite{Berezinskii1971,Kosterlitz1973,Kosterlitz1974} according to which the magnetization is suppressed by spin waves for $T<T_{KT}$ and by isolated vortices for $T>T_{KT}$. 
The validity of this scenario as well as its possible influence on the properties of two-dimensional XY spin fluids is, to our knowledge, a completely open question.
This matter is relevant for several physical systems, from ultracold polar atoms,~\cite{Jo2009} to certain ionic liquids.~\cite{Boudalis2017}

In this paper, we focus on a $2d$ ferromagnetic ideal spin fluid, constituted by repulsive disks dressed with short-range ferromagnetic pair interactions.
We shall show that, as in the $3d$ case, a ferromagnetism-induced phase separation (FIPS) develops, and provides this system with a non-trivial phase diagram. 
Our main goal is to obtain such phase diagram, and in particular to characterize the interplay between the magnetization crossover and the liquid-gas separation.  

We first perform Molecular Dynamics (MD) simulations in the microcanonical ensemble to identify the magnetic and structural properties of finite-size systems, focusing on the fluid phases.
In so doing, we recover a phase diagram qualitatively very similar to those obtained for planar spins in $3d$.~\cite{Omelyan2009}
Schematically, we find a phase separation between a paramagnetic gas and a ferromagnetic liquid.
The coexistence curve ends in a tricritical point, from which stems a magnetization Curie-like line which extends in the region of the phase diagram where the temperature is too high for liquid-gas phase separation to occur, also called the supercritical fluid region in the standard theory of simple liquids.~\cite{Hansen2006}
However, a finite-size scaling analysis suggests that the ferromagnetic phase escapes the BKT scenario.
This result is confirmed by the numerical study of the equilibration dynamics  using equilibrated initial conditions in which the particles are positioned on the vertices of a regular triangular lattice with a few free vortices, or rapid quenches from random high temperature initial conditions into the magnetized phase. 
No vortices survive either way. 
The magnetically ordered phases of finite-size systems are characterized by an extremely large correlation length, as expected for a critical system at its lower critical dimension: in the thermodynamic limit (i.e. when the system size becomes larger than the correlation volume), the ferromagnetic order is destroyed by low-energy spin wave excitations, and the Curie line becomes a crossover.
Regarding the phase separation, equilibration dynamics following quenches into the coexistence region show that the growth of the order parameter associated to the liquid-vapour phase separation does not follow standard algebraic scalings. 
Instead, its growth rate is forced to synchronize to that of the magnetization, suggesting that the gas-side spinodal line coincides with the Curie (crossover) line.

On the theoretical side, we adapt standard methods from the liquid-state theory to describe spin fluids, which carry an internal degree of freedom.
We thereby obtain the mean-field equation of state and phase diagram.
The latter qualitatively reproduces the phenomenology observed in the simulations.
In particular, the magnetization sets in along a Curie line, which is also the gas-side spinodal line.
Finally, our methods allow for a characterization of the role played by the magnetic interaction range and the softness of the repulsive interaction.

The paper is organized as follows. 
In Section~\ref{sec:Model}, we introduce the model. 
Section~\ref{sec:MD} presents phenomenology observed in molecular dynamics simulations performed in the microcanonical ensemble.
Section~\ref{sec:Lattice} describes an on-lattice approach that predicts phase separation and magnetization in the case of hard-core repulsion.
Section~\ref{sec:CurieWeiss} introduces a self-consistent approach, based on a Curie-Weiss-like approximation, coupled to two different approximations for the local structure of the spin fluid. 
Finally, in Section \ref{sec:Conclusion}, we present our conclusions and some perspectives for future research.

\section{The model \label{sec:Model}}

Throughout this paper, we will study systems of particles described by the Lagrangian
\begin{eqnarray}
 L &=& 
 \sum\limits_{i = 1}^{N} \frac{m}{2} \bm{\dot{r}}_i^2   
 + \sum\limits_{i = 1}^{N} \frac{I}{2} {\dot{\theta_i}}^2 
 \nonumber\\
 && 
 -  \frac{1}{2} \sum\limits_{k \neq i} U(r_{ik}) 
 + \frac{J_0}{2} \sum\limits_{k \neq i} J(r_{ik}) \cos\theta_{ik} 
 \; ,
\end{eqnarray}
where $m$ is the mass of each particle, $I$ their moment of inertia, $\bm{r}_i$ the position of the $i$th particle and $\theta_i$ the angle coding for the direction of the XY spin they carry. 
The last two terms represent the interactions. 
$U(r_{ik})$ is a purely repulsive potential, and $J(r_{ik})$ a finite-range ferromagnetic coupling, with $J_0$ its typical amplitude. 
Note that the spins here simply represent an internal anisotropy and follow \emph{precessional} dynamics with an associated rotational kinetic energy, making the present model a planar rotor model~\cite{Lepri2001,Wysin2005} and not a ferromagnet model \emph{stricto sensu}.
We will henceforth use an adimensionalized version of this Lagrangian, defined through the replacements $\bm{r}_i/\sqrt{I/m} \to \bm{r}_i$, $t/\sqrt{I/J_0}\to t$, $L/J_0 \to L$, $U/J_0 \to U$. 
The Hamiltonian associated to this adimensionalized dynamics can then be written in the usual way, 
\begin{eqnarray}
 H &=& \sum\limits_{i = 1}^{N} \frac{1}{2} \bm{p}_i^2 + 
       \sum\limits_{i = 1}^{N}\frac{1}{2} {\omega_i}^2 
  \nonumber\\
  && 
  + \frac{1}{2} \sum\limits_{k \neq i} U(r_{ik}) - \frac{1}{2} \sum\limits_{k \neq i} J(r_{ik}) \cos\theta_{ik}
  \; ,
\end{eqnarray}
where the canonical momenta are defined as $\bm{p}_i = \bm{\dot{r}}_i$ and $\omega_i = \dot{\theta_i}$. 
Finally, the Hamiltonian equations of motion,
\begin{eqnarray}
 \bm{\dot{p}_i} &=& \sum\limits_{k(\neq i)} \left(\frac{\partial J(r_{ik})}{\partial \bm{r}_i} \cos\theta_{ik} - \frac{\partial U(r_{ik})}{\partial \bm{r}_i} \right) 
 \; , \label{eq:Ham1} \\
 \dot{\omega}_i &=& \sum\limits_{k(\neq i)}  J(r_{ik}) \sin\theta_{ik} 
 \; , \label{eq:Ham2}
\end{eqnarray}
will be used in the simulations presented in the next section.
The first of these equations displays quite clearly why the physics of liquids of this kind could be interesting from a fundamental viewpoint: 
a ferromagnetic coupling ($J>0$) that decays with distance ($J'(r) < 0$) implies that particles with spins lying in the same half-plane ($\bm{s}_i\cdot\bm{s}_j >0$) are attracted to each other, while particles with spins lying in opposite 
half-planes ($\bm{s}_i\cdot\bm{s}_j <0$) are purely repulsive. Therefore, even for a purely repulsive $U$ coupling, spin-carrying particles feature a spin-mediated effective attraction, that could allow for a liquid-gas phase separation.

\section{Phenomenology from Molecular Dynamics simulations \label{sec:MD}}

\begin{figure}
    \centering
    \includegraphics[width = \columnwidth]{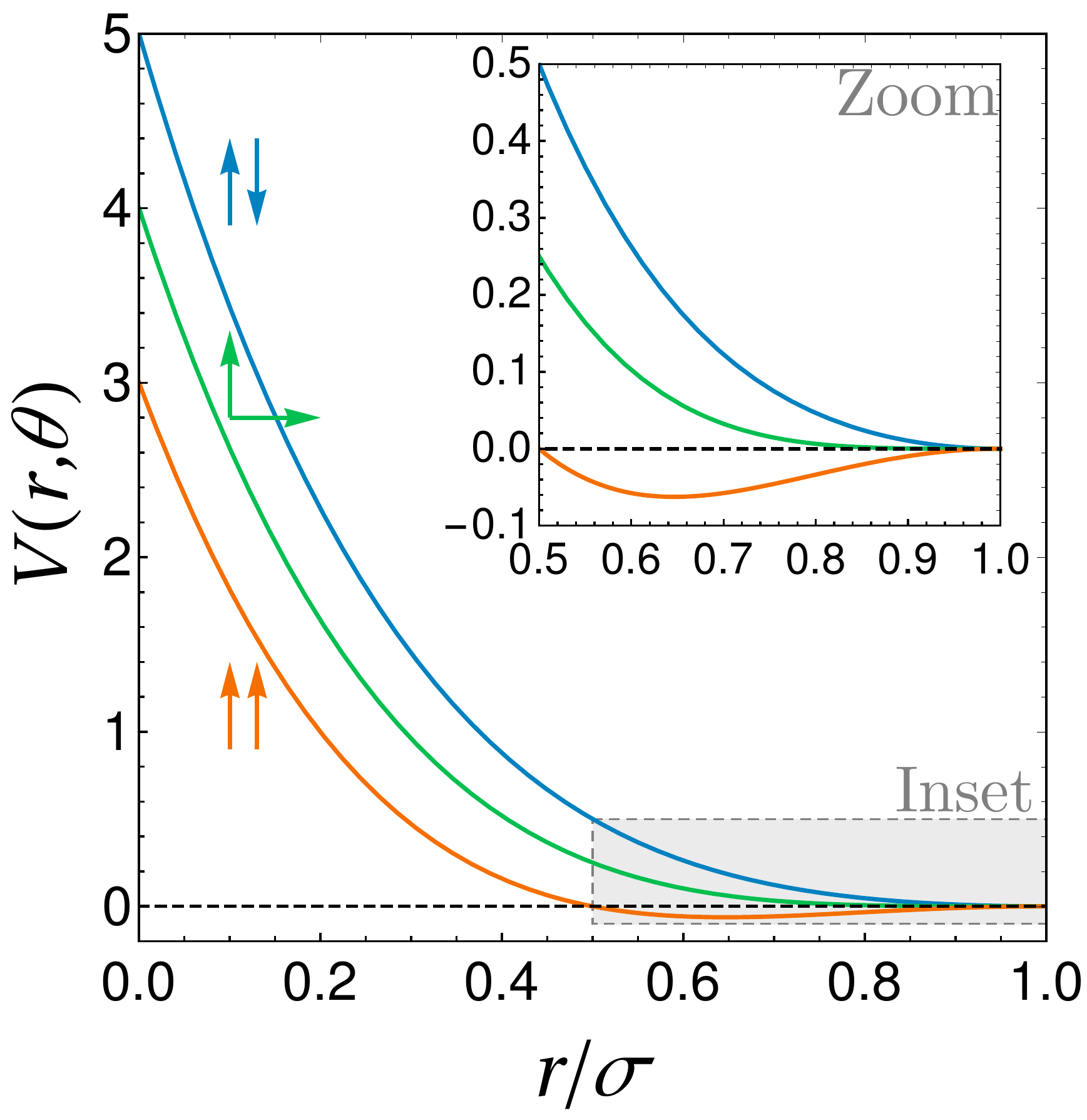}
    \caption{{\textbf{Effective pairwise interaction.}} 
    Total pairwise interaction potential $U(r) - J(r) \cos\theta$ used in simulations for aligned (orange), orthogonal (green) and anti-aligned (blue) spins.
    We highlight the $V(r,\theta) = 0$ line, which separates the repulsive ($V > 0$) and attractive ($V<0$) parts of the potential, with a dashed black line.
    In the inset, we show a zoom on the region indicated by a gray rectangle, in order to emphasize the attractive part in the case of aligned spins.}
    \label{fig:Effpot}
\end{figure}

We start by presenting the key phenomenology of the system using Molecular Dynamics (MD) simulations that rely on the above Hamiltonian equations of motion, and were performed in the microcanonical ensemble.
To ensure well-behaved dynamics, we use soft interaction potentials, given by
\begin{eqnarray}
 J(r) &=&     (\sigma - r)^2 \Theta(\sigma - r)  \; , \nonumber \\
 U(r) &=& U_0 (\sigma - r)^4 \Theta(\sigma - r) \; ,  \label{eq:Int}
\end{eqnarray}
where $\Theta$ is a Heaviside step function, $\sigma$ is a range that was fixed to $1$, and $U_0 = 4$ was chosen so that both potentials are equal at half-range. 
We show the associated effective pairwise interaction $V(r,\theta) \equiv U(r) - J(r) \cos\theta$ in the cases of aligned, anti-aligned and orthogonal spins in Fig.~\ref{fig:Effpot}.

We define the particle radius $r_0 = \sigma / 2$, the zero-temperature exclusion radius for fully-aligned spins, and use it to define the packing fraction $\phi$. 
We simulate the dynamics starting from random states with uniformly distributed $\left\{\bm{r}_i, \theta_i \right\}_{i=1..N}$ and $\left\{\bm{p}_i, \omega_i \right\}_{i=1..N}$ drawn from centered, reduced Gaussian distributions. 
Such initial states were placed into a square box with periodic boundary conditions and, after giving some time for the dynamics to settle in, are subjected either to a numerical annealing or to a high-rate quench. 
These procedures are implemented as follows.

Numerical annealings are performed by multiplying all rotational velocities by $\lambda_{A} = 0.9999$ every 100 time units in our adimensionalized variable, with an integration time step equal to $\delta t = 10^{-3}$ in the same units. 
This method enables us to reach low-energy states which, if the cooling is slow enough, should be equilibrium states of the system.

Quenches are carried out by multiplying all rotational velocities and momentum components by $\lambda_{Q} = 0.10$ once, at some initial time. 
This method violently takes the system away from equilibrium, thereby enabling us to study the subsequent equilibration dynamics. 

\subsection{Magnetization properties}

In this subsection, in order to decouple the discussion of the behaviour of the magnetization from the one of the structural properties as much as possible, we start by presenting the results obtained at a number density such that very little local density fluctuations can take place, namely $\rho = N/L^2 \approx 2.81$ where $L$ is the size of the system.
Equivalently, this density corresponds to a packing fraction $\phi = 0.55$.

\subsubsection{Equilibrium properties}

The system we are studying here is, from the magnetic point of view, an off-lattice version of a diluted $2d$ XY model. 
As such, true long-range magnetic order is forbidden by the Mermin-Wagner theorem: there is no ferromagnetic phase transition at finite temperature for continuous spins with isotropic and finite-range interactions in two dimensions.~\cite{Mermin1966,Mermin1967}
The square-lattice XY model is however known to present a static phase transition at a temperature $T_{KT}$ between a high-temperature disordered paramagnet and a low-temperature ferromagnetic phase with quasi-long-range order.~\cite{Berezinskii1971,Kosterlitz1973,Kosterlitz1974}
Here, we will demonstrate that this BKT scenario does not survive in our off-lattice setting.

Even in the absence of long-range order in the $L\to\infty$ limit, when the linear system size $L$ is smaller than the spin-spin correlation length $\xi$, a non-zero magnetization $m(T,L)$ is observed at low enough temperatures $T$.
Furthermore, $m(T,L)$, $\xi(T,L)$, and the magnetic susceptibility $\chi_m(T,L)$ obey scaling laws around a temperature $T_c(L)$ to be determined empirically.

For usual magnetization transitions, these scaling laws take power law forms: $m  \sim t^\beta$, $\xi \sim t^{-\nu}$, and $\chi_m \sim t^{-\gamma}$, where $t = T /T_c(L) - 1$ and $T_c(L)\to T_C$, the Curie temperature, in the thermodynamic limit $L\to \infty$.
In practice, infinite system size scalings lead to the following finite system size scalings of $m$ and $\chi_m$ with $L$,
\begin{eqnarray}
 m      &\propto& L^{-\beta/\nu}, \label{eq:mFSS}\\
 \chi_m &\propto& L^{\gamma/\nu}. \label{eq:chiFSS}
\end{eqnarray}

These scalings, however, only hold as long as the dimensionality of space, $d$, is high enough. 
For any magnetic system, there is a value $d_c$ of $d$, called the \emph{lower critical dimension}, below which no critical transition occurs at any finite temperature.
As indicated by the Mermin-Wagner argument, as well as field-theoretical calculations,~\cite{Nelson1977} for the XY model, $d_c = 2$.
As a consequence, the magnetization of the square-lattice XY model is only non-zero at $T = 0$ in the thermodynamic limit.
The way in which $m$ is suppressed with system size, however, undergoes a dramatic change depending on the value of the temperature.

In the limit $T\to 0$, the square-lattice XY magnetization modulus is suppressed with system size following the scaling~\cite{Tobochnik1979,Archambault1997}
\begin{eqnarray}
 \ln m &=&  -\frac{T}{8 \pi J}\ln\left(a N \right), \label{eq:SWscaling}
\end{eqnarray}
where $N$ is the number of sites, $J$ the ferromagnetic coupling constant, and $a$ a constant.
This unusual scaling is often called the spin-wave scaling, as it reflects the fact that the magnetization is slowly suppressed by low-energy plane waves, that can be described by a free field theory at low temperatures.~\cite{Amit1980a, Nelson1977}

As temperature grows, there is a finite temperature $T_c(L)$ at which the suppression of the magnetization becomes dramatically faster.
This is a sign of an essential singularity of the underlying theory, also called the BKT transition, characterized by the proliferation of free vortices above $T_c(L)$. 
Around this singularity, the correlation length diverges exponentially, $\ln\xi \sim t^{-\nu}$, with $\nu = 1/2$ and $T_c(L)\to T_{KT}$, which approaches a finite value, as the system size grows. 
Note that in this context, $T_c(L)$ is sometimes denoted $T^{\star}(L)$,~\cite{Bramwell1994} a notation we shall adopt henceforth.

Regardless of these differences, Eqs.~(\ref{eq:mFSS}) and~(\ref{eq:chiFSS}) still hold, so that the magnetization and magnetic susceptibility can still be related to the linear size of the system and to the critical exponent ratios.~\cite{Canova2014}
In particular, the magnetic susceptibility at the BKT transition universally~\cite{Kosterlitz1974} grows like $L^{7/4}$, although it does not feature a peak at $T_{KT}$, but keeps growing to its spin-wave value as temperature decreases.~\cite{Tobochnik1979}
Equivalently, one defines the anomalous dimension $\eta = 2 - \gamma/\nu$, which in the BKT universality class~\cite{Kosterlitz2016} takes the value $\eta = 1/4$.
Finally, the finite size scalings of $m$ and $\chi_m$ are linked by the hyperscaling relation,~\cite{Fisher1974} $2 \beta + \gamma = \nu d $ which, for $d = 2$, can be rewritten as $2\beta / \nu = \eta$.

We are now in a position to check whether the BKT scenario survives in $2d$ XY spin fluids.
Using MD simulations, we cool down systems with a number of particles ranging from 128 to 16384 with a slow simulated annealing (i.e., slow enough to ensure that equilibrium is reached), and measure their magnetization, and rotational and translational temperatures, defined as
\begin{eqnarray*}
 \bm{M}    &=&  \sum\limits_{i = 1}^{N} \bm{s}_i, \\
  T_R      &=&  \frac{1}{N} \sum\limits_{i = 1}^{N} {\omega_i}^2 - \left(\frac{1}{N} \sum\limits_{i = 1}^{N} {\omega_i} \right)^2, \\
  T_T      &=&  \frac{1}{N} \sum\limits_{i = 1}^{N} {p^2_{x/y,i}} - \left(\frac{1}{N} \sum\limits_{i = 1}^{N} {p_{x/y,i}} \right)^2.
\end{eqnarray*}
We then compute the averages of the modulus of the magnetization, the temperature $T$ and the usual magnetic susceptibility defined as
\begin{eqnarray*}
 m         &=& \frac{1}{N}\langle \left|\bm{M}\right| \rangle, \\
 T         &=& \langle T_R \rangle = \langle T_T \rangle, \\
 \chi_m    &=&  \frac{1}{N} \left(\langle \bm{M}^2\rangle - \langle \bm{M} \rangle^2\right),
 \end{eqnarray*}
 where $\langle \cdot \rangle$ denotes an average over independent configurations, here obtained by letting the dynamics run for sufficiently long times and using different initial conditions.
 Note that, having checked that $T_R = T_T$, we henceforth use the symbol $T$ for temperature, without further specification.
 
\begin{figure}
\centering
\includegraphics[width=.48\columnwidth,height=.48\columnwidth]{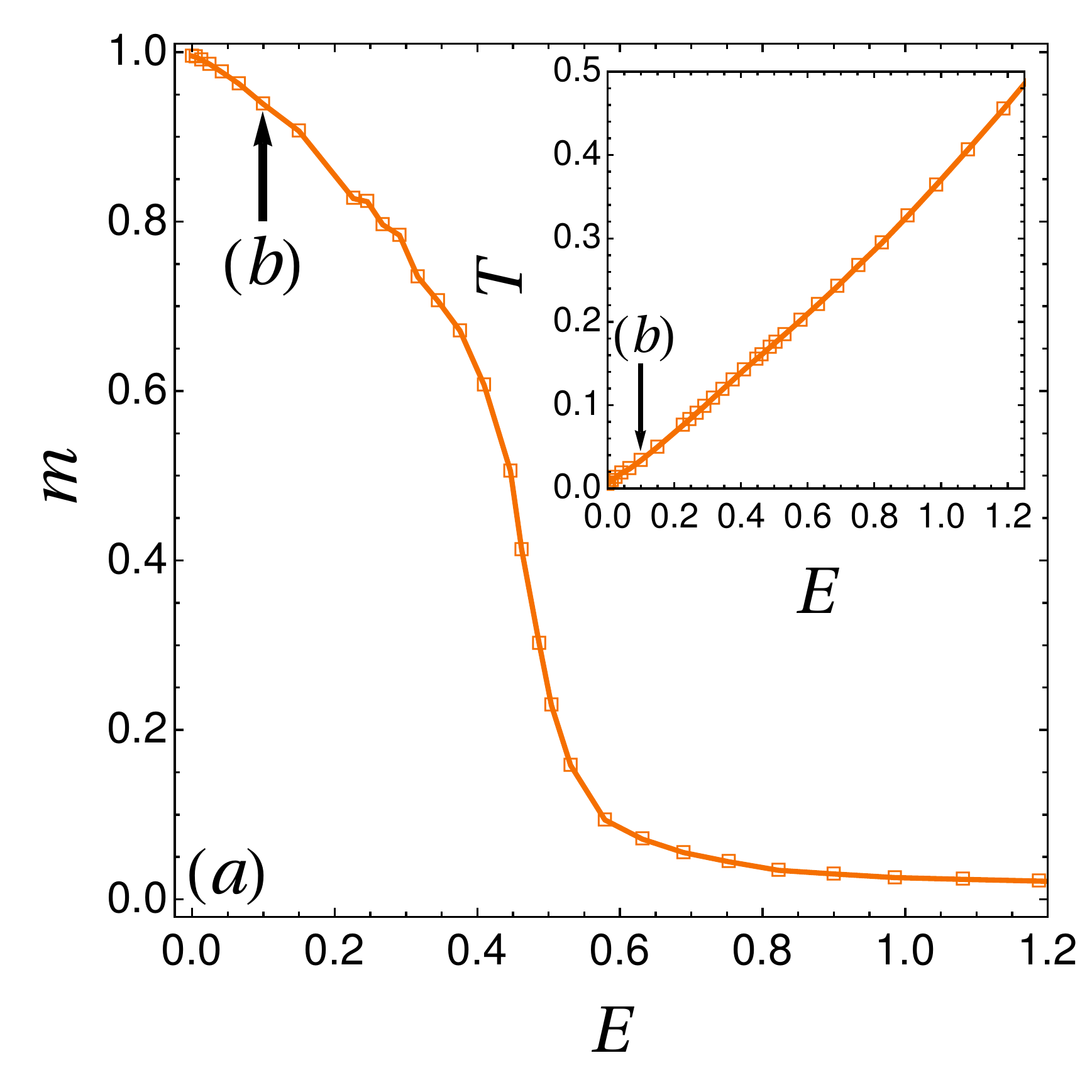} 
\includegraphics[width=.48\columnwidth,height=.48\columnwidth]{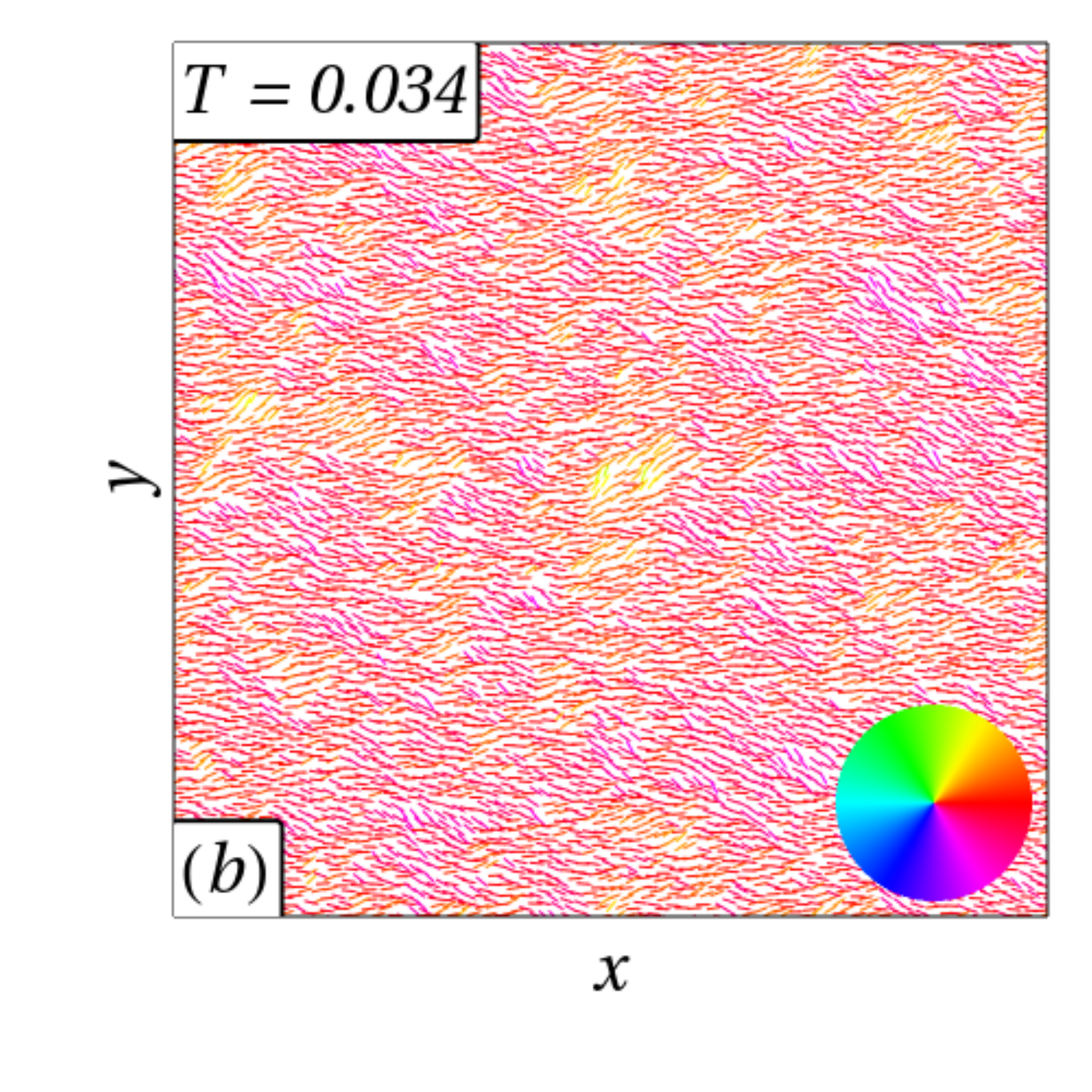}\\
\includegraphics[width=.48\columnwidth,height=.48\columnwidth]{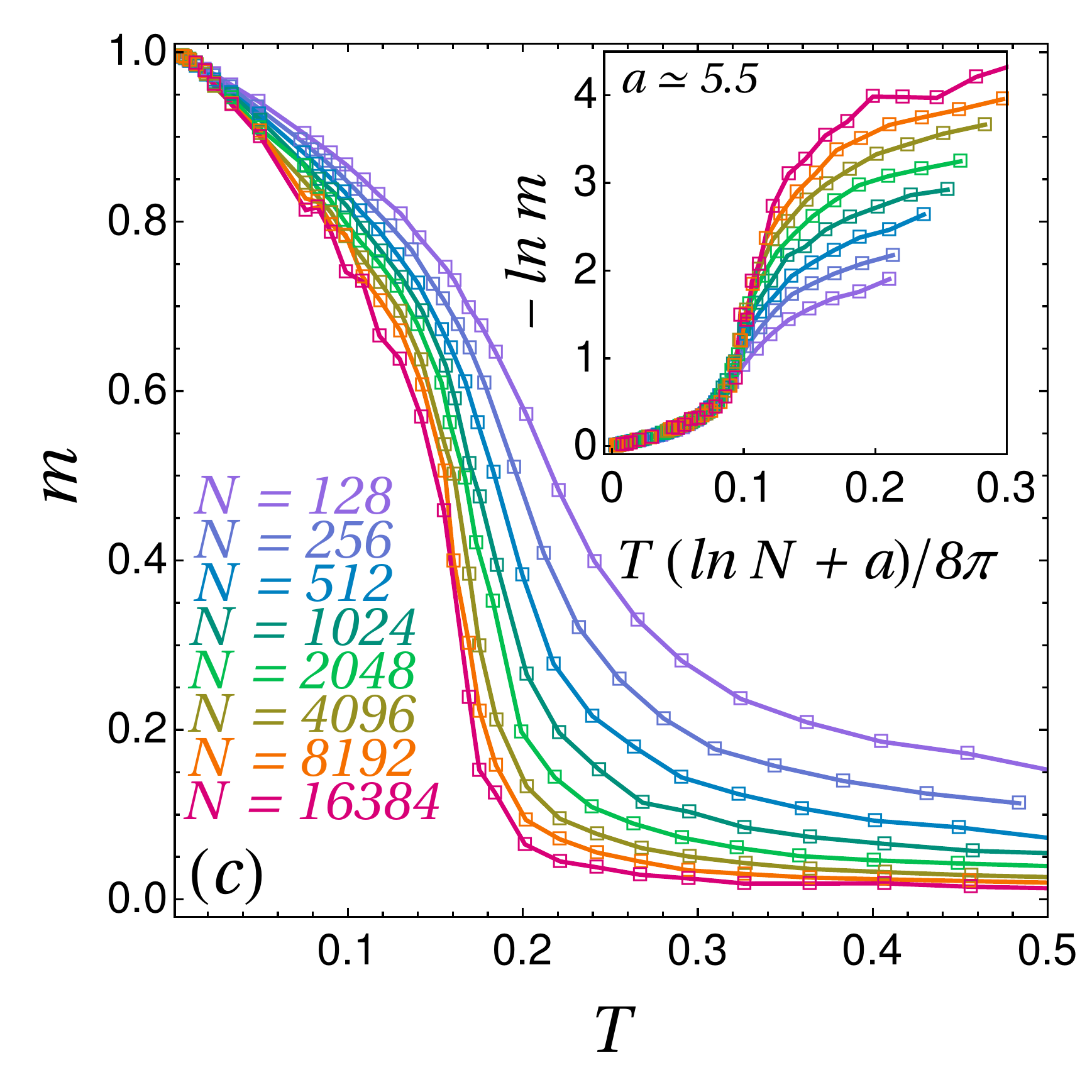}
\includegraphics[width=.48\columnwidth,height=.48\columnwidth]{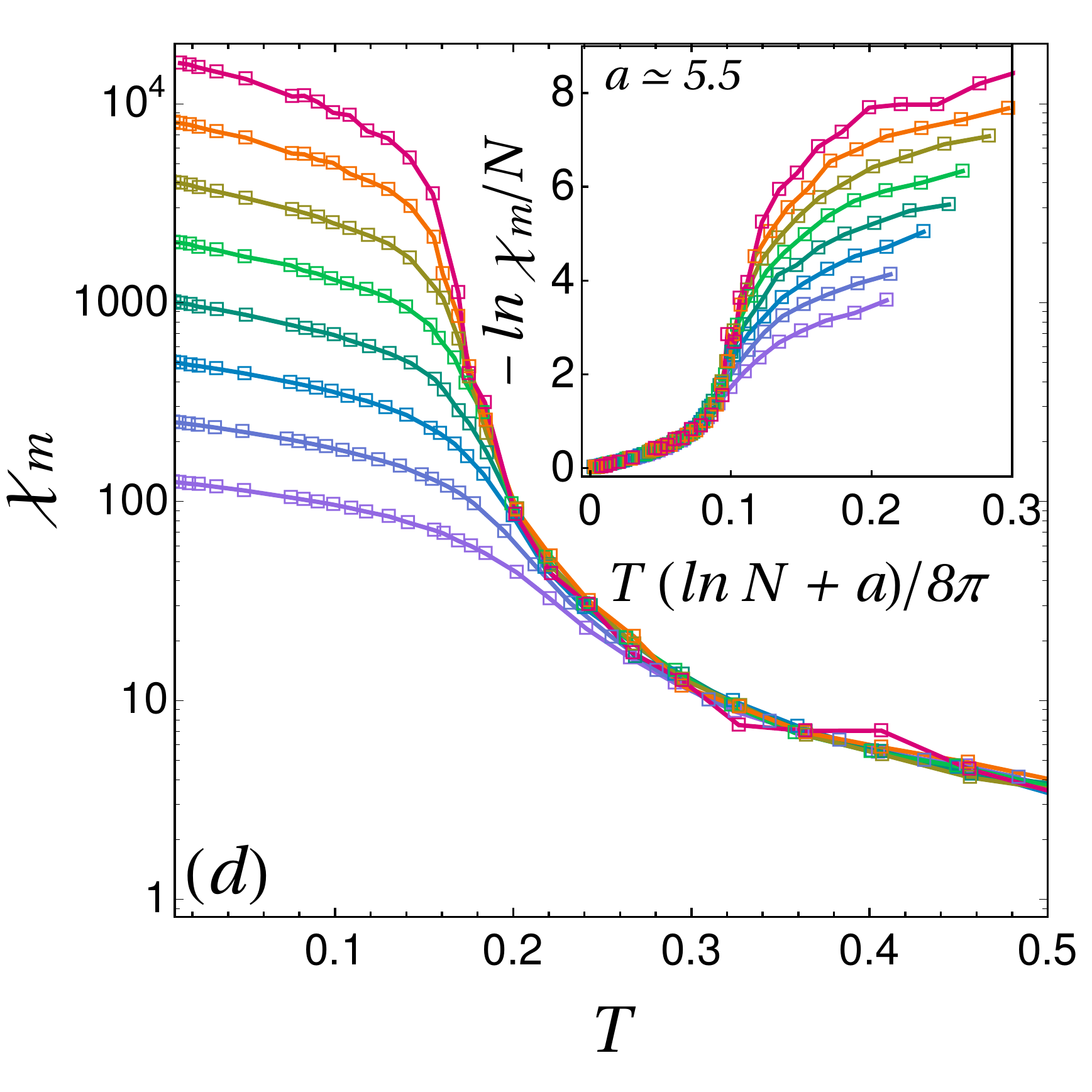}
\caption{{\bf{Magnetic Properties at Low Temperatures.}}
$(a)$ Magnetization modulus and temperature against the energy per particle for $N = 8192$, $(b)$ typical snapshot of a system of $N = 8192$ particles in the spin-wave regime $(c)$ magnetization modulus against temperature for growing system sizes, and $(d)$ magnetic susceptibility (in log scale) against temperature for growing system sizes, all at $\phi = 0.55$. 
In the inset of $(a)$, we report the measured mean temperature against the energy per particle.
Black arrows indicate where the snapshot shown in $(b)$ lies in phase space.
In the inset of $(b)$, we show the colour-code used for spins to represent their orientation.
In the insets of $(c)$ and $(d)$, we report the best collapses of $m$ and $\chi_m$, respectively, near $T = 0$ using the spin-wave scaling described in the main text. 
\label{fig:MagProp}}
\end{figure}
 
Curves obtained for $N = 8192$ particles are showcased in Fig.~\ref{fig:MagProp}, which displays $m(E)$ and $T(E)$ curves obtained by MD simulations after averaging over $10^2$-$10^3$ configurations for each point in panel $(a)$ and its inset.
At low energies and temperatures, we observe a non-zero magnetization, that crosses over to values close to zero at a finite energy and temperature.
In panel $(b)$, we show a typical snapshot of the configuration of spins in the high-magnetization regime for $N = 8192$ particles. 
In this snapshots as well as all the others in this section, spins are colour-coded depending on their direction, using the hue variable of the HSL colour code,~\cite{Hanbury2008} meaning that $\theta = 0$ is coded by pure red, $\theta = 2\pi / 3$ by pure green, and $\theta = 4\pi/3$ by pure blue, and that every intermediate colour is a linear interpolation between the nearest two primary colours.
This pictures shows that in the low-temperature regime $T\to 0$, the magnetization is suppressed by harmonic spin waves.
This is similar to the very low-temperature regime of the on-lattice XY model,~\cite{Bramwell1994} described by a massless theory.~\cite{Amit1980a, Nelson1977}

In panel $(c)$, we show the $m(T)$ curves obtained by combining $m(E)$ and $T(E)$ curves such as those shown panel $(a)$, this time varying the size of the system between $N = 128$ and $N = 16384$ at fixed packing fraction.
We observe that the magnetization is suppressed at all temperatures, and that the finite-size crossover from high- to low-magnetization is pushed to lower values of the temperature when increasing the system size, as expected from a model at its lower critical dimension.
In the inset of $(c)$, we check that the low-temperature part of the magnetization curves follows the spin-wave scaling given in Eq.~\ref{eq:SWscaling}.

In panel $(d)$, we show the corresponding $\chi_m(T)$ curves, with a logarithmic vertical scale.
At low temperatures, this susceptibility grows as the system size increases, and seems to saturate, as expected in the spin-wave regime where the system is smaller than its correlation length.
In the inset of $(d)$, we plot $\chi_m / N$ against the same rescaled temperature as in the inset of $(c)$, and thereby show that the zero-temperature susceptibility is extensive. 
This is a check that, as predicted by spin-wave calculations,~\cite{Tobochnik1979} the magnetic properties of the system at low temperature are controlled by a Gaussian fixed point at $T=0$, so that $\eta\to 0$ as $T\to 0$.

Let us now discuss the nature of the finite-size crossover between the low- and high-magnetization regimes.
In order to do so, let us define a modified susceptibility that features a maximum at a finite temperature,~\cite{Archambault1997} defined through the fluctuations of the modulus of the magnetization,
\begin{equation}
    \chi_{|m|} = \frac{1}{N} \left(\langle \left|\bm{M}\right|^2\rangle - \langle \left|\bm{M}\right| \rangle^2\right).
\end{equation}

\begin{figure}
\centering
\includegraphics[width=.48\columnwidth,height=.48\columnwidth]{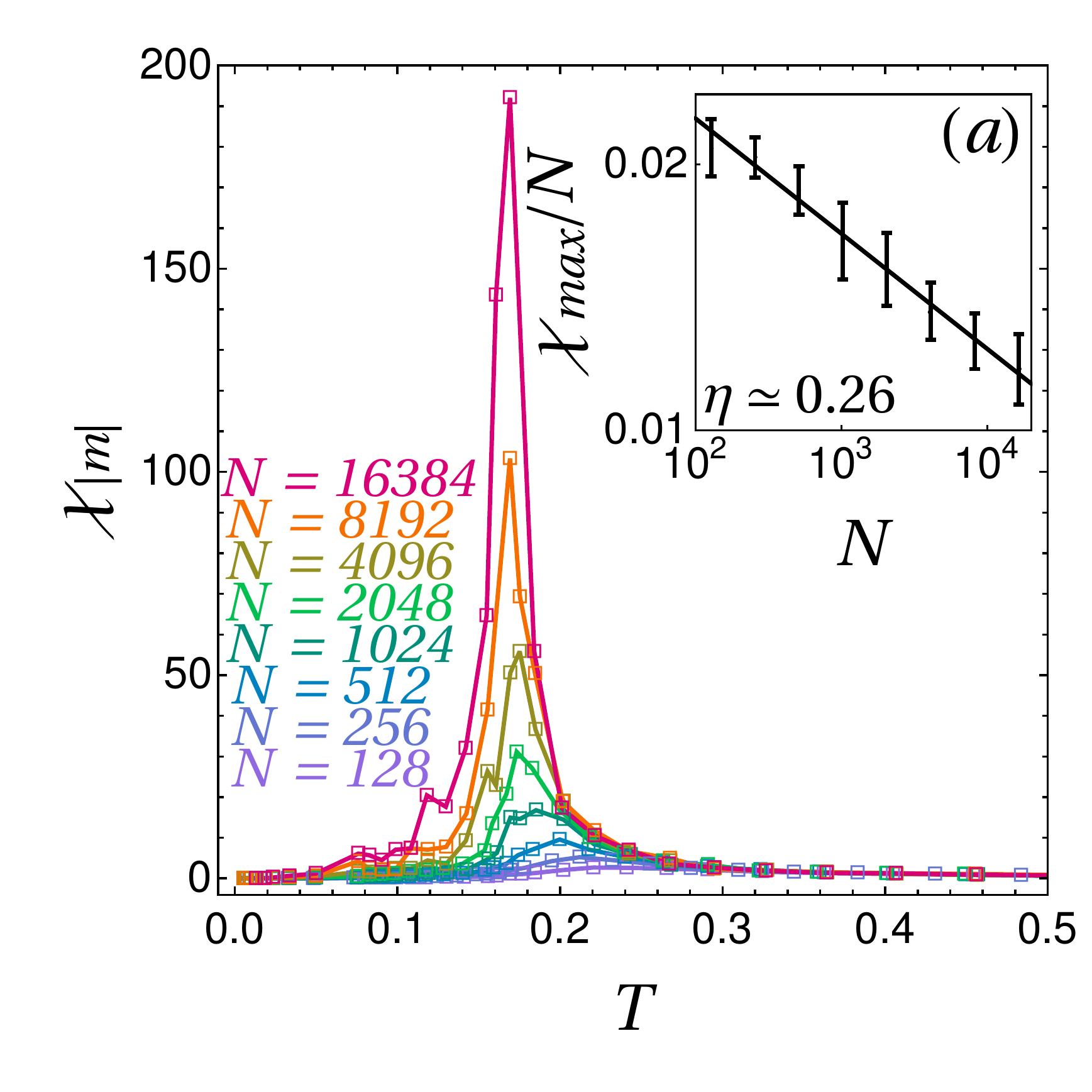} 
\includegraphics[width=.48\columnwidth,height=.48\columnwidth]{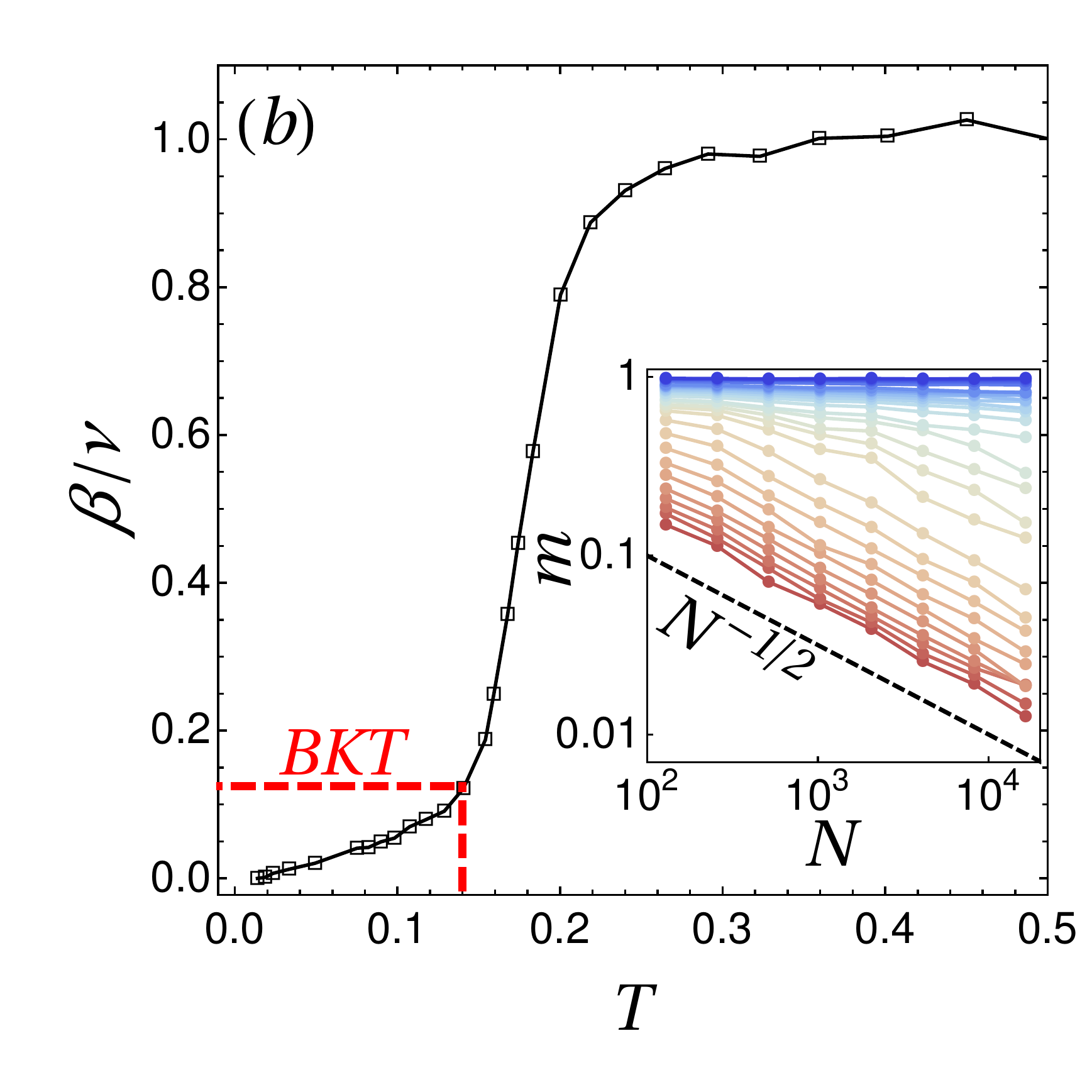}\\
\includegraphics[width=.48\columnwidth,height=.48\columnwidth]{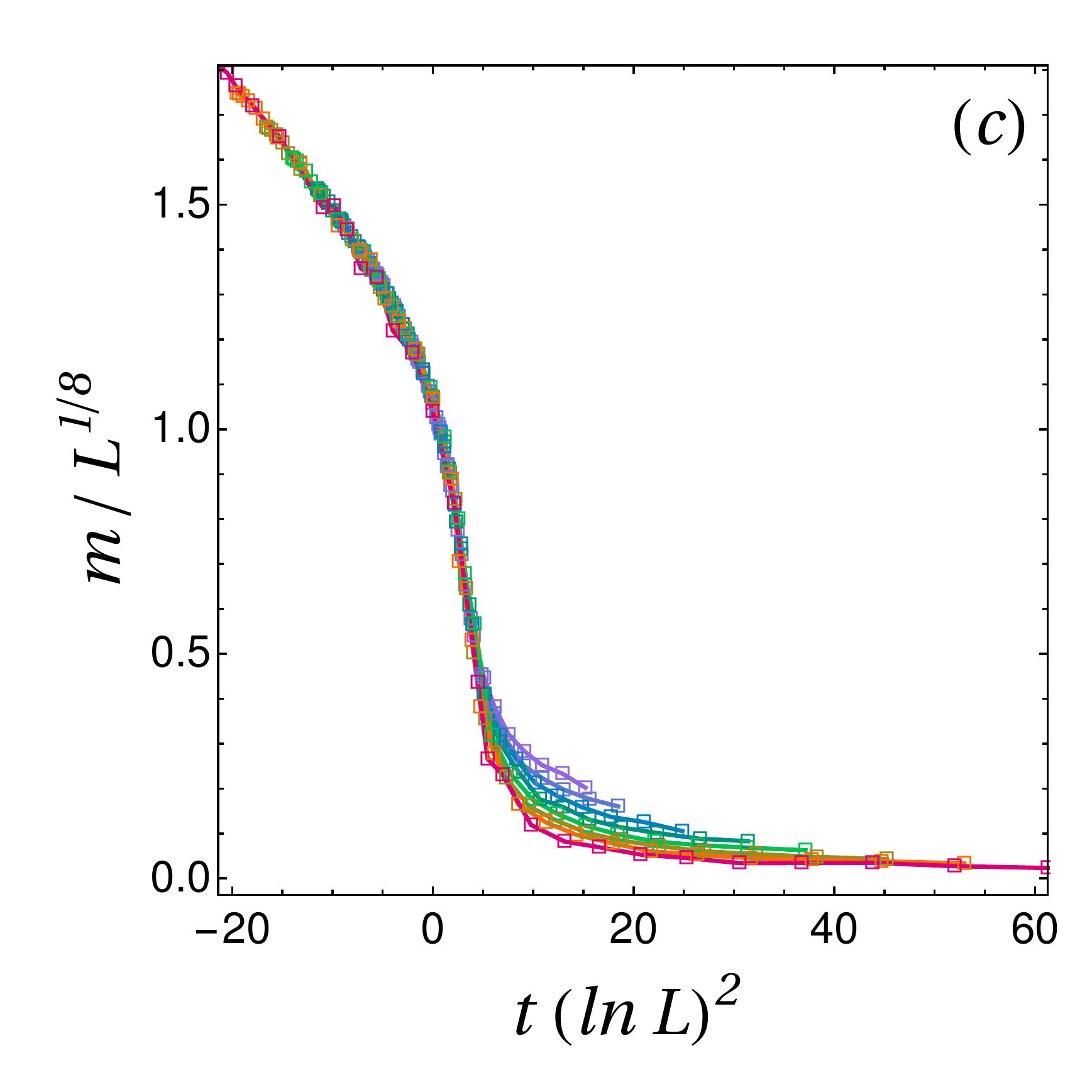}
\includegraphics[width=.48\columnwidth,height=.48\columnwidth]{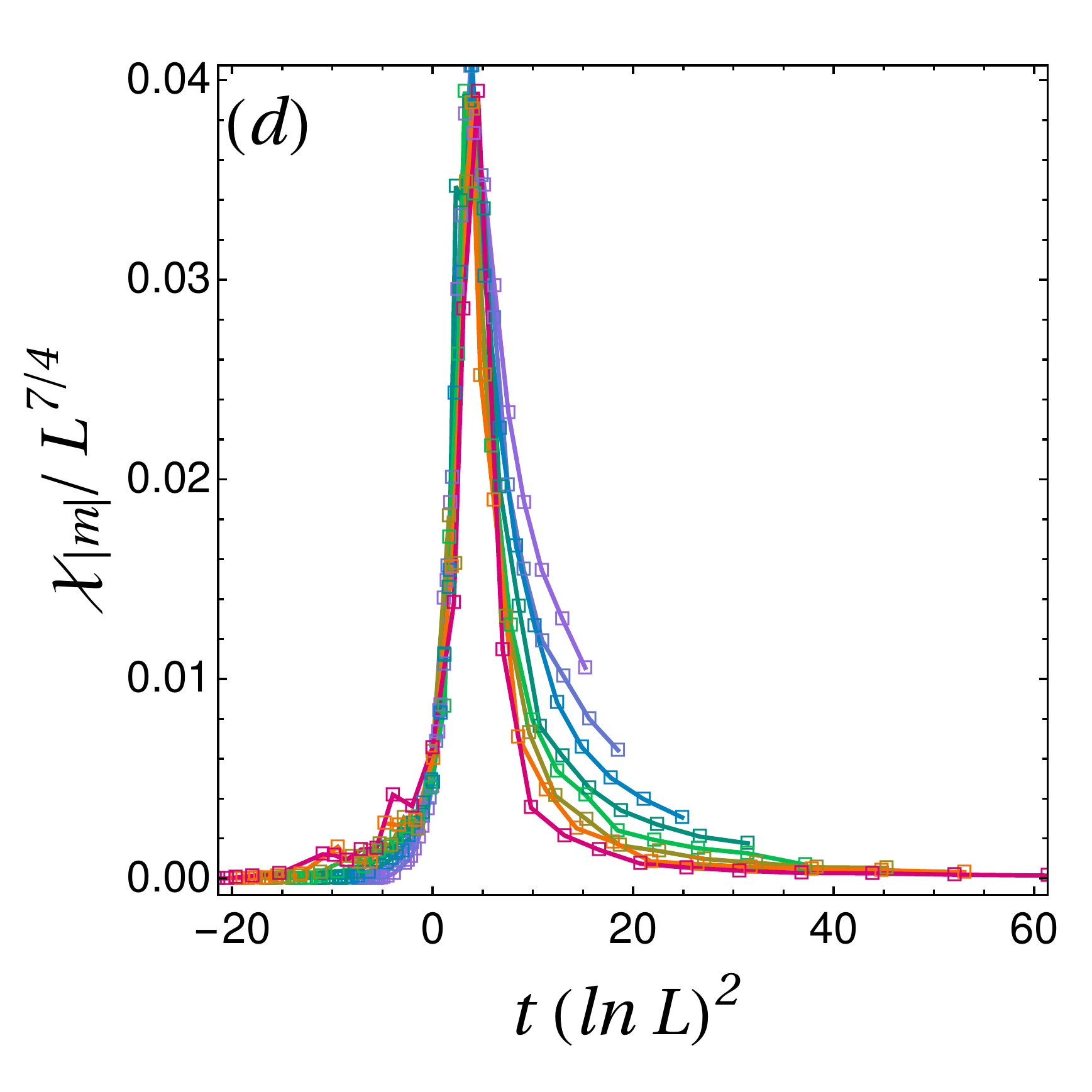} \\
\includegraphics[width=.48\columnwidth,height=.48\columnwidth]{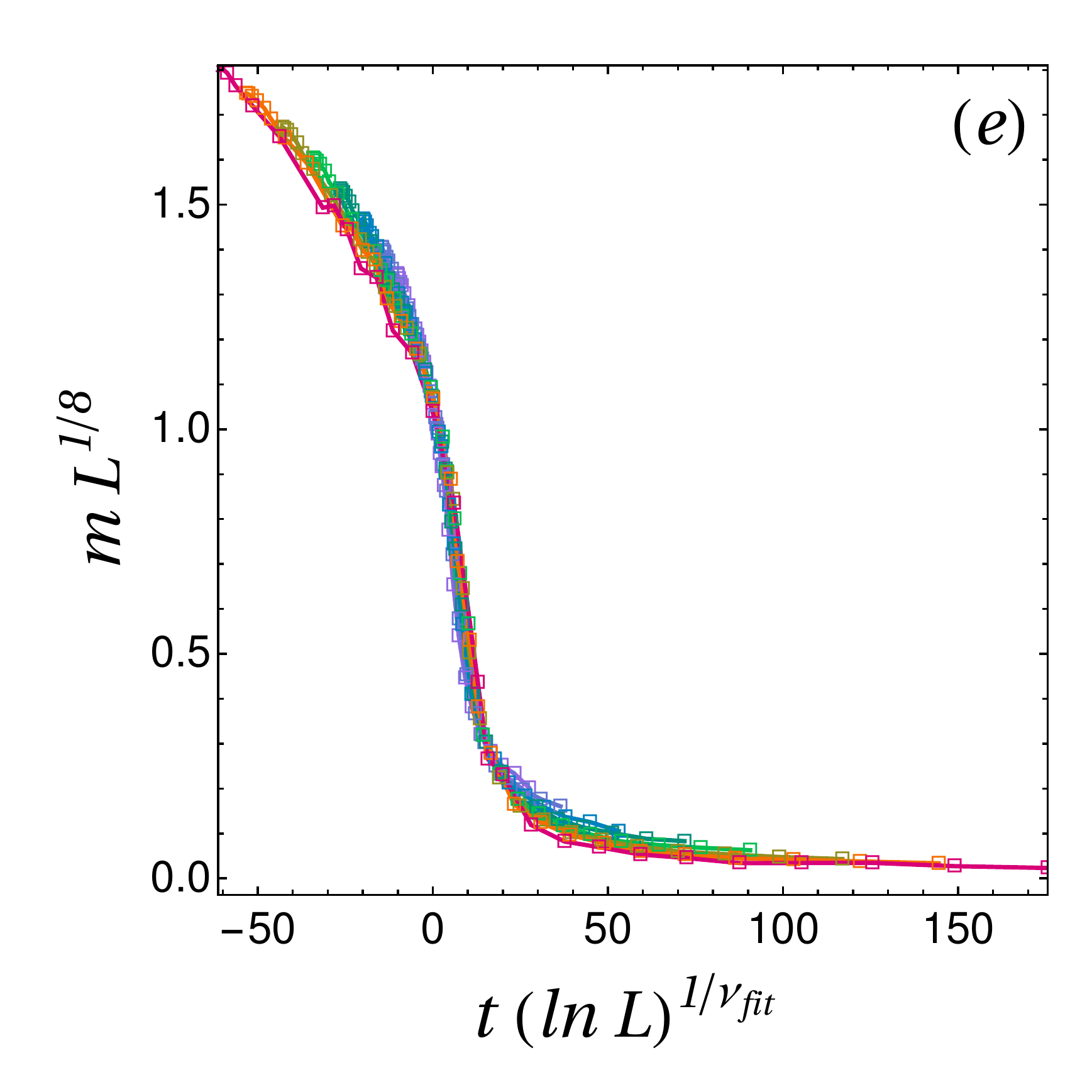}
\includegraphics[width=.48\columnwidth,height=.48\columnwidth]{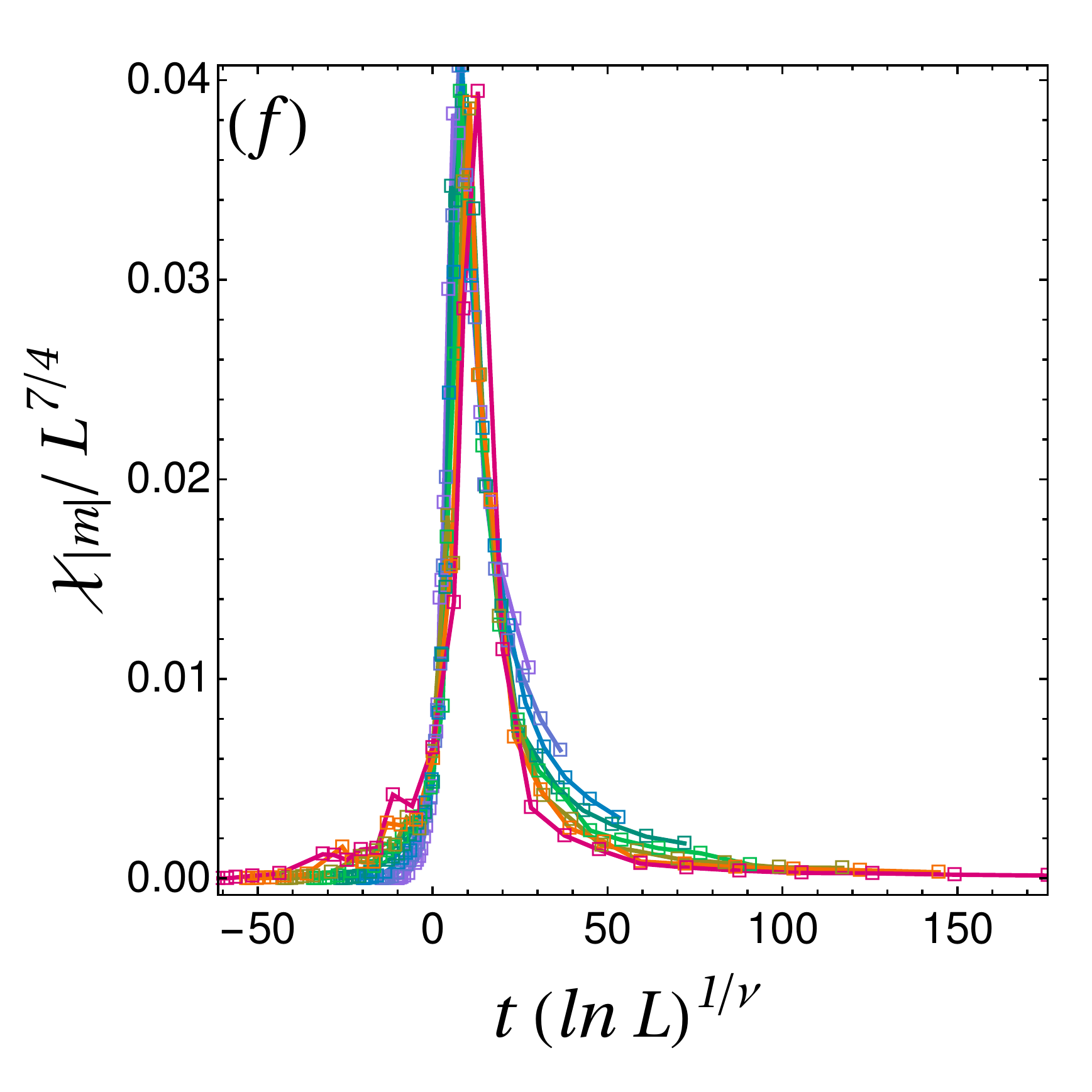}\\
\includegraphics[width=.48\columnwidth,height=.48\columnwidth]{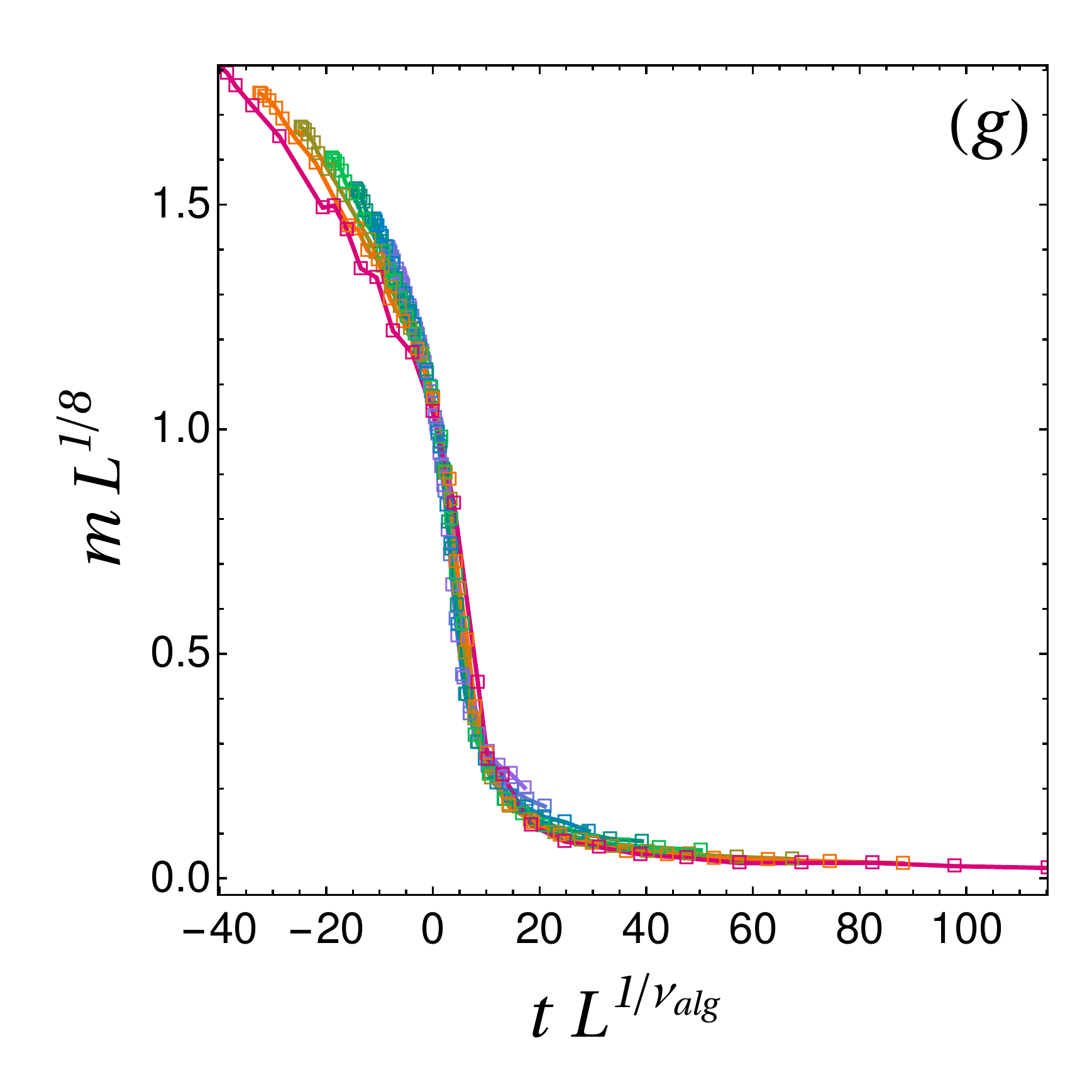}
\includegraphics[width=.48\columnwidth,height=.48\columnwidth]{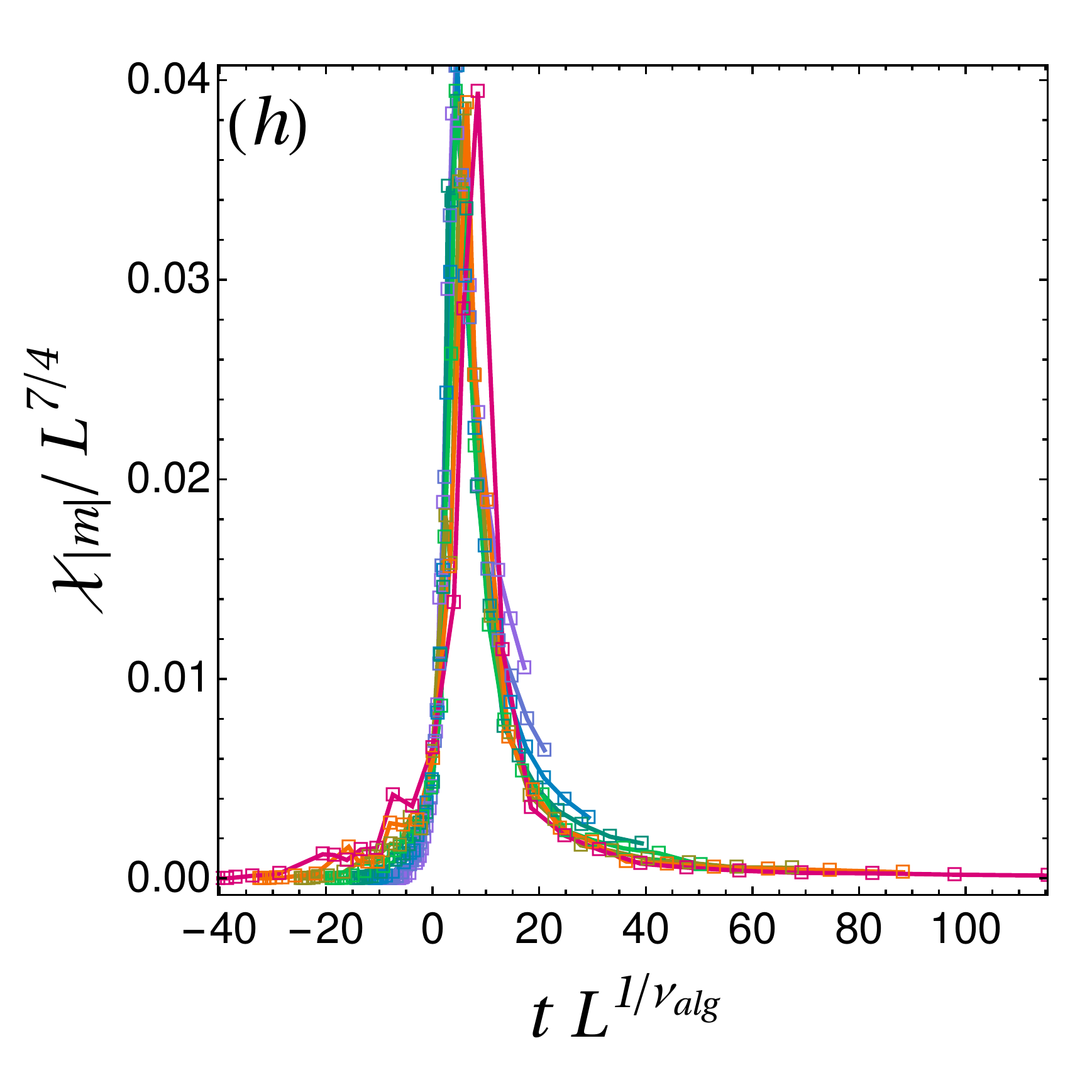}
\caption{{\bf{Finite-Size Scalings Near the Crossover.}}
$(a)$ Susceptibilities of the modulus of the magnetization for different system sizes, $(b)$ algebraic scaling exponent of the modulus of the magnetization with size, $(c)-(d)$ logarithmic rescaling of the magnetization and modulus susceptibility curves with BKT exponents, $(e)-(f)$ best logarithmic rescaling of the magnetization and modulus susceptibility with a non-BKT $\nu \approx 0.372$, $(g)-(h)$ best algebraic rescaling of the magnetization and modulus susceptibility with $\nu_{alg} \approx 1.3$.
In the inset of $(a)$, we report the measured maxima of $\chi_{|m|}/N$ versus $N$ in log-log representation, with error bars showing a 5\% confidence interval around each value, and the best power law fit of it, corresponding to an exponent $\eta \approx 0.26$.
The red line in $(b)$ indicates the location of the expected BKT value of $\beta/\nu = 1/8$ at $T_{BKT}$, which is here very close to the apparent transition temperature.
In the inset of $(b)$, we report $m(N)$ in log-log scales at various temperatures, going from blue (low temperatures) to red (high temperatures). 
The dashed black line shows the $N^{-1/2}$ limit observed at high temperatures. 
In $(c)-(h)$, $t = T/T_{KT}^\star - 1$ is the reduced temperature, where $T_{KT}^\star$ is the putative BKT temperature found in $(b)$.
\label{fig:BKTFSS}}
\end{figure}

This modified susceptibility is plotted against the temperature for different system sizes in Fig.~\ref{fig:BKTFSS}$(a)$.
As expected, it features a maximum at a finite temperature that decreases with the system size, reflecting the crossover that $m$ undergoes.
This maximum, $\chi_{max}$, grows higher and sharper as the system size grows.
As shown in the inset of Fig.~\ref{fig:BKTFSS}$(a)$, where we plot $\chi_{max}/N$ against the system size in log-log scale, we measure $\chi_{max} \propto L^{2 - \eta}$ with $\eta \approx 0.26 \pm 0.01$.
This value is reminiscent of the BKT exponent $\eta = 0.25$, although it is no proof of BKT behaviour alone.

To investigate further, in Fig.~\ref{fig:BKTFSS}$(b)$, we plot the exponent $\beta/\nu$ obtained at each temperature by using the scaling law $m(L,T) \propto L^{-\beta/\nu}$.
The corresponding log-log plot of $m$ against the system size is shown in the inset of this panel.
We find that $\beta/\nu$ smoothly goes from $0$ at zero-temperature (where $m = 1$ regardless of the system size) to $1$ at high temperatures (where the magnetization is simply a sum of independent random variables).
The value $\eta\approx 0.26$, if it corresponds to a BKT-like critical point, can be associated to the critical value of $\beta/\nu$ using the hyperscaling law $2 \beta/\nu = \eta$.
We can therefore evaluate a candidate value for $T_{KT}$ from Fig.~\ref{fig:BKTFSS}$(b)$.
Following this strategy, we find $T_{KT}^\star \approx 0.14$, and we define the reduced temperature $t = t/T_{KT}^\star - 1$.
Moreover, recalling the on-lattice spin-wave scaling given in Eq.~(\ref{eq:SWscaling}),
at low temperatures, we expect $\beta/\nu = T/(4\pi J_{sw}).$
As a result, the initial slope of $\beta/\nu (T)$ enables us to estimate the effective coupling felt by the low-temperature spin waves.
We find $J_{sw} \approx 0.13$.
Interestingly, in the BKT scenario, the critical temperature seems to follow the RG prediction $T_{KT} \approx 1.35 J$.~\cite{Archambault1997,Janke1991}
Here, however, we find that $T^\star_{KT} \approx 0.14$ is significantly smaller than $1.35*J_{sw} \approx 0.19$, which is a first clue that the present system might not follow the BKT scenario.

Having determined the candidate values of the temperature $T_{KT}^\star$, and of the exponents $\eta$ and $\beta/\nu$, we still have to check whether the observed value of $\nu$ is compatible with BKT physics.
In order to do so, we try three different strategies to rescale the magnetization and susceptibility curves.
\begin{itemize}
    \item[(i)] Recalling the BKT scaling of the correlation length, $\ln\xi \sim t^{-1/2}$, we rescale the reduced temperature by multiplying it by $(\ln L)^2$, and plot the rescaled magnetization $m L^{1/8}$ and susceptibility $\chi_{|m|} / L^{7/4}$ in Fig.~\ref{fig:BKTFSS}$(c)-(d)$.
    \item[(ii)] Assuming that the system does not follow BKT scalings but another similar scaling $\ln\xi \sim t^{-\nu}$, we determine the value of $\nu$ such that the curves of $m L^{1/8}$ and $\chi_{|m|} / L^{7/4}$ against $t (\ln L)^{1/\nu}$ collapse best near the crossover. The resulting curves, obtained for $\nu = 0.37$ (below the BKT value $\nu = 1/2$), are shown in Fig.~\ref{fig:BKTFSS}$(e)-(f)$.
    \item[(iii)] Assuming that, for the range of sizes explored in this paper, the crossover can be described by an effective algebraic rescaling of the reduced temperature, we seek the value $\nu_{alg}$ that leads to the best collapse of $m L^{1/8}$ and $\chi_{|m|} / L^{7/4}$ against $t L^{1/\nu_{alg}}$ near the crossover. The resulting curves, shown in Fig.~\ref{fig:BKTFSS}$(g)-(h)$, are obtained for $\nu_{alg} = 1.3$.
\end{itemize}

We find that out of these three strategies, the BKT rescaling yields the poorest collapse at temperatures near, but above the crossover.
As in the case of an on-lattice XY model the susceptibility is known to be very close to the exact RG predictions,~\cite{Archambault1997} this is a sign that the finite-size crossover of the magnetization might, in fact not follow the BKT scenario here.
However, a thorough proof of this result using only finite size scalings would be numerically very tedious, as several decades of $\log N$ would be required for a precise determination of the value of $\nu$ in the hypothesis $\ln\xi \sim t^{-\nu}$.
A recent example of these difficulties is the study of the closely related problem of melting of hard disks in $2d$, which was shown (after a 50-year long controversy) to follow a two-step melting scenario.~\cite{Bernard2011}

\begin{figure}[b!]
\centering
\includegraphics[width=.48\columnwidth,height=.48\columnwidth]{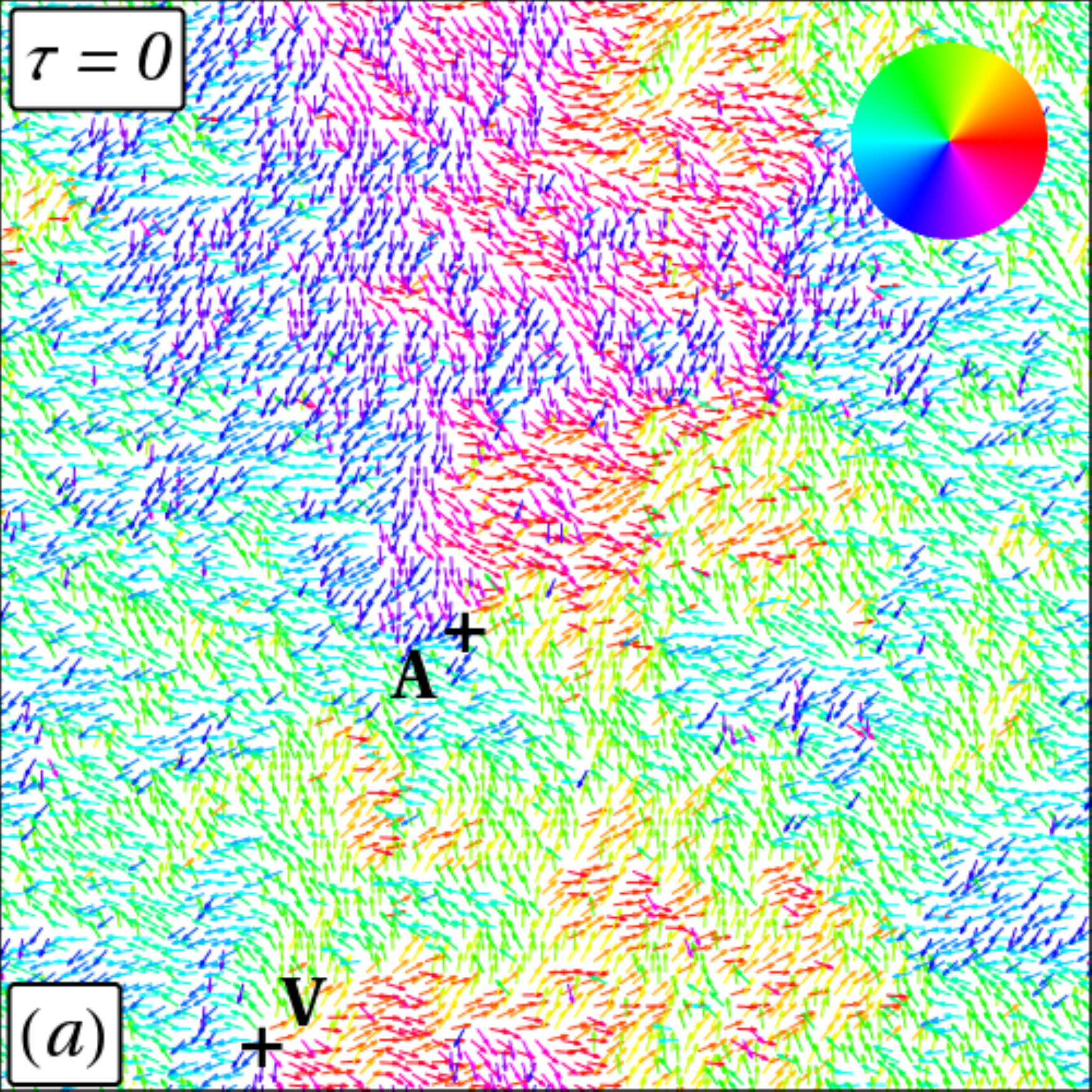} 
\includegraphics[width=.48\columnwidth,height=.48\columnwidth]{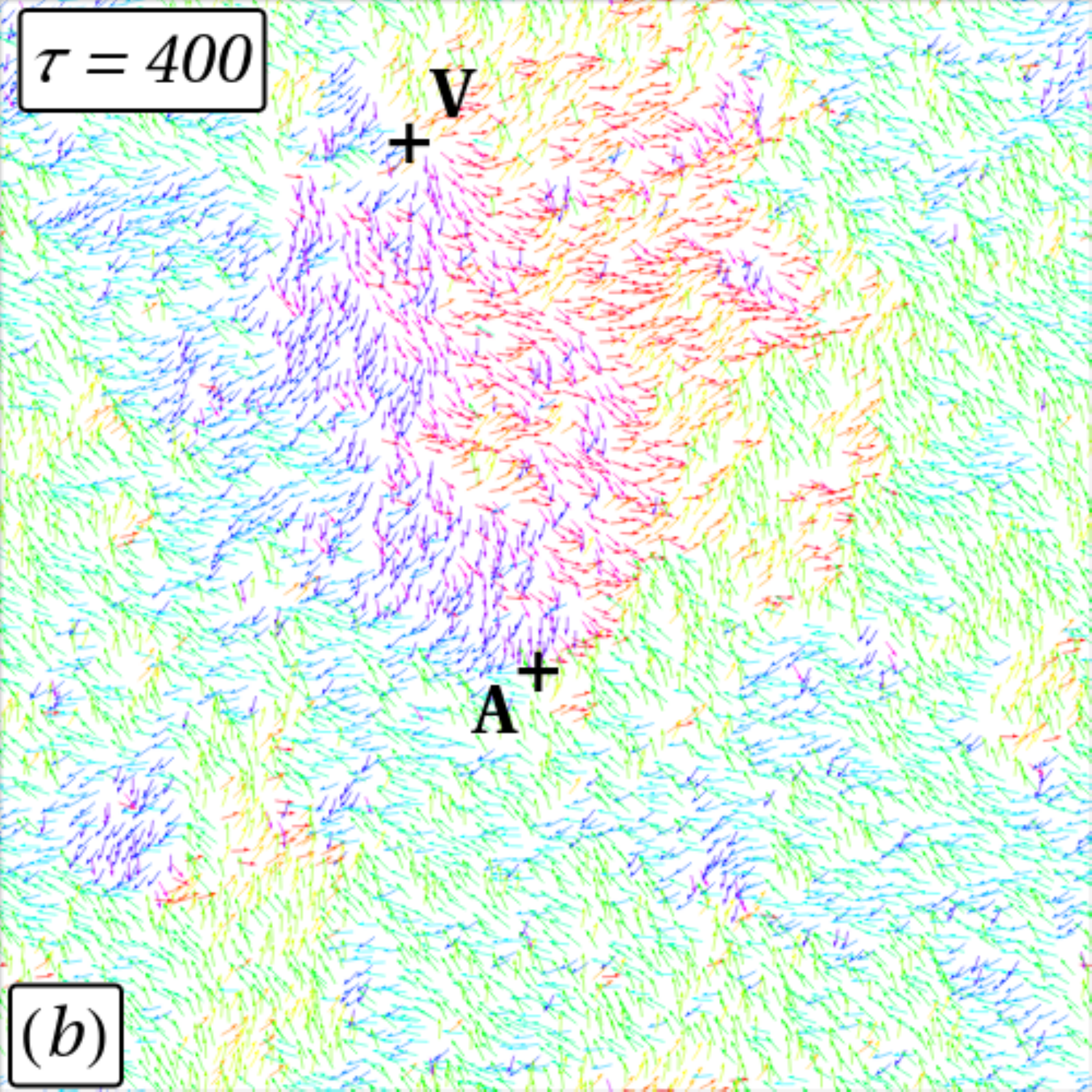}\\
\includegraphics[width=.48\columnwidth,height=.48\columnwidth]{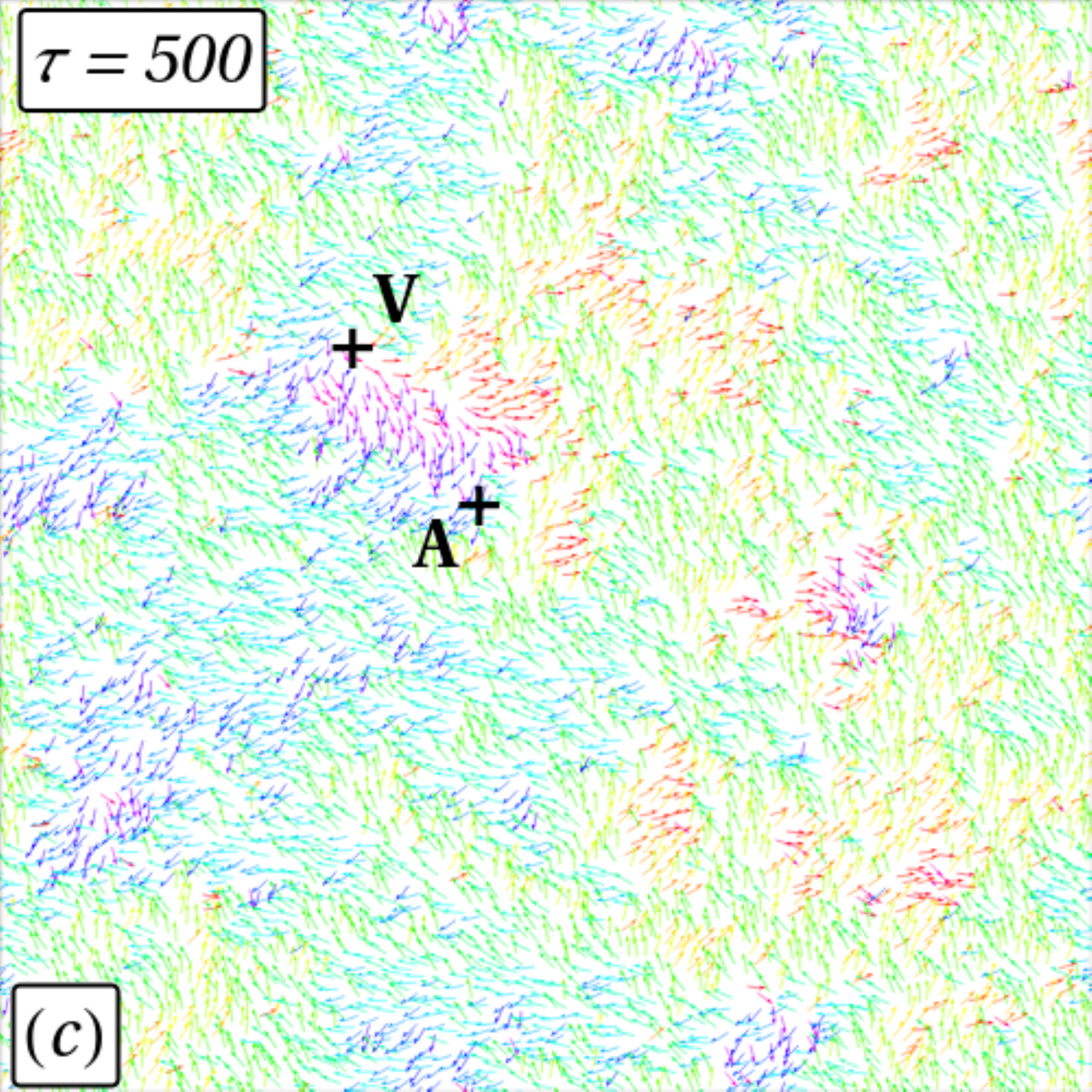}
\includegraphics[width=.48\columnwidth,height=.48\columnwidth]{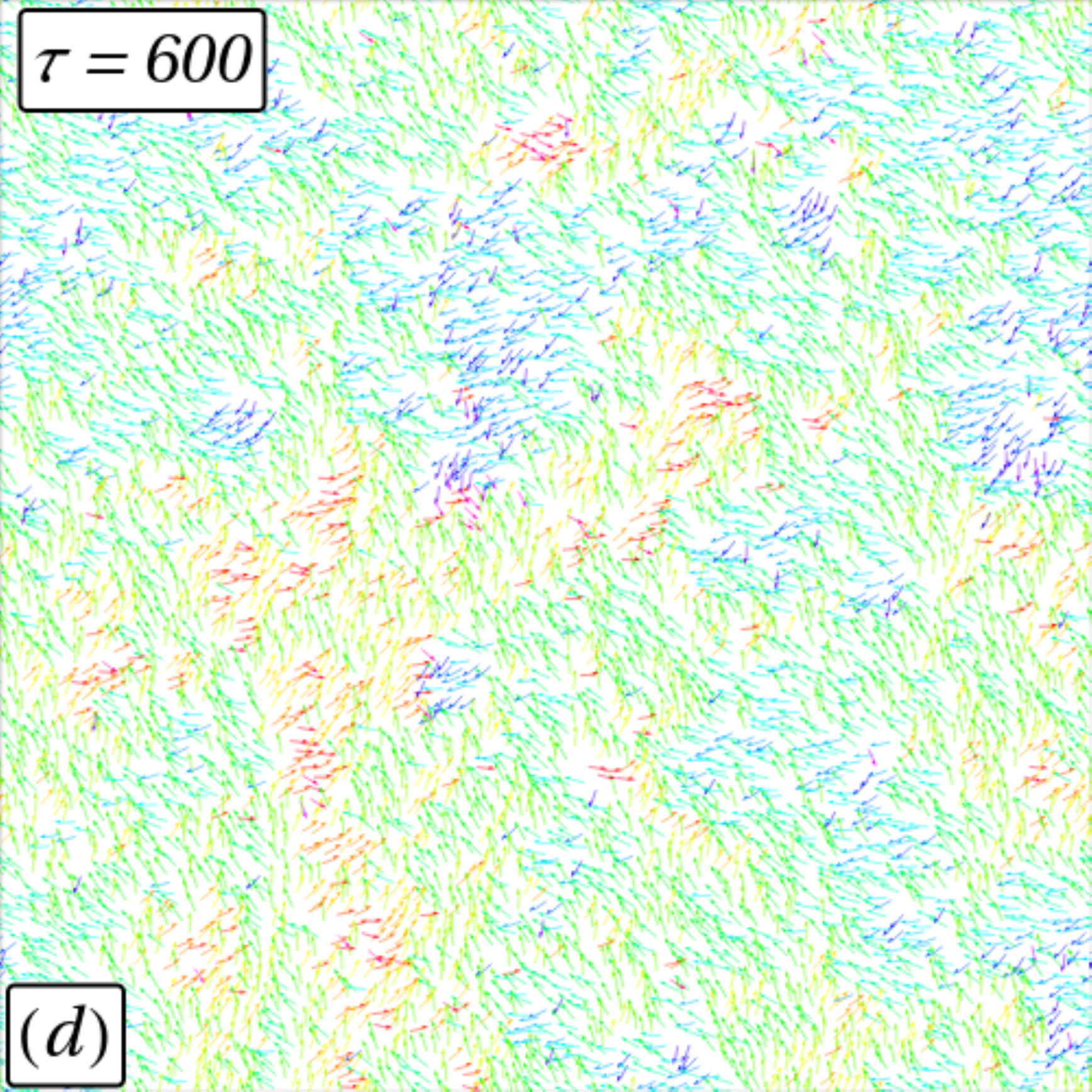}
\caption{{\bf{Dynamics starting with free vortices in the initial condition.}}
We run the dynamics of the spin fluid, starting from the initial condition $(a)$, that was prepared by equilibrating 8192 spins pinned on a triangular lattice at $T \approx 0.15$. In $(b)-(d)$, we show snapshots taken at times $\tau = 400$, $500$ and $600$.
The position of the two free point defects is indicated by a cross, as well as a letter reflecting its nature: V for vortex and A for antivortex.
\label{fig:VortexFree}}
\end{figure}

That is why we now focus on another aspect of the on-lattice BKT phenomenology, the unbinding of pairs of topological defects at the crossover.
In order to do so, we simulate a system of $N = 8192$ particles, interacting through the same $J(r)$ and placed in a box with the same linear length as heretofore, but with particles pinned  on the nodes of  a regular triangular lattice.
We then cool it down using MD simulations, and taking the same annealing rate as before, but without updating the particles' positions.
As expected in an on-lattice setting because the BKT scenario holds there, at temperatures close to but above the finite-size crossover temperature, we observe free vortices and antivortices in the system.
A typical example is shown in Fig.\ref{fig:VortexFree}$(a)$, which was obtained at a temperature $T \approx 0.15$, and where we highlight with black crosses and letters an unbound vortex (V) - antivortex (A) pair.
In order to check the stability of this structure in the off-lattice setting, we use this configuration as an initial condition for particle positions, spins and rotational velocities, draw their velocity components from Gaussian distributions with variance $T$, and let the dynamics run.
As shown in Fig.\ref{fig:VortexFree}$(b)-(d)$, in which we show snapshots separated from the initial condition by a time $\tau$ each time indicated in the top left corner, this leads to the annihilation of the topological defects when they meet.

The annihilation of free topological defects suggests that this kind of defects is made unstable by the coupling to fluid motion and attracto-repulsion, so that this system does not follow the usual BKT scenario of vortex unbinding at finite temperatures.
This suggestion is made stronger by the fact that free vortices similar to those shown in Fig.\ref{fig:VortexFree}$(a)$ are {\it{never}} clearly observed at any step during the annealing, even for sizes and temperatures at which they have been reported~\cite{Tobochnik1979} and should be expected if the BKT scenario were realized.
Instead, the suppression of magnetization is seemingly still caused by (anharmonic) spin waves.

All in all, the behaviour observed here is reminiscent of the finite size 1d Ising model,~\cite{Rulquin2016} which features an exponentially growing correlation length at low temperatures analogous to spin waves and a finite-size crossover to low magnetization at higher temperatures.
If the analogy holds, the crossover happens because, at the lower critical dimension, the correlation length $\xi$ grows exponentially with temperature and diverges.~\cite{Nelson1977}
As a consequence, finite size systems are in practice always smaller than $\xi$ at a finite temperature, and therefore behave in a mean-field-like way at small enough temperatures.

\subsubsection{Dynamics after a quench}

\begin{figure}
\includegraphics[width=.45\columnwidth]{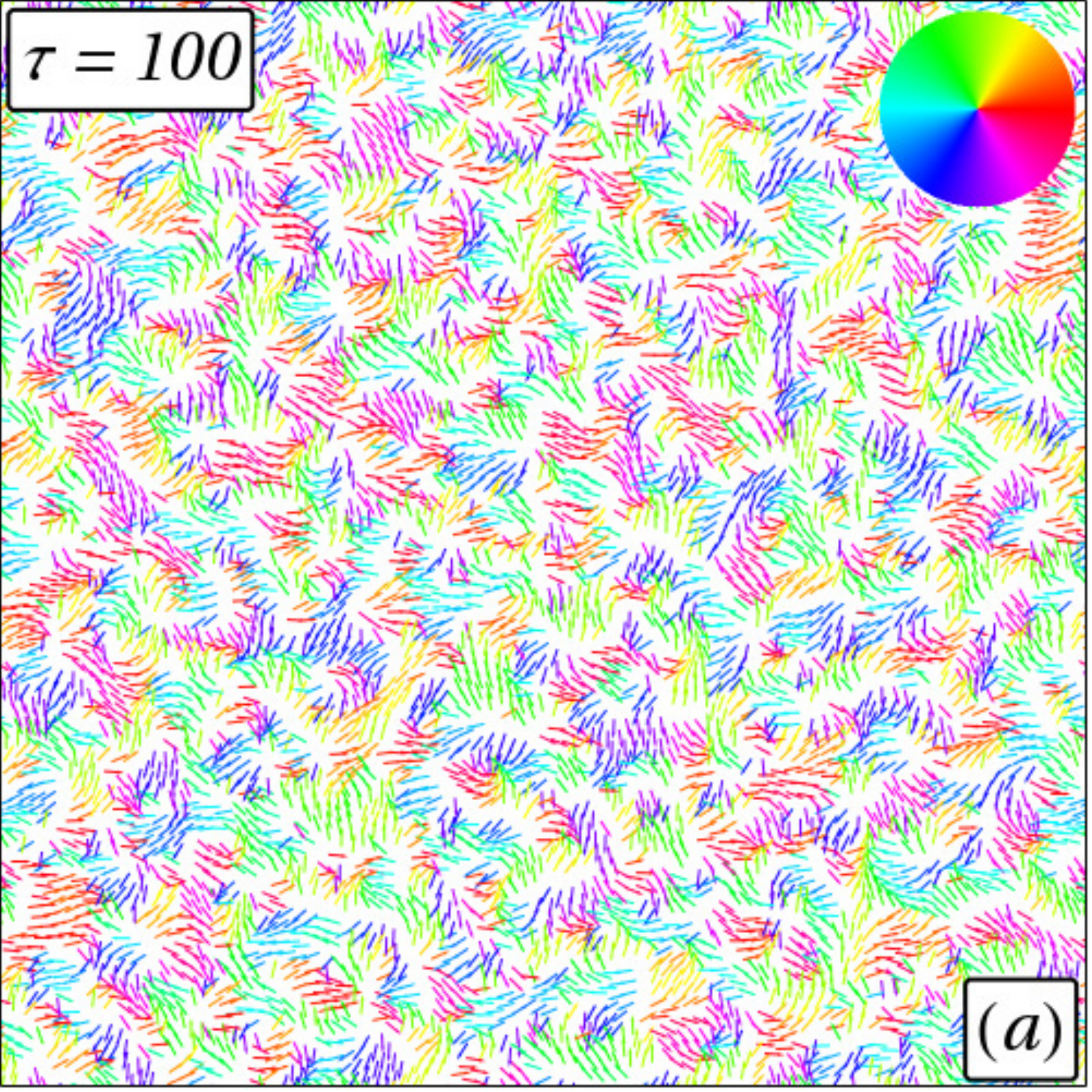}
\includegraphics[width=.45\columnwidth]{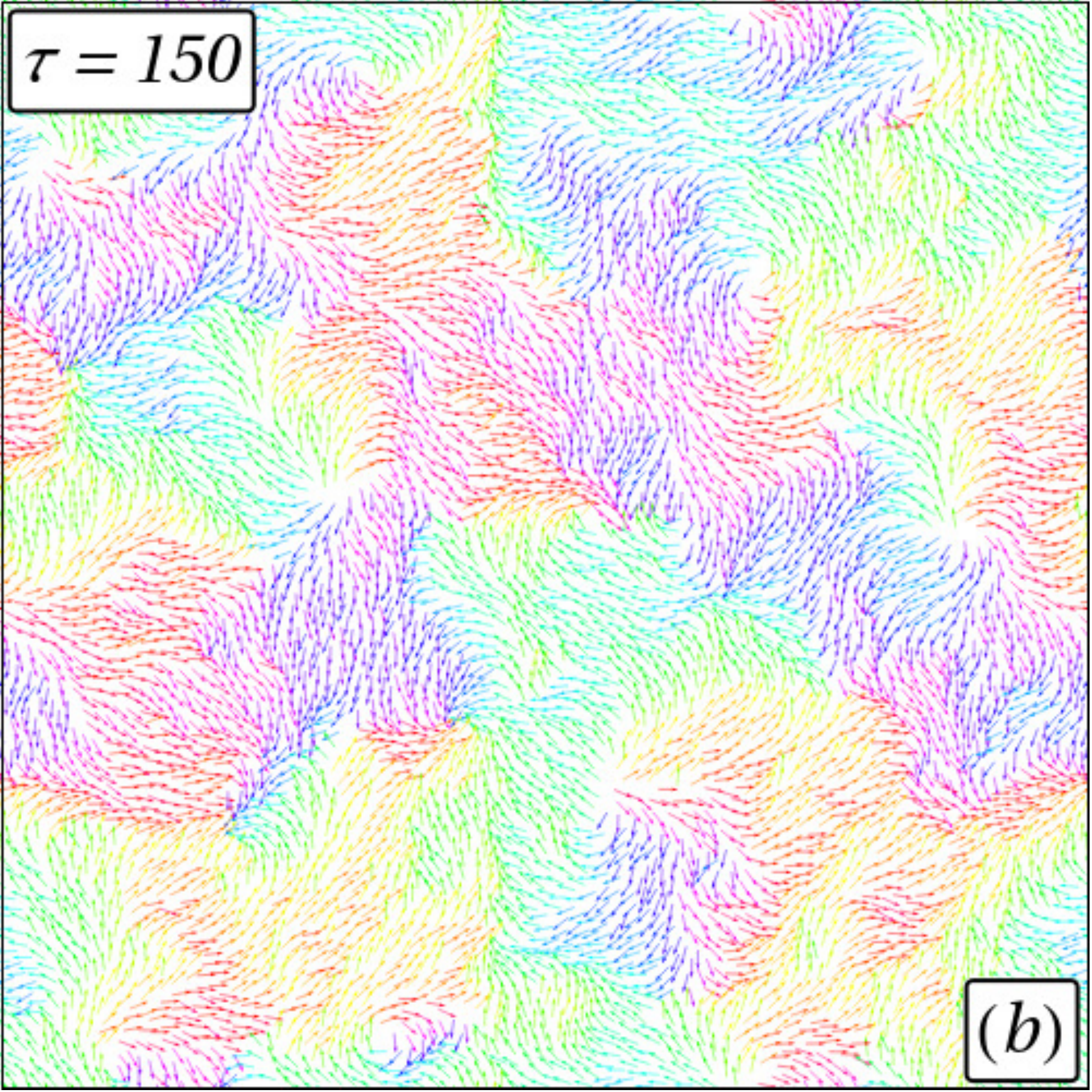}\\
\includegraphics[width=.45\columnwidth]{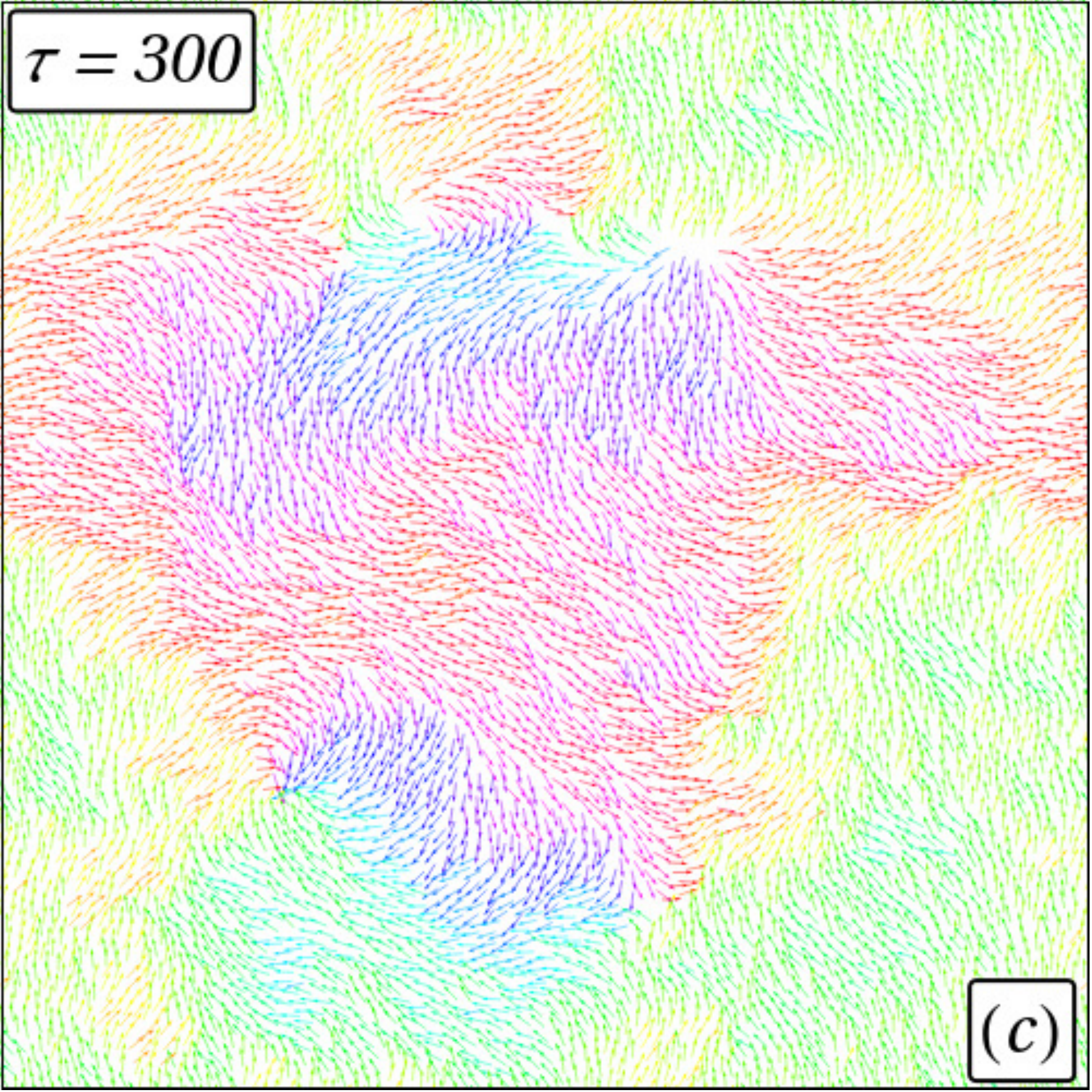}
\includegraphics[width=.45\columnwidth]{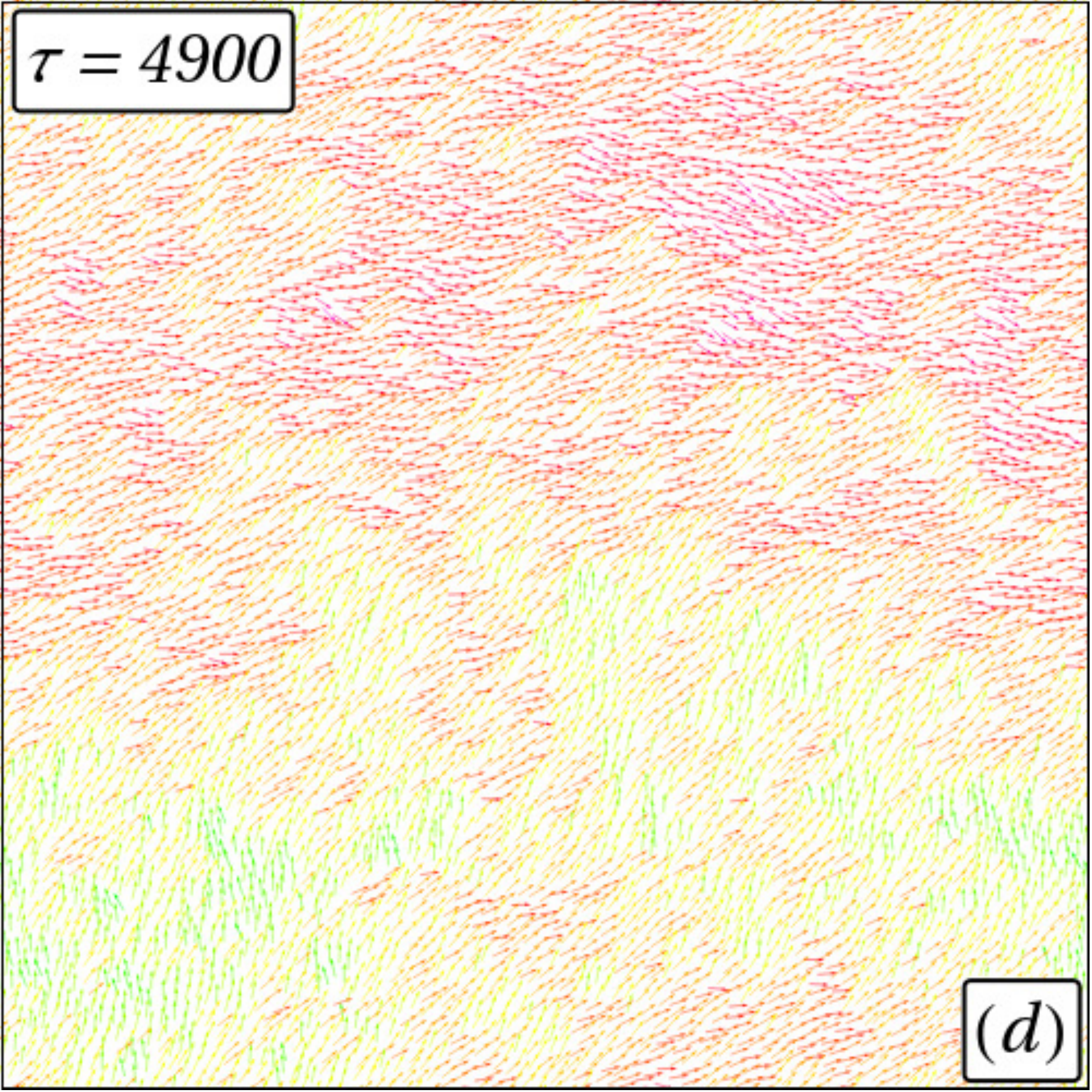}
\caption{{\bf{Magnetic equilibration after a quench.}}
Snapshots of a system composed of 8192 particles, in a square box with periodic boundary conditions, after a quench from an average temperature $T_0 / T_{KT}^\star \approx 15$ to an average temperature $T_Q / T_{KT}^\star \approx 1.7\,10^{-1}$, at 4 different times $\tau$ after the quench, $(a)$ $\tau = 100$ $(b)$ $\tau = 150$ $(c)$ $\tau = 300$ $(d)$ $\tau = 4900$. 
Spins are color-coded on a circle, which is reminded in panel $(a)$.
\label{fig:QuenchSnaps}}
\end{figure}

A final confirmation of the absence of vortices at equilibrium is provided by following the non-equilibrium relaxation after a fast quench to a very low temperature (see Fig.~\ref{fig:QuenchSnaps}).
At short times, rather inhomogeneous states with a lot of vortices develop.
These vortices rapidly annihilate, leading at longer times to the emergence of large domains with homogeneous magnetizations. 
The short lifetime of the vortices, is coherent with the equilibrium results: in the absence of a BKT transition, the vortices created by the non-equilibrium dynamics die out during equilibration. 
This contrasts with the quench dynamics of the square-lattice XY model~\cite{Jelic2011} at very low temperatures, where exceeding vortex-antivortex pairs annihilate rather slowly and a finite density of paired vortices survives.

Altogether we have obtained evidence that our $2d$ XY spin fluids do not belong to the BKT universality class.~\cite{Kosterlitz2016}
This important feature is another manifestation of the fragility of the BKT behaviour against changes in the underlying lattice,~\cite{Baek2007} the interaction symmetry,~\cite{Holm1997} the addition of quenched non-magnetic impurities,~\cite{Wysin2005} or the form of the potential.~\cite{Domany1984,VanEnter2002}
In the present case, the coupling between the spin alignment and the attraction-repulsion between particles makes point defects even costlier than usual.

\subsection{Phase Separation and Domain Growth}

We now concentrate on the stability properties of the fluid phases at different packing fractions and temperatures.
Like previously done for the magnetic properties, one can study them either after a slow annealing or a fast quench, thus focusing on the static equilibrium properties or the equilibration dynamics.
Note that we choose not to discuss the solid phases of the system here.

\subsubsection{Equilibrium properties}

\begin{figure*}
\begin{minipage}{.60\textwidth}
\includegraphics[width=.99\textwidth,height=.99\textwidth]{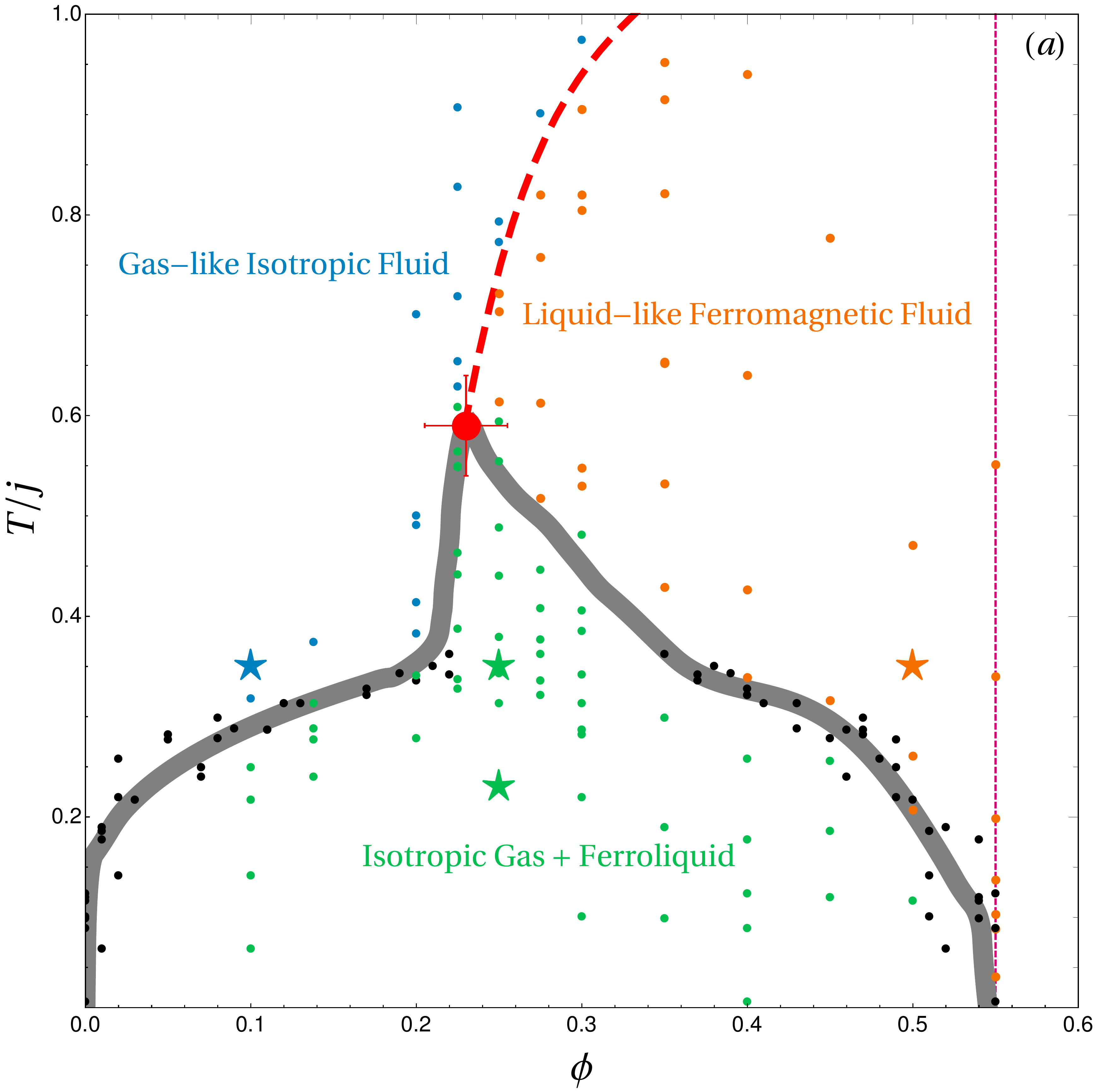} 
\end{minipage}
\vspace{\fill}
\begin{minipage}{.30\textwidth}
\includegraphics[width=.99\textwidth,height=.99\textwidth]{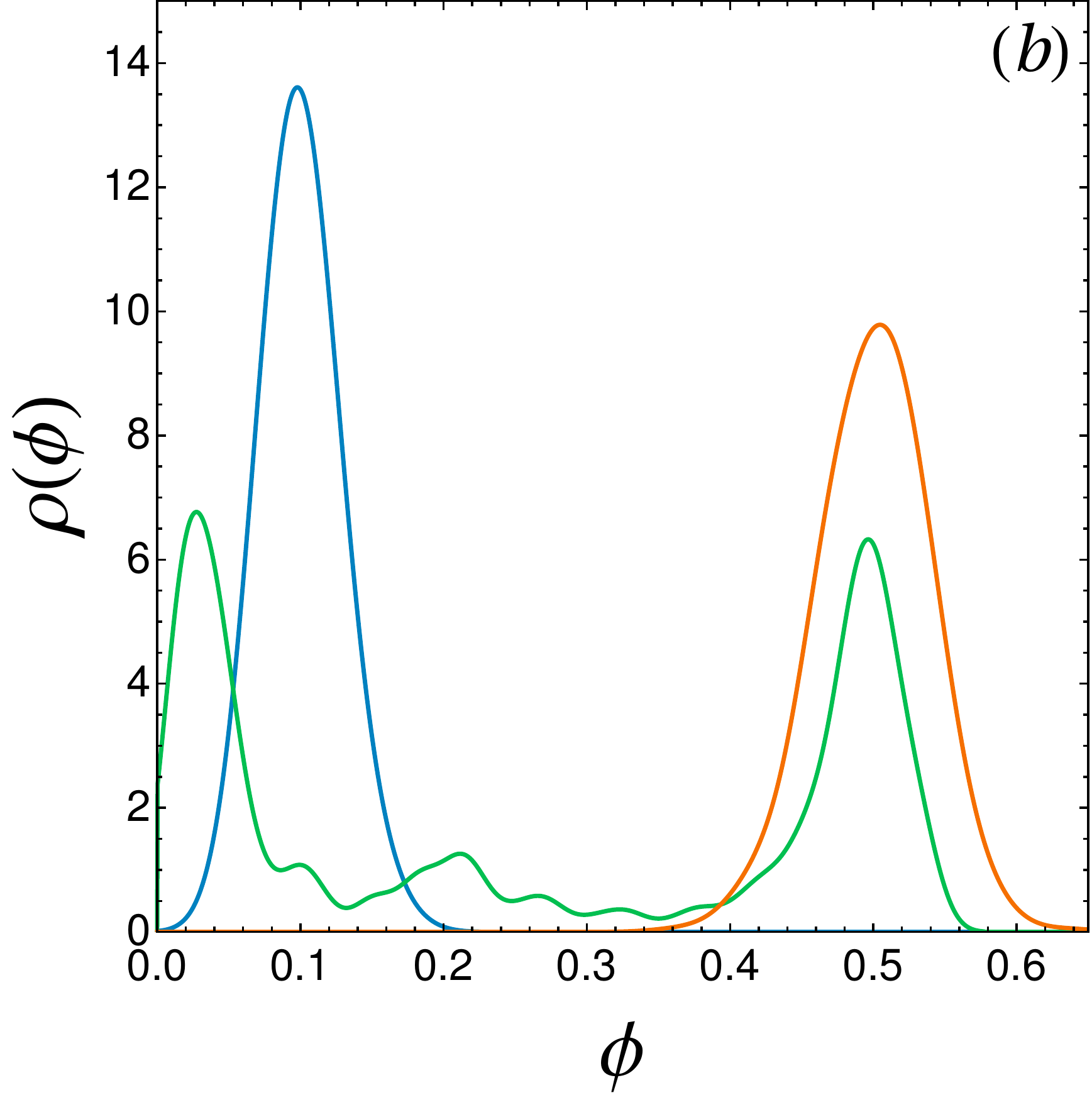} \\
\includegraphics[width=.99\textwidth,height=.99\textwidth]{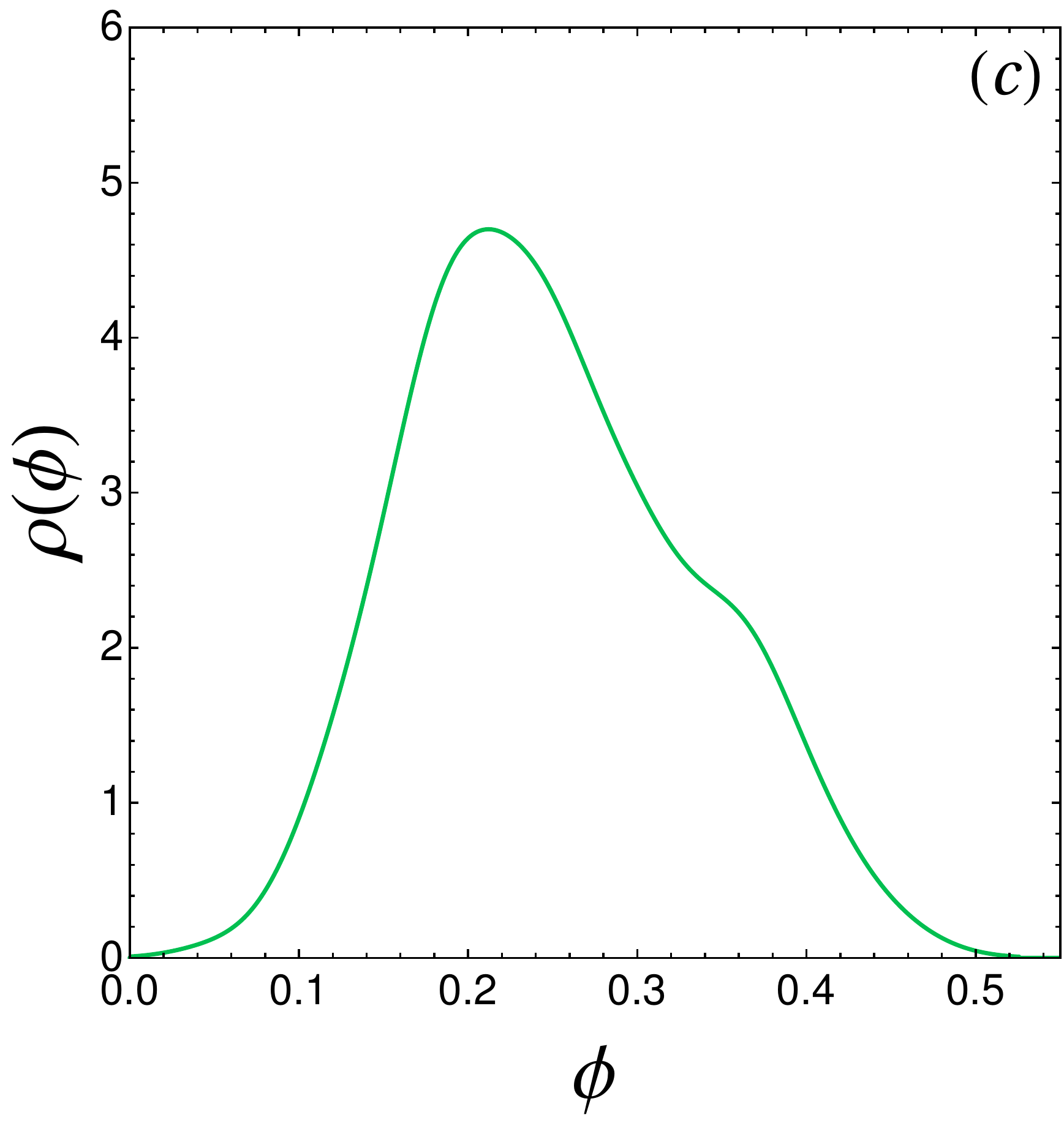}
\end{minipage}
\includegraphics[width=.24\textwidth,height=.24\textwidth]{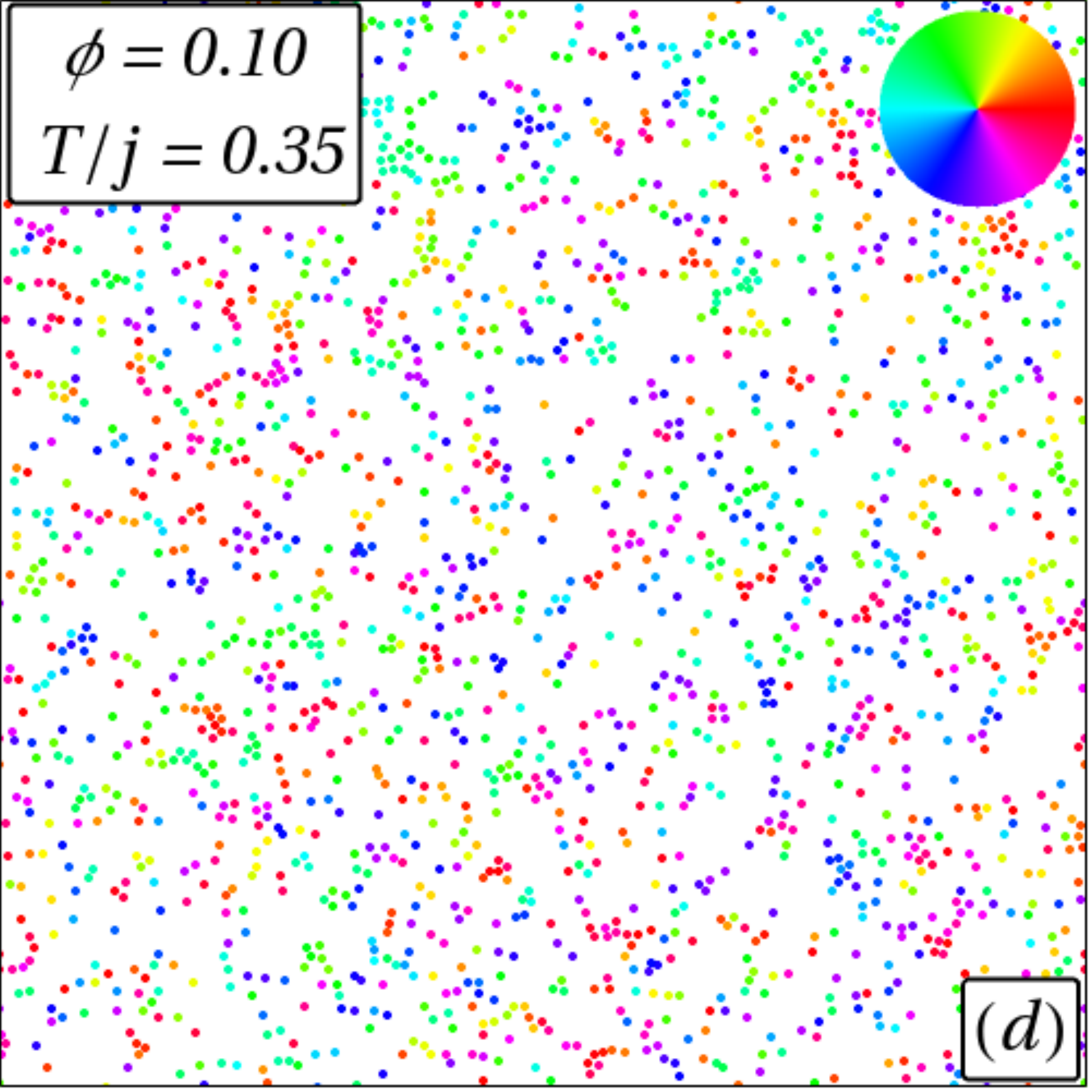}
\includegraphics[width=.24\textwidth,height=.24\textwidth]{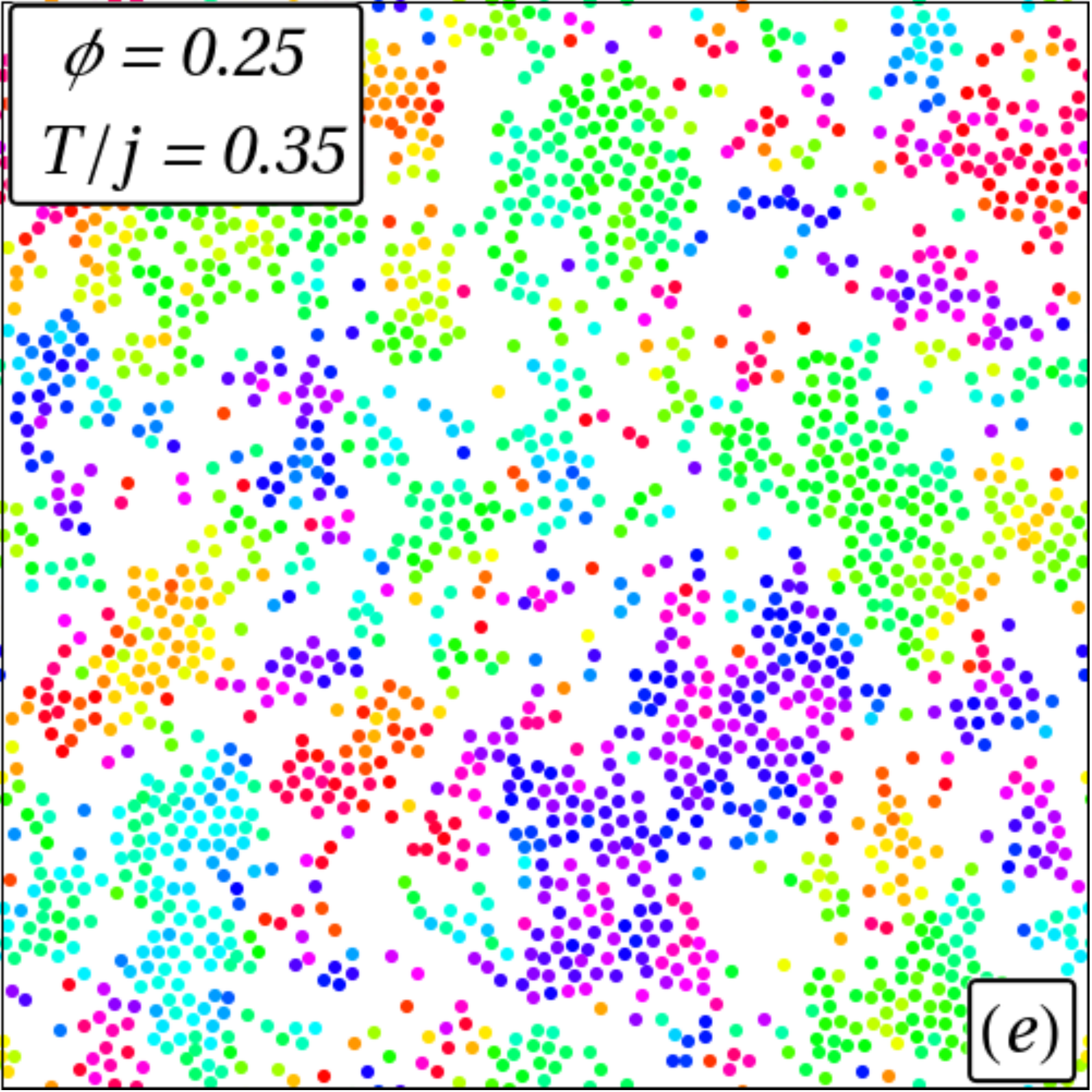}
\includegraphics[width=.24\textwidth,height=.24\textwidth]{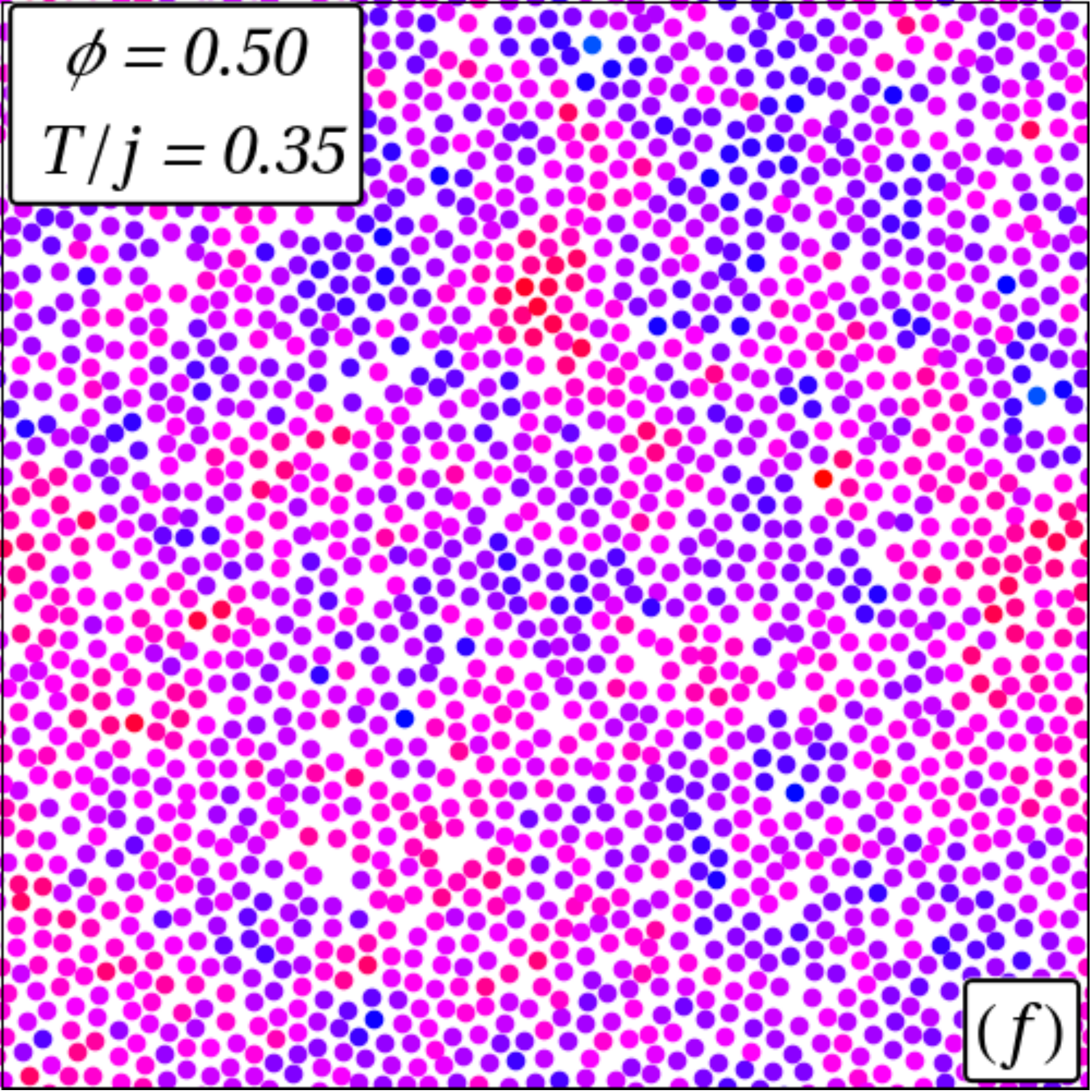}
\includegraphics[width=.24\textwidth,height=.24\textwidth]{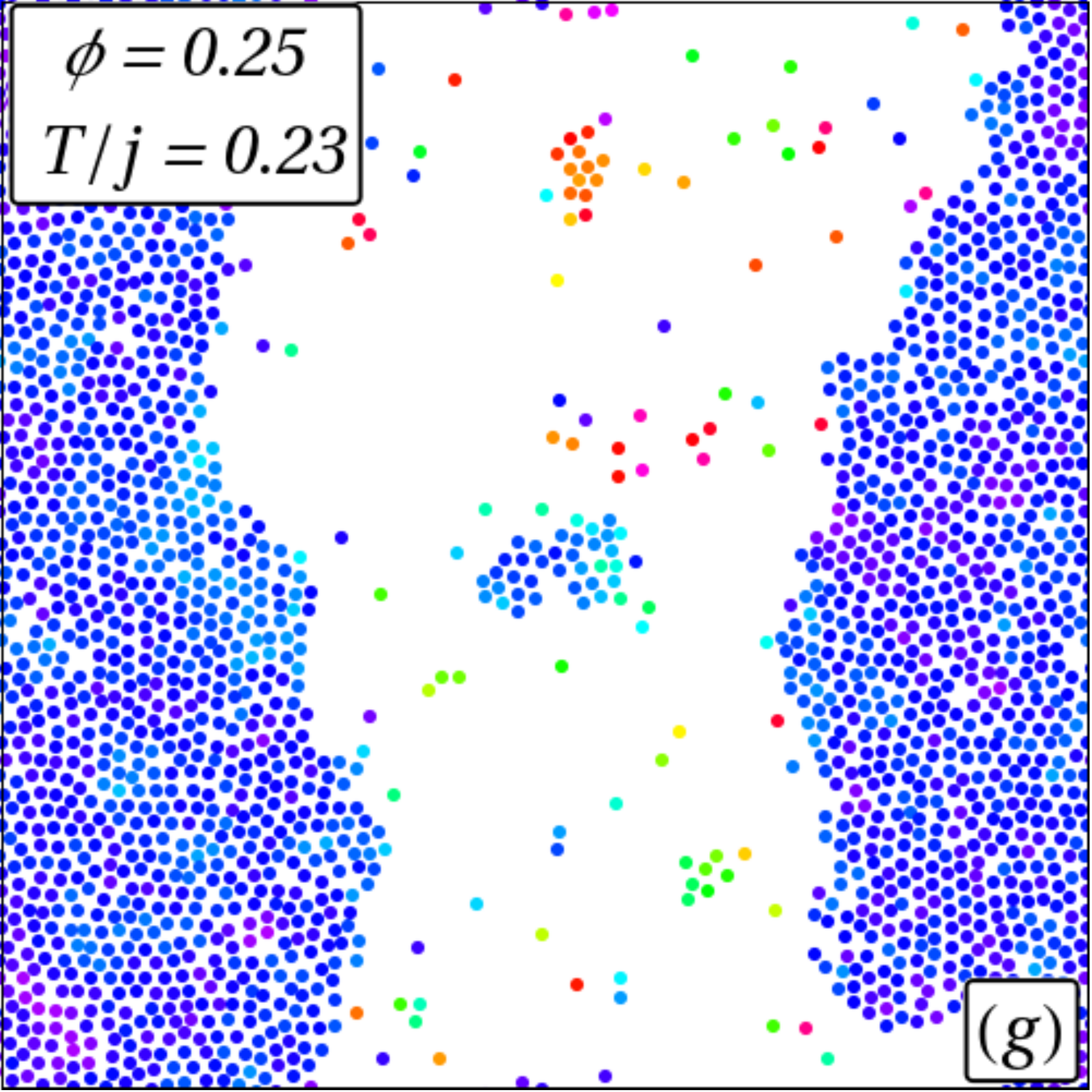}
\caption{{\bf{Numerical phase diagram.}} $(a)$ Points on the phase diagram close to the putative finite-size Curie line (dashed red) and coexistence curve (gray) below the liquid-gas critical point (red dot). 
We used orange disks for homogeneous magnetized states, blue disks for homogeneous isotropic states, and green disks for phase separated states. Whenever possible, we determined the densities of the two fluids in equilibrium: the corresponding liquid and gas densities are plotted as black disks. 
We spot by a magenta dashed line the packing fraction $\phi = 0.55$ that was used throughout Figs.~\ref{fig:MagProp}-\ref{fig:QuenchSnaps}.
Stars are points for which we show density distributions in $(b)$ and $(c)$. 
$(b)$: well-resolved density peaks corresponding to each part of the phase diagram, that correspond to the snapshots $(d)$, $(f)$ and $(g)$.
$(c)$ is an example of the density distribution of a phase-separated state, illustrated in the snapshot $(e)$, for which peaks cannot be resolved with the system size we used. 
It seems that the Curie line meets the coexistence curve exactly at the liquid-gas critical point, here found at $\phi \approx 0.23$ and $T/j \approx 0.6$, meaning that it is a tricritical point. 
A sharp feature of the coexistence line, that looks like a cusp, is visible on this line.
Error bars for the tricritical point were roughly determined by visual inspection of density inhomogeneities in that region of the phase diagram.
 \label{fig:SimPD}}
\end{figure*}

A liquid-gas phase separation is expected in systems such that interactions have an attractive part, and is rather robust against the precise shape of the potential.~\cite{Hansen2006,Rovere1990} 
It is associated to a line of first-order phase transitions terminating in a critical point that belongs to the Ising universality class, both on-lattice and in continuous space.~\cite{Lee1952} 
The symmetry associated to the transition is the discrete $\mathbb{Z}_2$ symmetry, and the liquid-gas critical point is thus also expected in $2d$. 

A slow annealing, just like the one described in the previous section, can be repeated for other densities, which allow for larger local density fluctuations. 
We annealed systems with packing fractions $\phi \in \left[0.10;0.55\right]$ and $N = 2048$ particles.
We do observe phase-separated states (Fig~\ref{fig:SimPD}$(g)$), meaning that a spin-mediated effective attraction is sufficient for the separation to occur, without an explicit attractive part in the interaction potential.
This phase separation takes place between an isotropic gas and a ferromagnetic liquid, as described in previous works for Heisenberg and planar spins in three dimensions.~\cite{Tavares1995,Lomba1998,Omelyan2009} 
As a result, the liquid-gas critical point lies exactly on the finite-size Curie line, where the crossover to finite magnetization takes place in the supercritical fluid. 
This is reminiscent of the tricritical point, observed in higher dimension of space.~\cite{Tavares1995,Lomba1998,Omelyan2009} 
Such tricritical points are always accompanied by a cusp of the coexistence lines.~\cite{Widom1977}
This is clearly observed in Fig.~\ref{fig:SimPD}$(a)$ where the phase diagram indicates the domains of stability for the magnetically isotropic and ferromagnetic homogeneous fluids, together with the coexistence region between the two.
Note that the temperature has been rescaled by an averaged value of the ferromagnetic coupling,
\begin{eqnarray}
    j   &=& \frac{2}{\sigma}\int_{\sigma/2}^\sigma dr J(r) = 1/12.
\end{eqnarray}
In this phase diagram, the coexistence line is obtained by computing the coarse-grained density probability distribution.
The typical aspect of these distributions is shown in Fig.~\ref{fig:SimPD}$(b)$ and $(c)$, along with the corresponding snapshots in panels $(d)$ through $(g)$.
The homogeneous phases are clearly identified by a single peak in the distribution.
Deep in the coexistence regime, two well-identified peaks enable us to pinpoint the densities of the two coexisting phases.
Closer to the top of the coexistence region, the peaks are less separated but the distribution is still clearly not unimodal (Fig.~\ref{fig:SimPD}$(c)$).
This, together with the visual inspection of the system, enable us to infer the shape of the coexistence region closer to the critical point.

\subsubsection{Dynamics after a quench}

When quenching the system deep into the coexistence region, one expects both magnetic domain and liquid droplet growths.
We here investigate the joint dynamical evolution of these growth processes when the system relaxes to equilibrium.
More specifically, it is interesting to see how the droplet growth, associated to a correlation length $\xi_l$, is linked to the growth of magnetized domains, associated to an {\it{a priori}} different correlation length $\xi_m$. 

We define the spatial magnetic correlation function at time $\tau$, 
\begin{equation}
    C(r,\tau) = \langle \bm{s}_i (\tau) \cdot \bm{s_j} (\tau) \rangle_{r_{ij} = r},
\end{equation}
where the average is computed over all particle pairs separated by a distance $r$ and random initial conditions.
Similarly, we define the density-density correlation function at time $\tau$, 
\begin{equation}
    h(r,\tau) =\langle \delta\rho(0,\tau) \delta\rho(r,\tau) \rangle,
\end{equation}
related to the standard radial distribution function $g(r,\tau)$ through $h = g - 1$.
At short range, $h$ typically features exponentially damped oscillations that correspond to the microscopic structure of the liquid,~\cite{Fisher1969}  while longer-range structures code for the typical size of liquid domains in phase-separated states.~\cite{Das2015,Onuki2004}

Concerning the magnetic ordering, when the BKT scenario holds, the spatial correlation following a quench into the critical phase obeys the scaling law~\cite{Jelic2011}
\begin{equation}
    C_{BKT}(r,\tau) \sim r^{-\eta(T)} f_{BKT}\left(\frac{r}{\xi_m(\tau)}\right), \label{eq:CdynXY}
\end{equation}
where $\tau$ is the time after the quench, $f_{BKT}$ is a scaling function, and $\eta(T)$ is the static XY exponent. 
The temperature dependence of $\eta(T)$ reflects the fact that the whole low temperature phase is critical in the $2d$ XY model.
In the present system, we have provided evidence that the BKT scenario does not hold, so that we actually expect a non-critical scaling of the correlation function
\begin{equation}
    C(r,\tau) \sim f\left(\frac{r}{\xi_m(\tau)}\right), \label{eq:Cdyn}
\end{equation} 
typical of conventional coarsening systems.
Likewise, after a quench into a phase-separated region, the density-density correlation function $h$ is expected to follow the scaling law~\cite{Das2015,Onuki2004}
\begin{equation}
    h(r,\tau) \sim f_h\left(\frac{r}{\xi_l(\tau)}\right), \label{eq:hdyn}
\end{equation}
where $f_h$ is another scaling function. 

The correlation lengths $\xi_m$ and $\xi_l$ typically grow algebraically on short time scales,
\begin{equation}
    \xi \sim \tau^{1/z},
\end{equation}
where $z$, the dynamical exponent, defines the universality class of the dynamics.
Following the nomenclature introduced by Hohenberg and Halperin,~\cite{Hohenberg1977} well-known examples are that of model A with $z = 2$, when the order parameter is locally non-conserved, and that of model B with $z = 3$, when the order parameter is locally conserved.~\cite{Bray1994}

\begin{figure}
    \centering
    \includegraphics[width = .96\columnwidth]{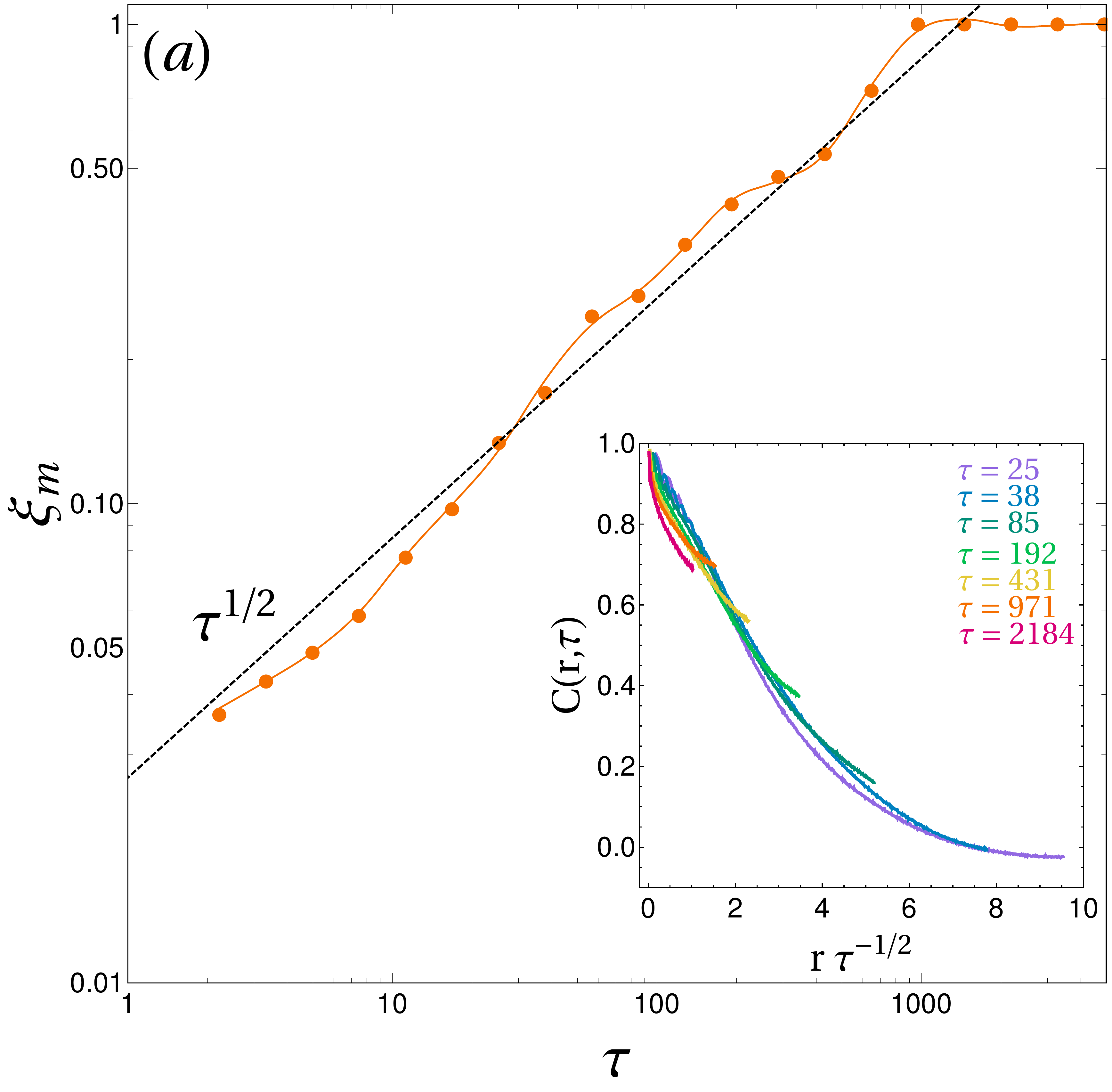}\\
    \includegraphics[width = .96\columnwidth]{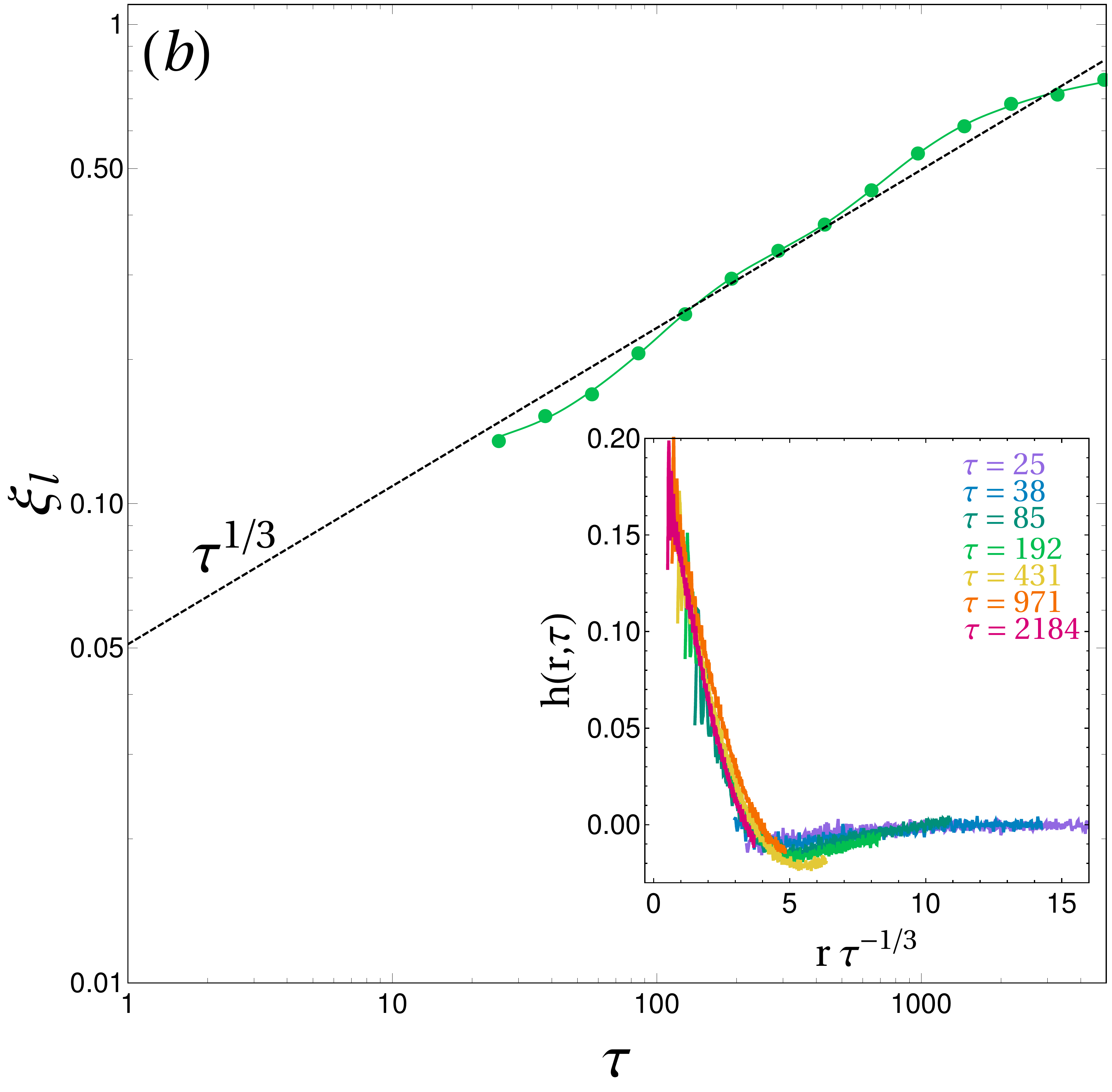}
    \caption{{\bf{Correlation lengths growth after a quench.}}
    $(a)$ Magnetic correlation length $\xi_m$, here obtained from $C(\xi_m(\tau),\tau) = 0.5$, which is in good agreement with a $\tau^{1/2}$ growth (dashed black line). 
    In the inset of $(a)$, we plot $C$ against the rescaled distance $r \tau^{-1/2}$, which leads to a rather good collapse.
    $(b)$ Liquid correlation length $\xi_l$, obtained from the correlation function $h = g -1$ using $h(\xi_l(\tau),\tau) =0$, follows a $\tau^{1/3}$ growth (dashed black line).
    In the inset of $(b)$, we plot $h$ against the rescaled distance $r \tau^{-1/3}$ after taking out the short distance oscillations, thus collapsing the curves. 
    Both correlation lengths are normalized by $L/2$, the upper bound of measurable correlation lengths in a periodic square box of linear length $L$.
    All the curves presented here were obtained for $N = 8192$ particles, for $\phi = 0.35$ and each $g$ or $C$ was averaged over 10 to 20 realizations.
    }
    \label{fig:zscaling}
\end{figure}

Here, we compute the correlation functions $C(r,\tau)$ and $h(r,\tau)$ for $\tau \in \left[ 5 ; 5000\right]$ for systems of $N = 8192$ particles, and average over 10 to 20 independent quenches.
Regarding the magnetization, the correlation length $\xi_m(\tau)$ is extracted through the definition $C(\xi_m (\tau),\tau) = 0.5$.
As can be seen on Fig.~\ref{fig:zscaling}$(a)$, $\xi_m(\tau)\sim \tau^{1/2}$, that is $z_m = 2$.
The inset of the figure confirms the scaling for the correlation function.
As for the density fluctuations, it is convenient to define the correlation length $\xi_l(\tau)$ as the first value beyond the oscillatory regime such that $h(\xi_l (\tau),\tau) = 0$.
It also features a power-law dependence on $\tau$ (Fig.~\ref{fig:zscaling}$(b)$), although with a different exponent: $\xi_l(\tau)\sim \tau^{1/3}$, so that $z_l = 3$.
Here also, the inset confirms the scaling property of the correlation function.

We conclude that the local conservation laws associated with the order parameters (density is conserved while magnetization is not) govern the dynamical scalings: the magnetization (respectively density) fluctuations follow the prescriptions of model A (respectively model B).

\begin{figure}[b!]
    \centering
    \includegraphics[width=.93\columnwidth]{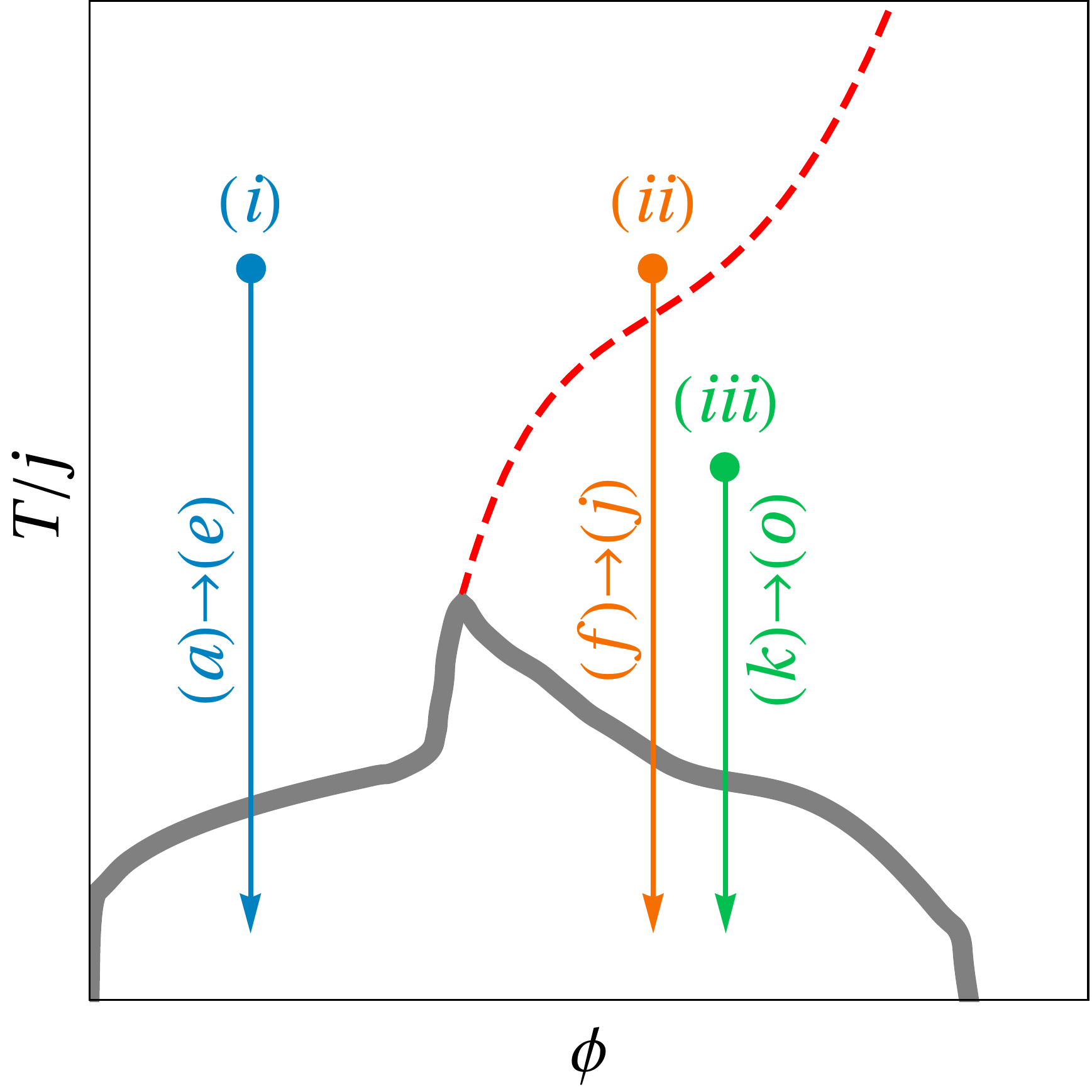}
    \caption{{\textbf{Sketch of the three kinds of considered quenches.}} The Roman numbers correspond to the definitions given in the main text. The letters indicate the corresponding panels of Fig.~\ref{fig:Quenches}.}
    \label{fig:sketch}
\end{figure}

However, the order parameters themselves are expected to grow concomitantly since the effective attraction is mediated by the local magnetization.
We shall investigate this focusing on three different quench protocols, as sketched on Fig.~\ref{fig:sketch}:
\begin{enumerate}
    \item[$(i)$] a quench at a density lower than that of the tricritical point (or critical density for short), starting from the high-temperature isotropic gas, and across the liquid-gas coexistence line, 
    \item[$(ii)$]  a quench at a density higher than the critical density, starting from an isotropic supercritical fluid, and across {\emph{both}} the Curie line and the liquid-gas coexistence line, and
    \item[$(iii)$] a quench at a density higher than the critical density, starting from a magnetized supercritical fluid and crossing only the liquid-gas coexistence curve. 
\end{enumerate}

\begin{figure*}\centering
\includegraphics[width=.25\textwidth]{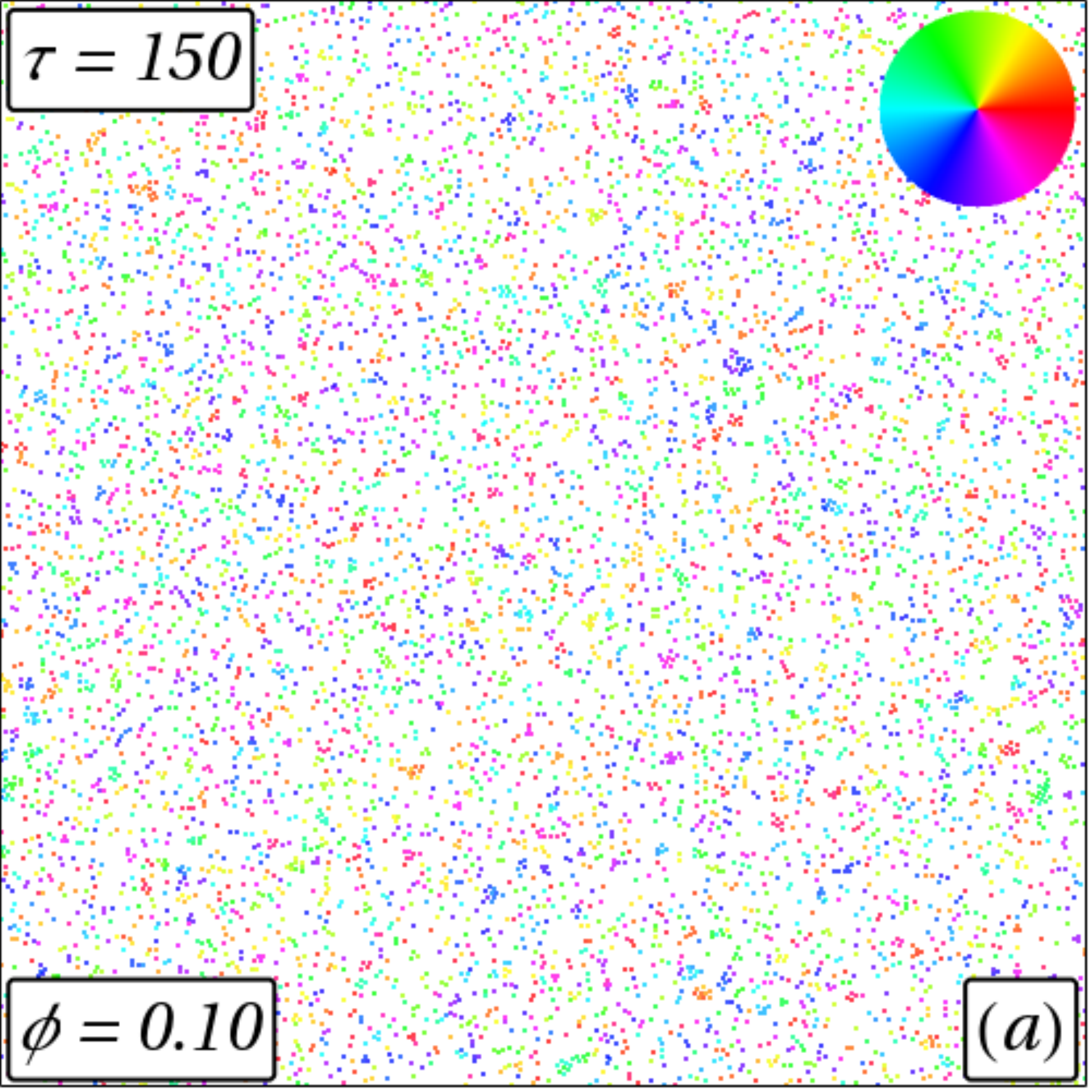}
\includegraphics[width=.25\textwidth]{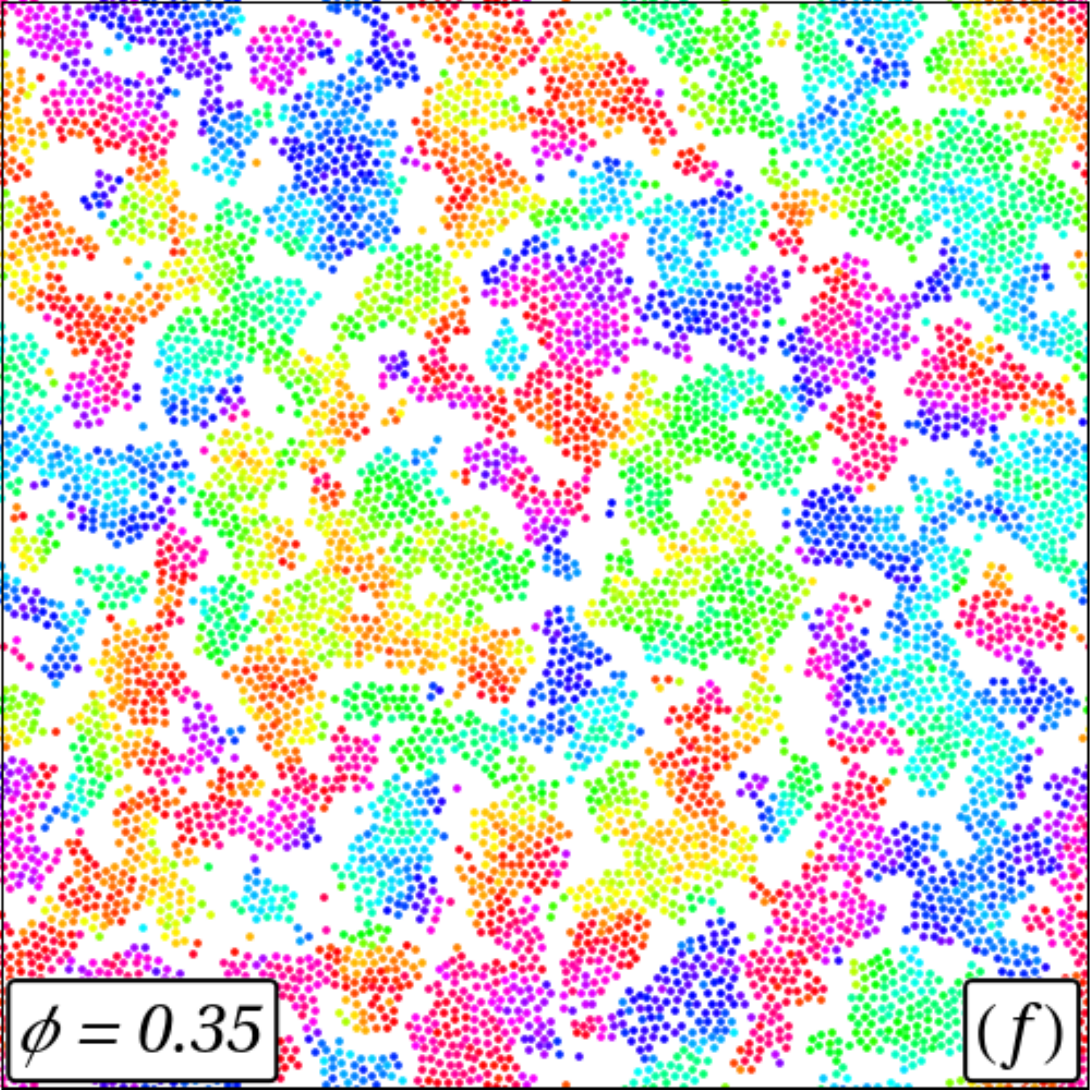}
\includegraphics[width=.25\textwidth]{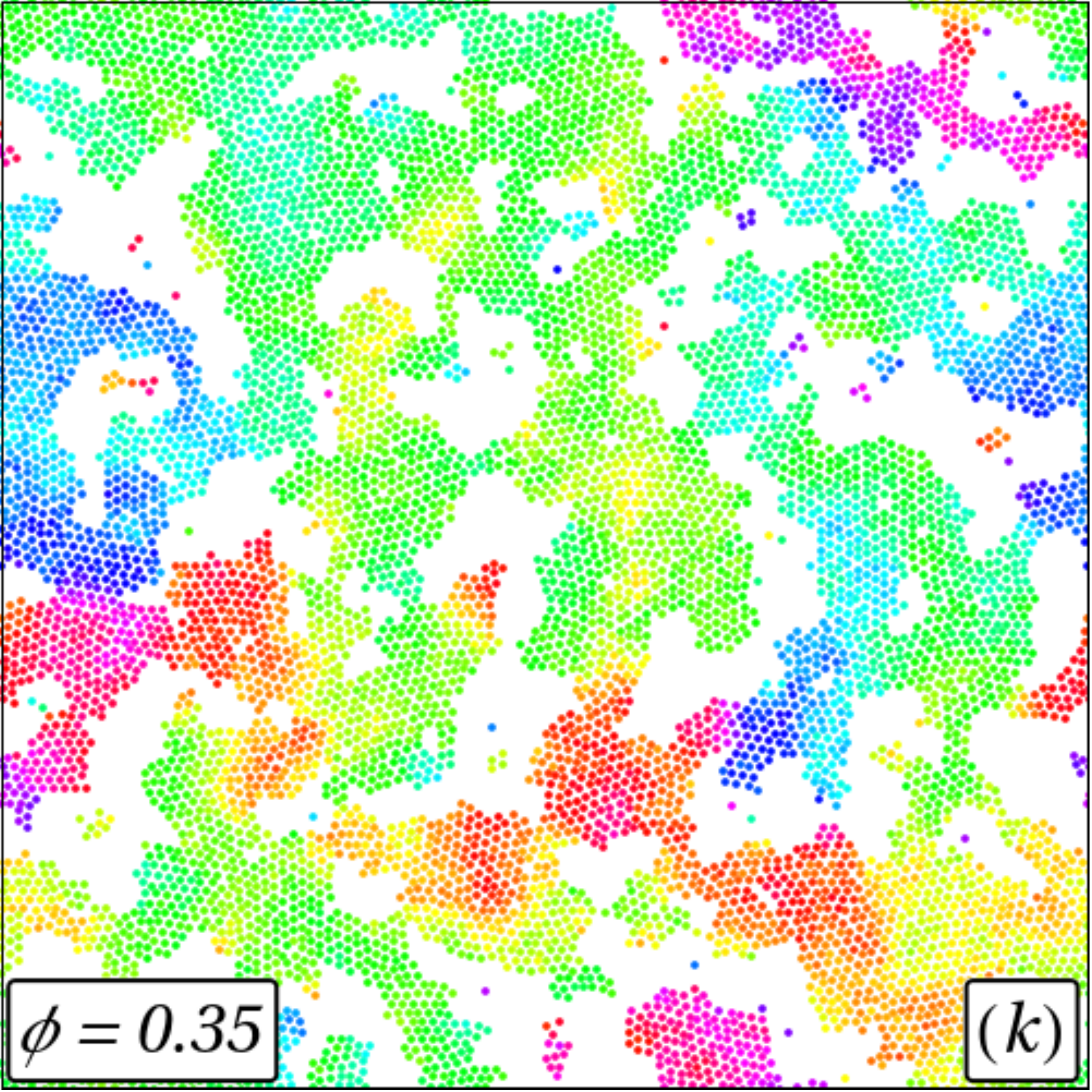} \\
\includegraphics[width=.25\textwidth]{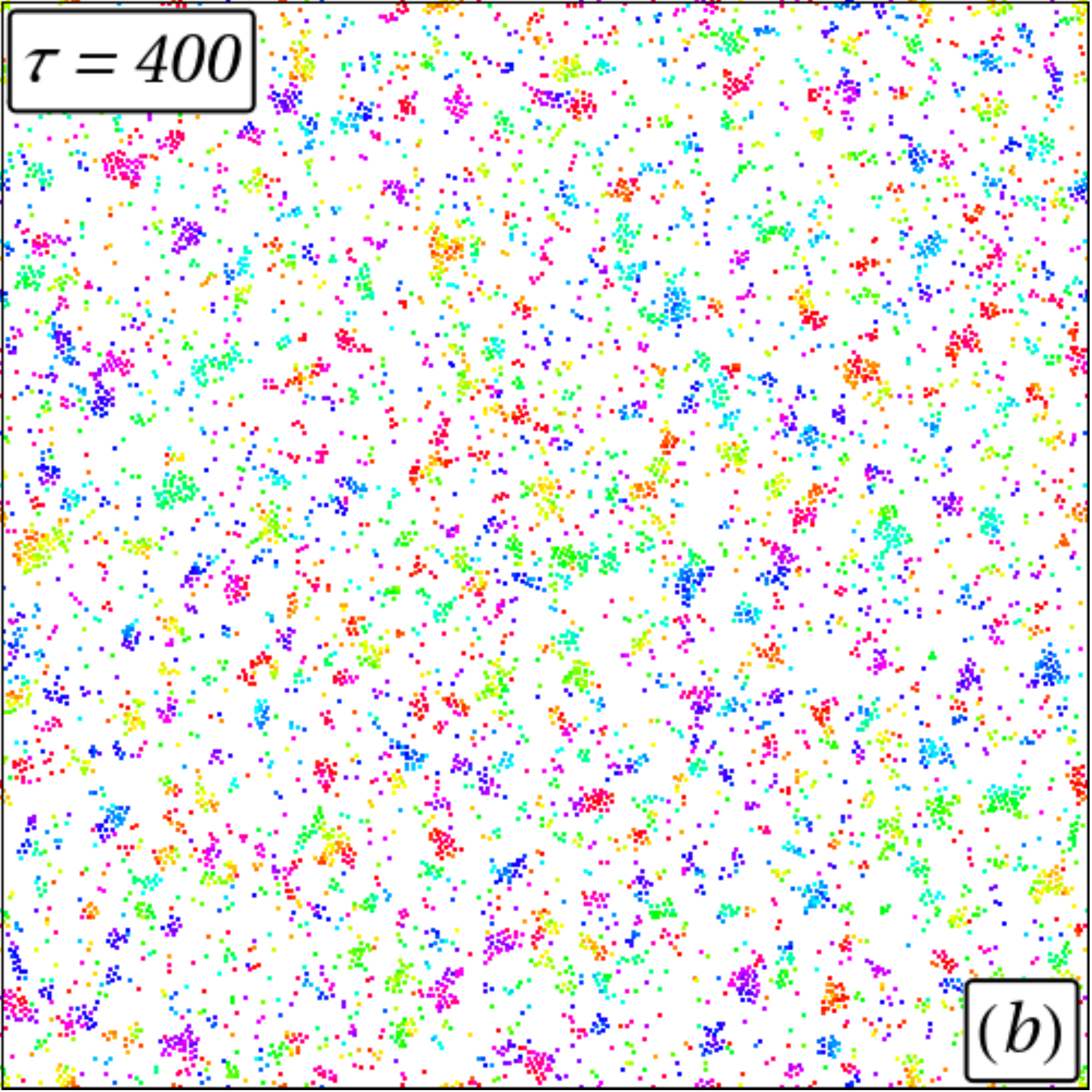}
\includegraphics[width=.25\textwidth]{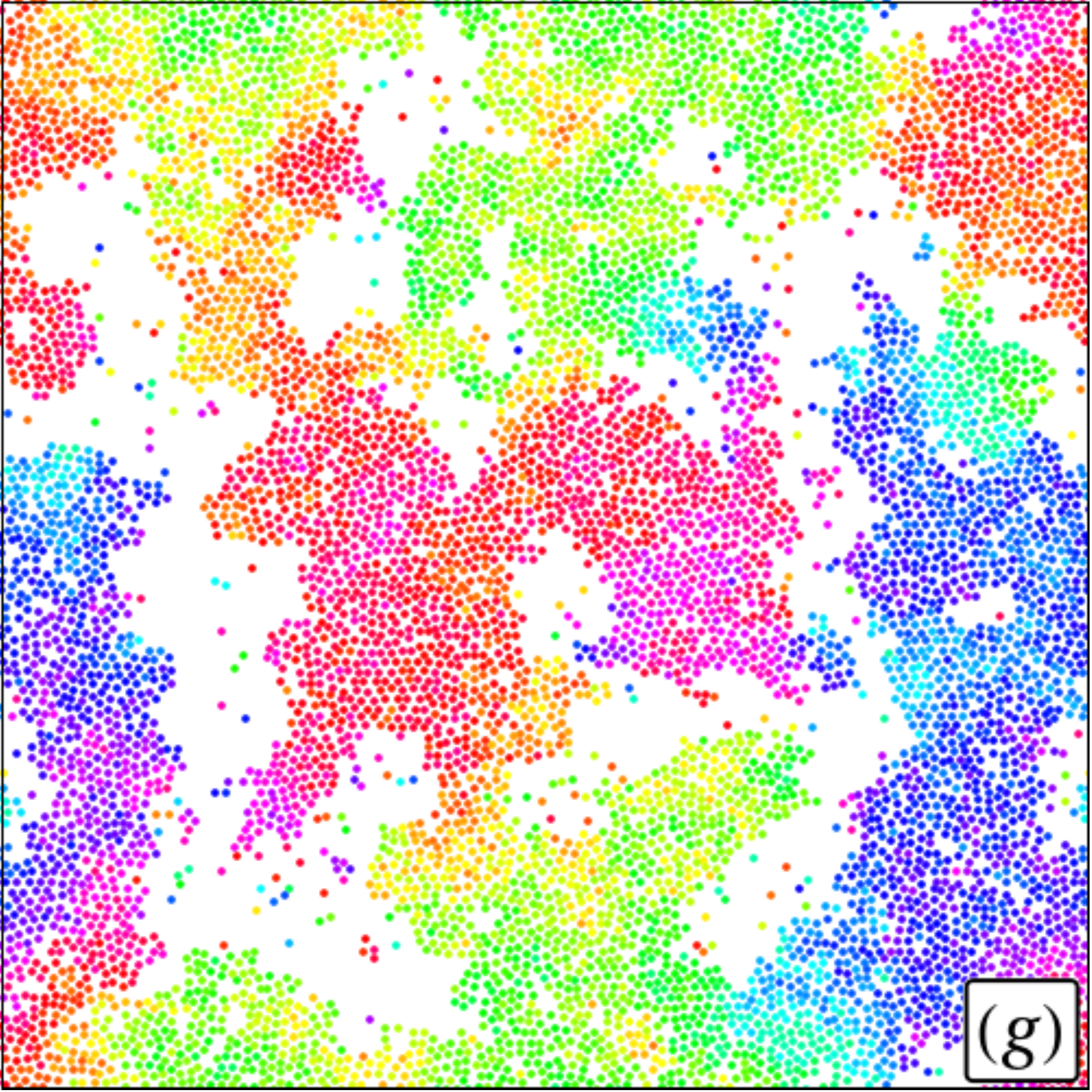}
\includegraphics[width=.25\textwidth]{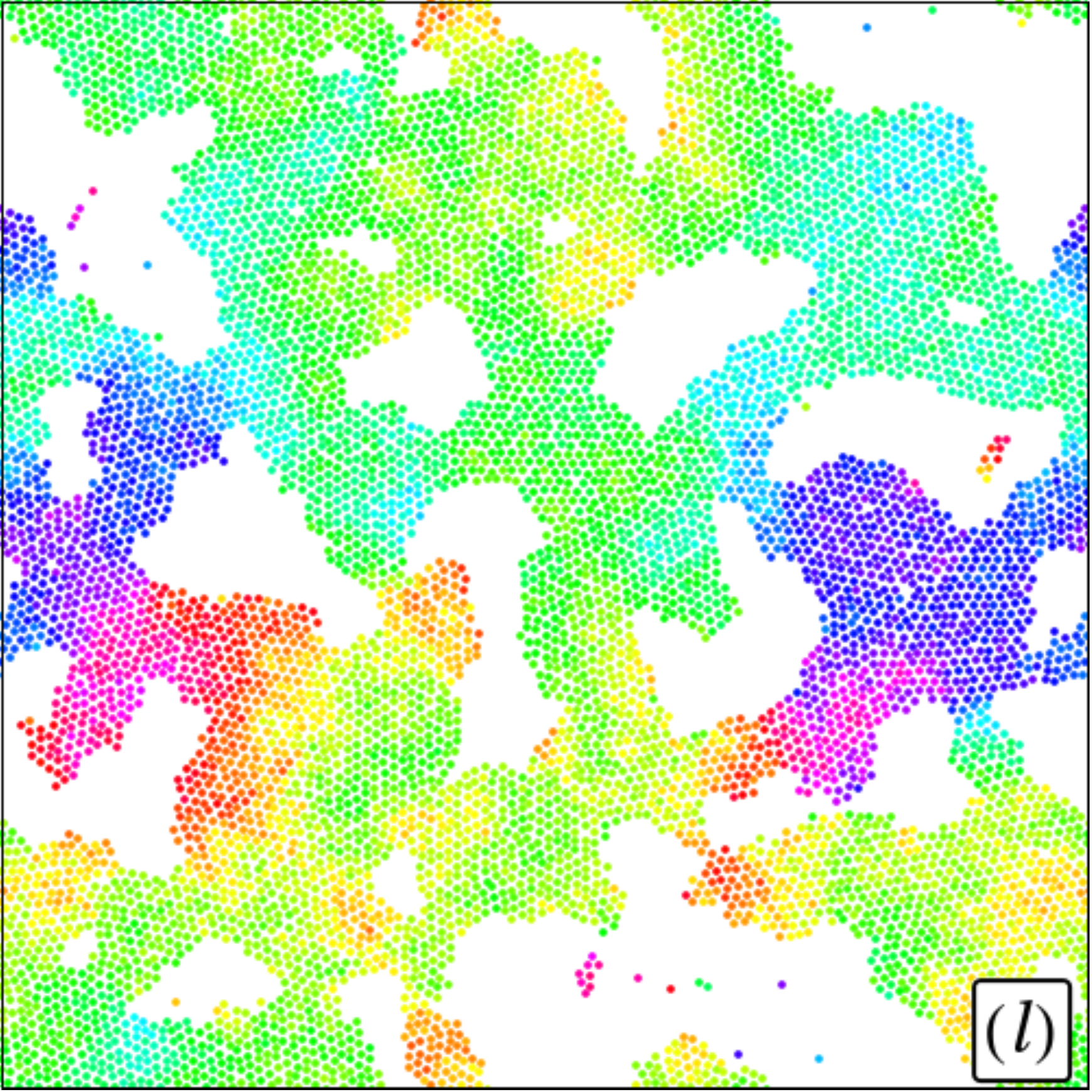} \\
\includegraphics[width=.25\textwidth]{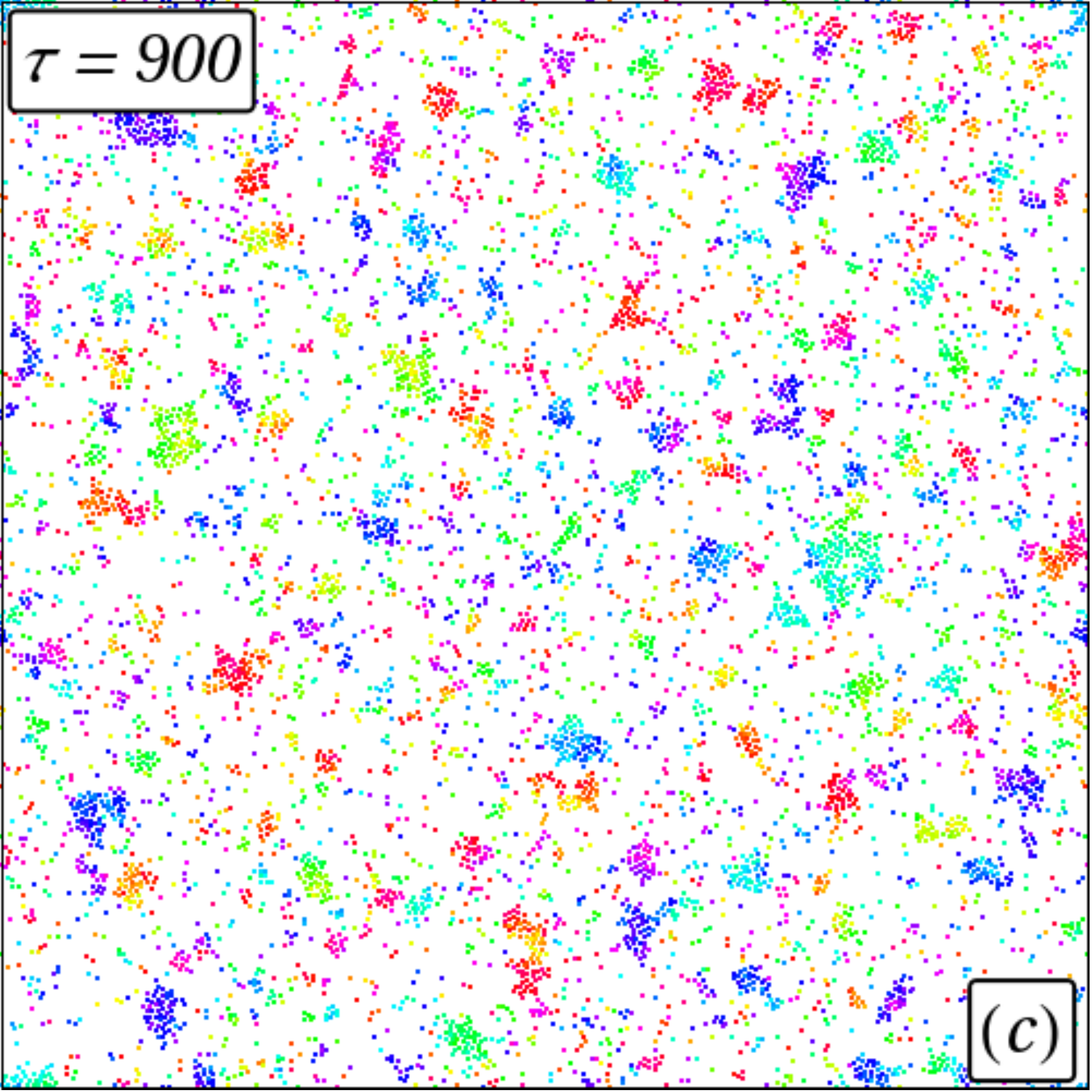}
\includegraphics[width=.25\textwidth]{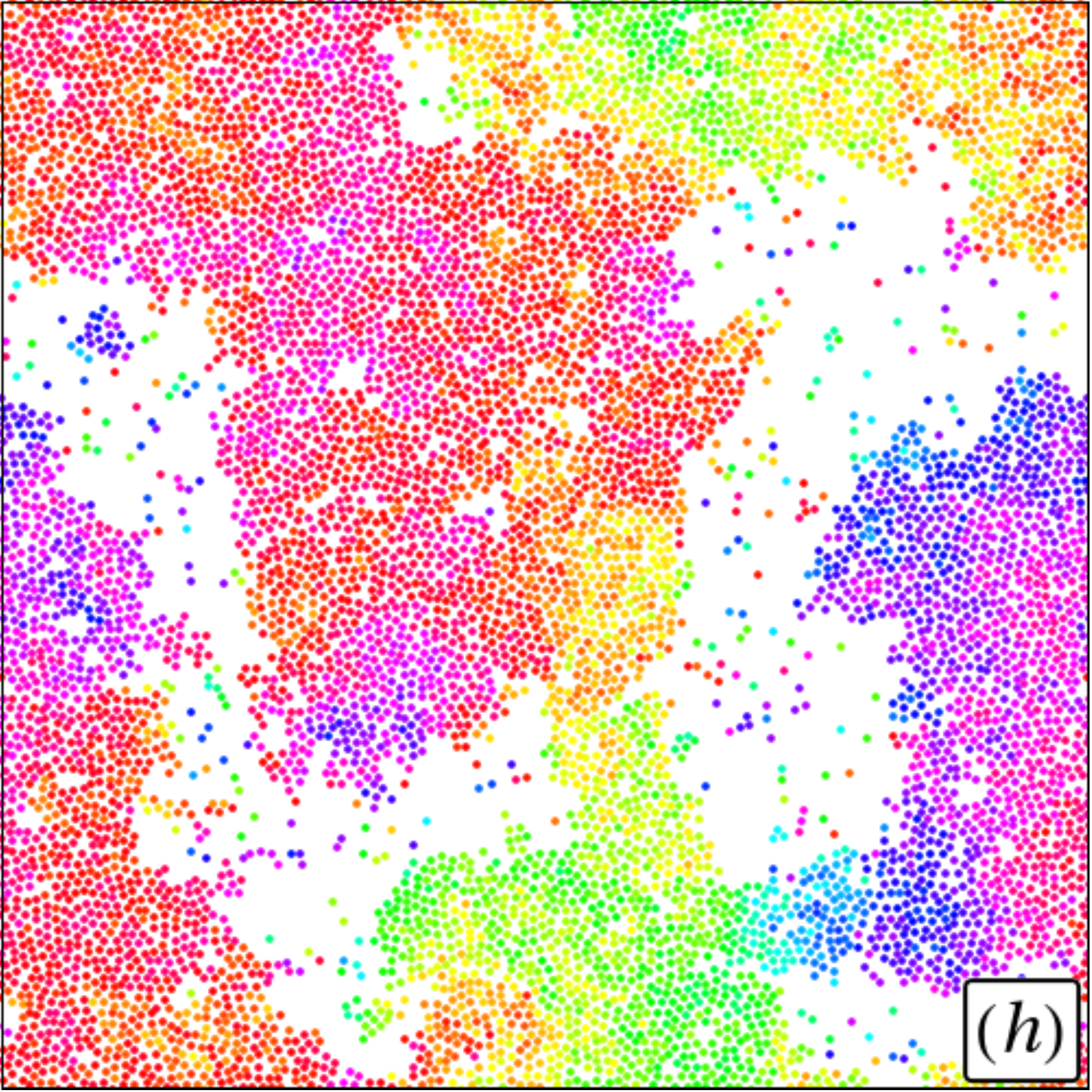}
\includegraphics[width=.25\textwidth]{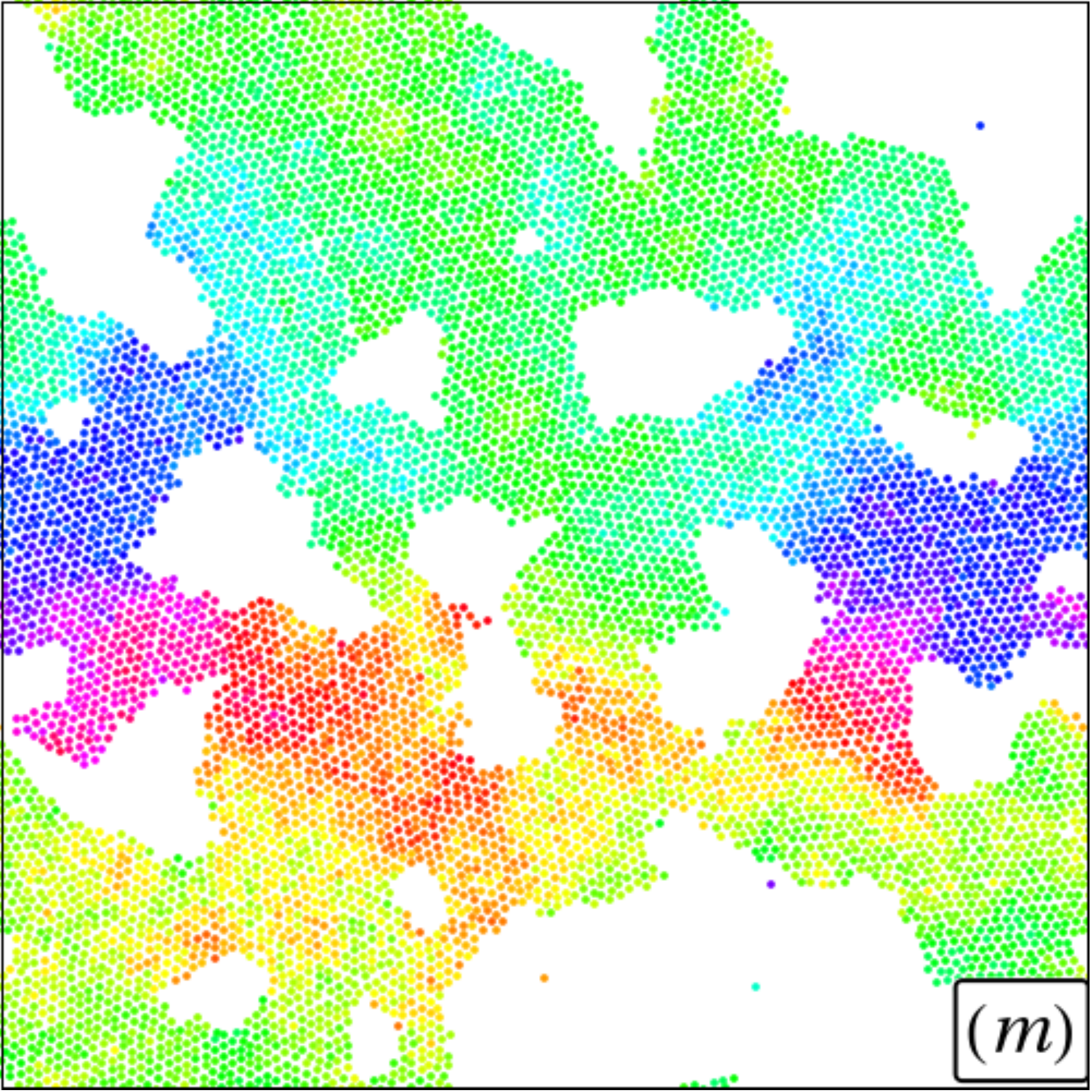} \\
\includegraphics[width=.25\textwidth]{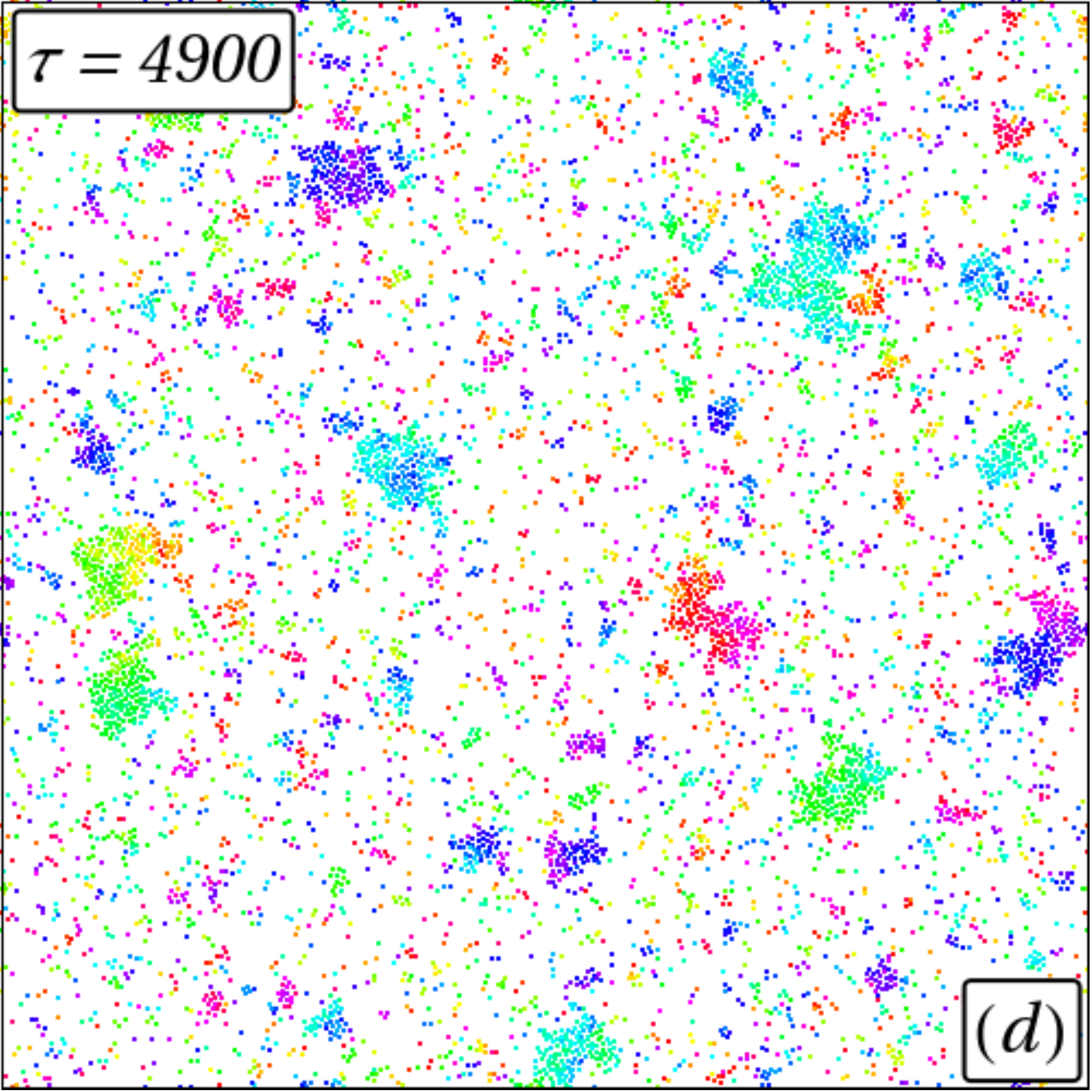}
\includegraphics[width=.25\textwidth]{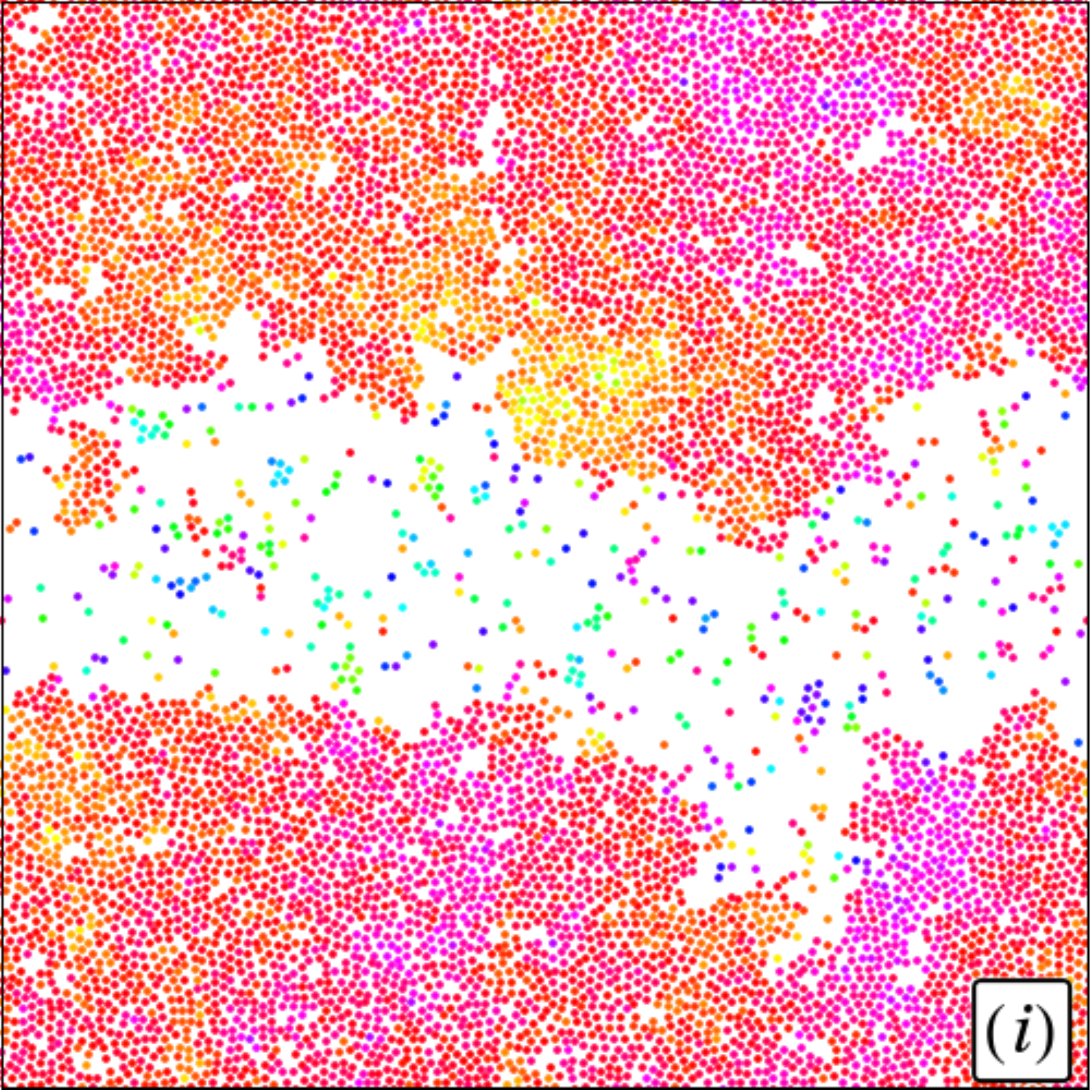}
\includegraphics[width=.25\textwidth]{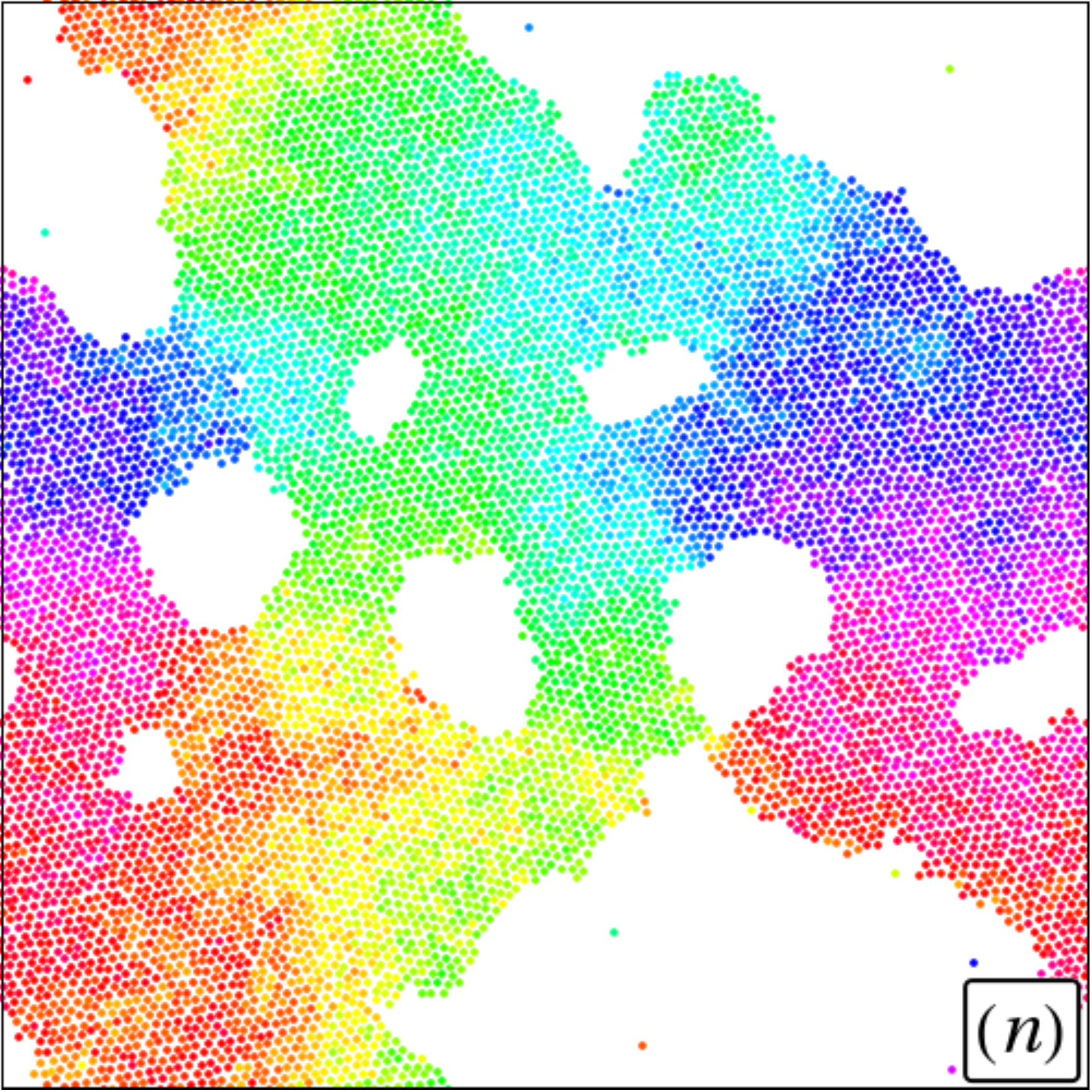} \\
\includegraphics[width=.25\textwidth,height=.25\textwidth]{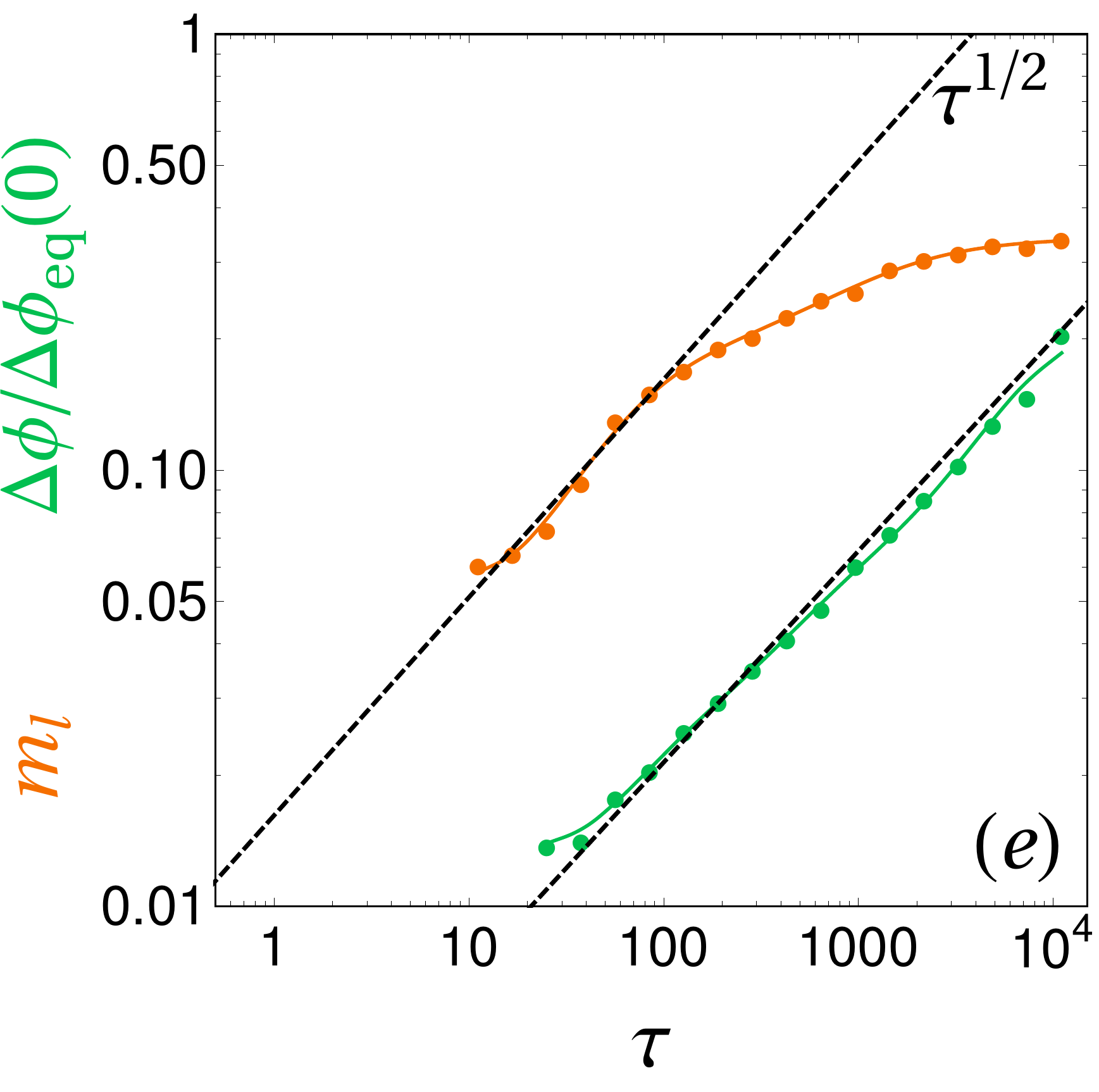}
\includegraphics[width=.25\textwidth,height=.25\textwidth]{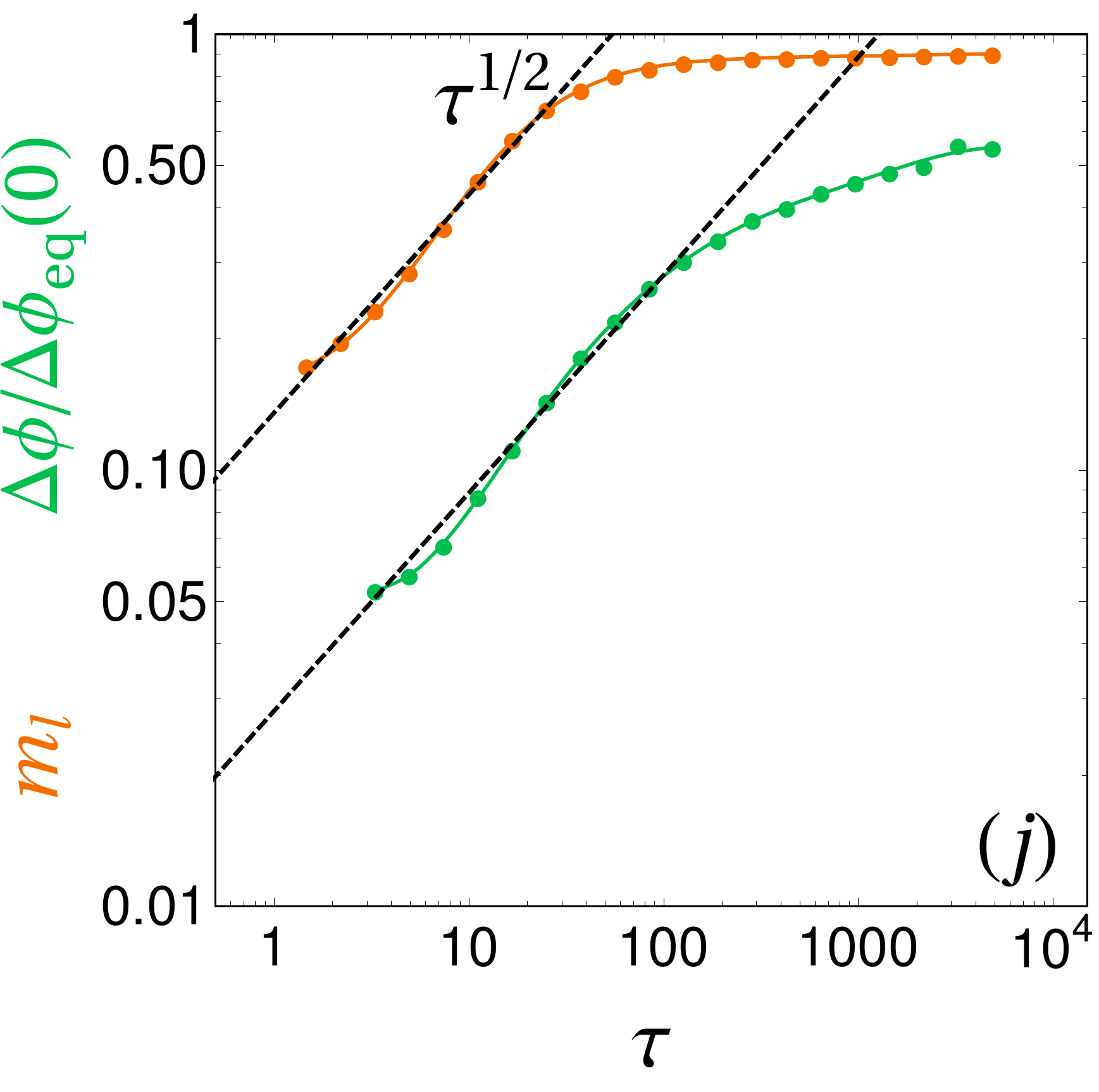}
\includegraphics[width=.25\textwidth,height=.25\textwidth]{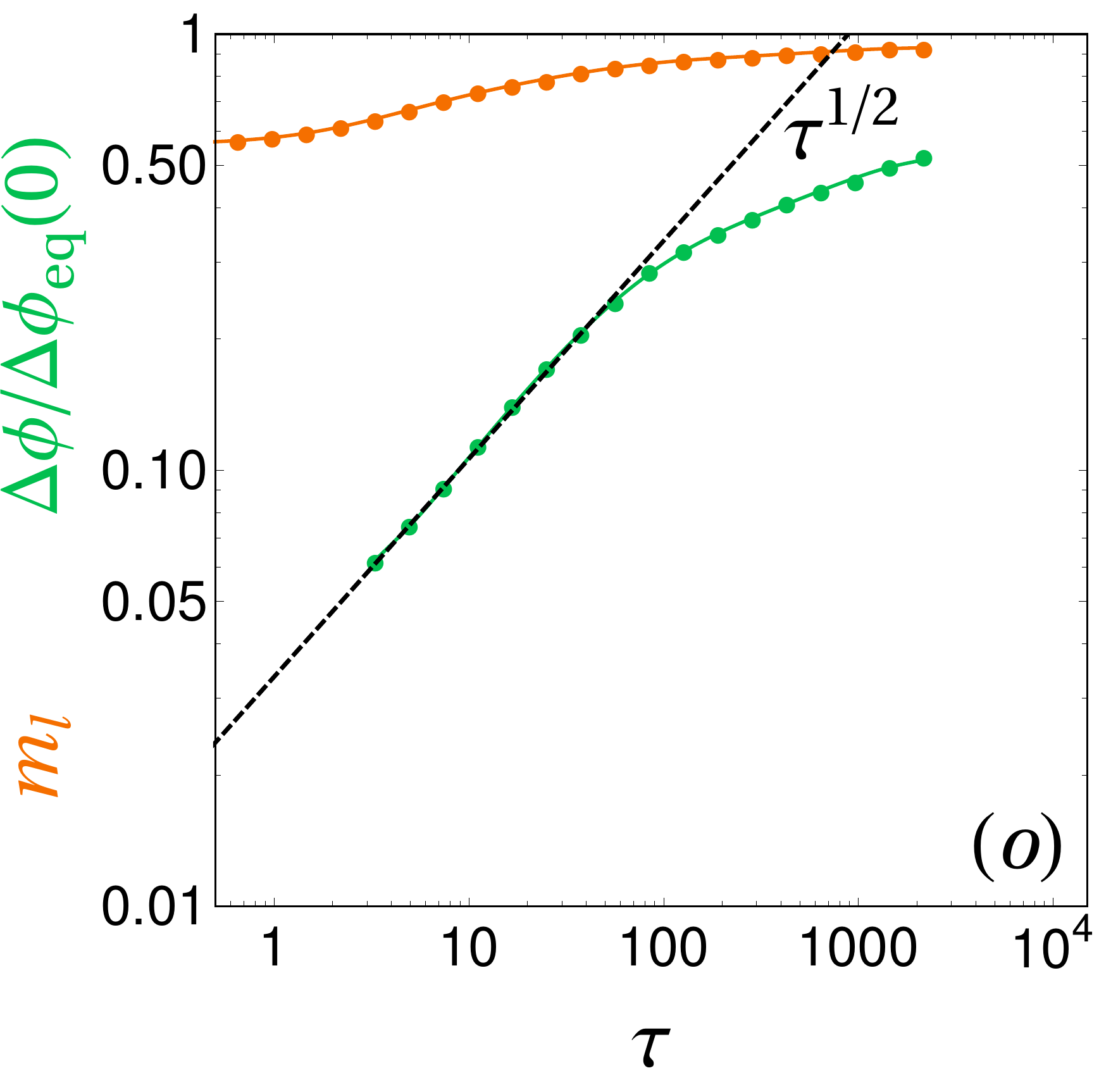}
\caption{{\bf{Dynamics after a quench through the liquid-vapour coexistence curve.}}
Quenches in the three cases discussed in the main text and schematically represented in the phase diagram in Fig~\ref{fig:sketch}. 
Four snapshots are shown in each case in panels $(a)-(d)$,$(f)-(i)$,$(k)-(n)$. 
We show the growth of the local order parameters $m_l$ and $\Delta\phi = \phi_l - \phi_g$, defined in Eqs.~(\ref{eq:liqmag}) and (\ref{eq:phigl}), in panels $(e)$, $(j)$ and $(o)$, using the same order as in the snapshots.
In order to make the snapshots easier to peruse, the times after the quench are all indicated in the leftmost column, and the packing fractions in the top row. 
The colour-code used for the spins is reminded in panel $(a)$. For all snapshots, we used $N = 8192$ particles.
\label{fig:Quenches}}
\end{figure*}
Figure~\ref{fig:Quenches} displays four successive snapshots illustrating the coarsening dynamics for each of the above cases.
In case $(i)$ (first column), the system is initially in a paramagnetic gas phase.
After the quench, magnetized liquid domains grow by attracting nearby particles whose spins are aligned with the magnetization of the domain.
In case $(ii)$ (second column), the system starts from a paramagnetic supercritical fluid.
Crossing the Curie line, the magnetization should set in.
However, crossing the coexistence line at a density larger than the critical one, density inhomogeneities develop in the form of gas pockets across which the ferromagnetic alignment cannot carry over.
In case $(iii)$ (third column), the system is initially in a ferromagnetic supercritical fluid with a finite magnetization.
Accordingly, the crossing of the coexistence curve leads to the coarsening of gas pockets, but within a magnetization pattern that is already established.

To be more quantitative, one needs to introduce proper order parameters measured within spatial scales dictated by the coarsening process.
The simulation box is divided into $(l/L)^2$ boxes $\mathcal{B}_a$ with linear size $l$.
We first introduce a local magnetization: 
\begin{eqnarray}
 m_a &=& \frac{1}{n_a}\left|\sum\limits_{\bm{r}_i\in\mathcal{B}_a} \bm{s}_i\right|,
\end{eqnarray}
where $n_a$ is the number of particles in box $\mathcal{B}_a$.
Since we are interested in the magnetization inside liquid droplets, we define a mean local magnetization modulus weighted by the local density so as to give less statistical weight to empty regions,
\begin{eqnarray}
m_l    &=& \frac{\overline{n_a m_a}}{\overline{n_a}}, \label{eq:liqmag}
\end{eqnarray}
where the overline symbolizes an average over all boxes.
The analogue of the magnetization for the liquid-gas transition is the difference $\Delta \phi = \phi_l - \phi_g$ between the average packing fraction of the liquid, $\phi_l$, and the one of the gas, $\phi_g$.~\cite{Lee1952}
These packing fractions are estimated from the list of local packing fractions $\phi_a = n_a \pi r_0^2 / l^2$ in box $\mathcal{B}_a$ by computing
\begin{subequations}
\begin{align}
 \phi_l  &= \frac{\overline{\phi_a \Theta\left(\phi_a - \overline{\phi_a}\right)}}{\overline{\Theta\left(\phi_a - \overline{\phi_a}\right)}}, \\
 \phi_g  &= \frac{\overline{\phi_a \Theta\left(\overline{\phi_a}- \phi_a \right)}}{\overline{\Theta\left( \overline{\phi_a} - \phi_a\right)}}, 
\end{align} \label{eq:phigl}
\end{subequations}
where $\Theta$ is the Heaviside step function.

By construction, $\Delta\phi$ is close to $0$ if the distribution of densities is unimodal and grows over time to its equilibrium value $\Delta\phi_{eq}(T)$ given by the coexistence curve.
We therefore normalize $\Delta\phi$ by the zero-temperature width of the numerical phase diagram shown in Fig.~\ref{fig:SimPD}$(a)$ to get a quantity that is bounded between $0$ (homogeneous phases) and $1$ ($T = 0$ phase-separated state at equilibrium). 

Just like the correlation length growth, a local order parameter, say $\Delta\phi$, is expected to follow an algebraic scaling $\Delta\phi \sim t^{\lambda_l/z}$ for short times following a quench across a critical transition.
The exponent $\lambda_l$ relates the growth of $m$ to the corresponding correlation length $\xi_l$: $\Delta\phi \sim \xi_l^{\lambda_l}$.~\cite{Bray1990,Bray1992} 
As seen on Fig.~\ref{fig:Quenches}$(e)$ and $(j)$ at short times, for both cases $(i)$ and $(ii)$, we observe
\begin{align}
    m_l &\sim t^{1/2}, \\
    \Delta\phi &\sim t^{1/2}.
\end{align}
Their dynamics are synchronized.
Note that $m_l$ is not strictly the order parameter associated with the correlation function $C(r,\tau)$, as the modulus in Eq.~(\ref{eq:liqmag}) erases long-range correlations between spin orientations. 
However, assuming that $m_l$ behaves like the local vector magnetization in the liquid at short times (i.e. when correlations are short-ranged), and together with the values $z_l = 3$ and $z_m = 2$, we respectively find $\lambda_l \simeq 3/2$ and $\lambda_m \simeq 1$ for the liquid and magnetic growths.

The value $\lambda_m \simeq 1$ is the expected analytical value for $O(n)$ vector models in $2d$ with $n\to\infty$,~\cite{Bray1990} and is therefore coherent from the magnetic point of view.
The $\lambda_l$ we find, however, is more surprising, as the standard liquid-gas separation belongs to the same universality class as the $2d$ Ising model with locally conserved order parameter and should thus exhibit $\lambda = 2$, following general scaling arguments.~\cite{Majumdar1994}
We attribute this discrepancy to the coupling of the density field to the locally non-conserved magnetization field via the spin-mediated effective attraction.
Coincidentally, the selected growth rate, $\tau^{1/2}$, is the slower of the two {\it{a priori}} expected values, namely $\tau^{1/2}$ for a standard $O(n)$ model in $2d$ and $\tau^{2/3}$ for a standard liquid-gas phase separation.

In case $(iii)$, following a quench through the coexistence curve only (i.e. starting from below the Curie line), Fig.~\ref{fig:Quenches}$(o)$ shows that $\Delta\phi$ retains the scaling $t^{1/2}$ while the magnetization has already set in before the quench and does not obey any algebraic scaling.

Altogether, the short-time dynamics of the order parameters following a quench is dominated by the slowest-growing one, namely the magnetization, thereby altering the standard scalings for the growth of the density order parameter.
An intuitive way to think about it is that the liquid-gas phase separation has to wait for the magnetization to set in in order to induce the effective attraction.
This interpretation means that the limit of stability of the gas, or gas-side spinodal line, and the continuation of the Curie line under the coexistence curve are one and the same.
If this were the case, both the compressibility and the magnetic susceptibility should diverge at the exact same values of temperature and density.
This result is, in fact, consistently recovered in the theoretical approaches developed in the following sections.

\subsection{Summary: a Ferromagnetism-Induced Phase Separation}

Altogether, the above numerical study shows that
\begin{enumerate}
    \item in spatially homogeneous states, the finite-size magnetization crossover escapes the BKT scenario as it is not accompanied by vortex unbinding, and is akin to a usual critical system studied at its lower critical dimension.
    \item a phase separation between a paramagnetic gas and a ferromagnetic liquid takes place as a result of the spin-mediated effective attraction.
    \item the finite-size Curie line hits the liquid-gas coexistence curve exactly at the critical point, and is accompanied by a cusp in the coexistence line.
    \item the relaxation dynamics following a quench into the coexistence region suggests that the finite-size Curie line and the gas-side spinodal coincide.
\end{enumerate}

In the following, we shall recover the above observations using different theoretical approaches.
Starting with a Bethe-lattice description, we obtain the mean-field phase diagram, which already captures the main aforementioned features.
This approach deals with the magnetic and liquid properties at the same mean-field level of approximation.
An alternative off-lattice approach would be to follow previous works on $3d$ spin fluids~\cite{Tavares1995,Lomba1998,Omelyan2009} and write down a set of integro-differential equations obtained from the Ornstein-Zernike equation and Born-Green-like closures.
Here, we rather develop lighter approaches, taking advantage of the mean-field-like behaviour of the magnetization, while retaining a finite-dimensional description of the liquid structure.
In short, we propose a Curie-Weiss like scheme for the magnetization, where the number of neighbours is found from the pair correlation function of the liquid.
Apart from confirming the general properties of the phase diagram, it enables us to capture the influence of both the softness of the repulsion, and the interaction ranges.

\section{Bethe-lattice description \label{sec:Lattice}}

First, let us discuss an on-lattice description of the equilibrium properties of our spin fluid.
In order to model our continuous-space interactions, we take inspiration from the Blume~\cite{Blume1966}-Capel~\cite{Capel1966} (BC) model, defined as a lattice model of ferromagnetic Ising spins with vacancies.
It is described by the Hamiltonian
\begin{equation}
    H_{BC} = - J \sum\limits_{<i,j>} n_i n_j S_i S_j  + U \sum\limits_{i} n_i^2,
\end{equation}
which contains ferromagnetic alignment of strength $J$, an on-site repulsion of amplitude $U$, acting on Ising spins $\left\{S_i = \pm 1\right\}$ and occupation numbers $\left\{n_i = 0,1\right\}$.
\begin{figure}
\includegraphics[width=.49\columnwidth]{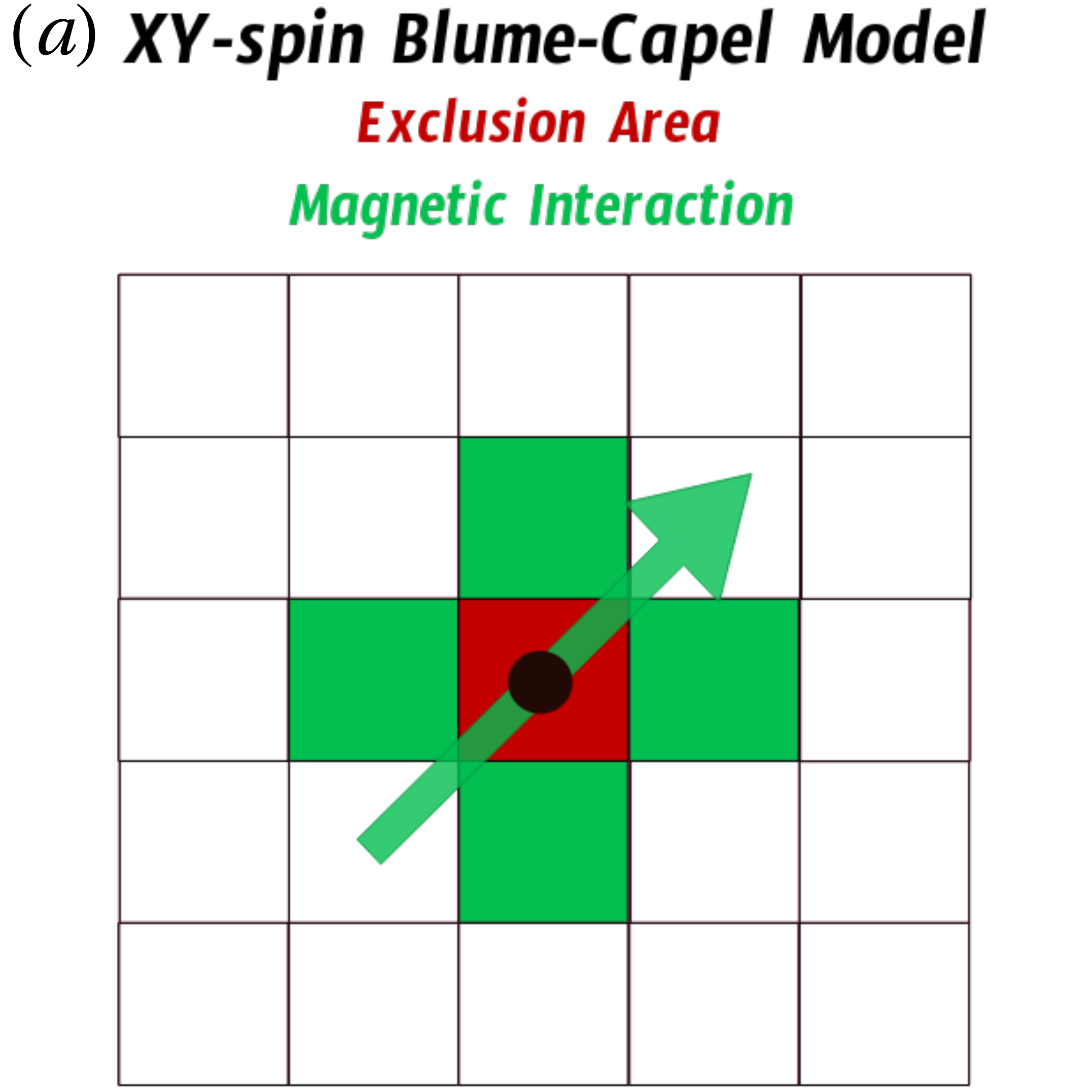} 
\includegraphics[width=.49\columnwidth]{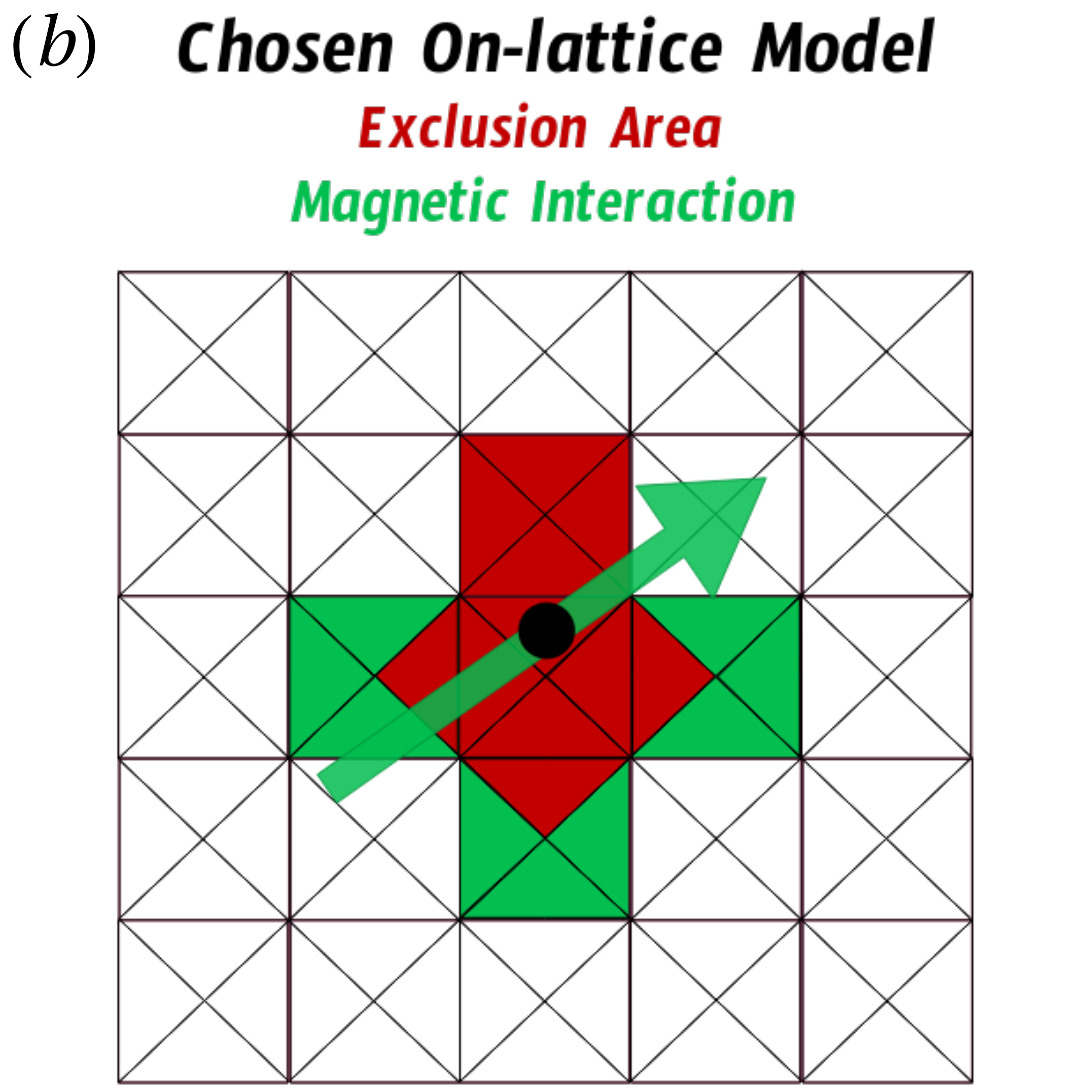} \\
\includegraphics[width=.95\columnwidth]{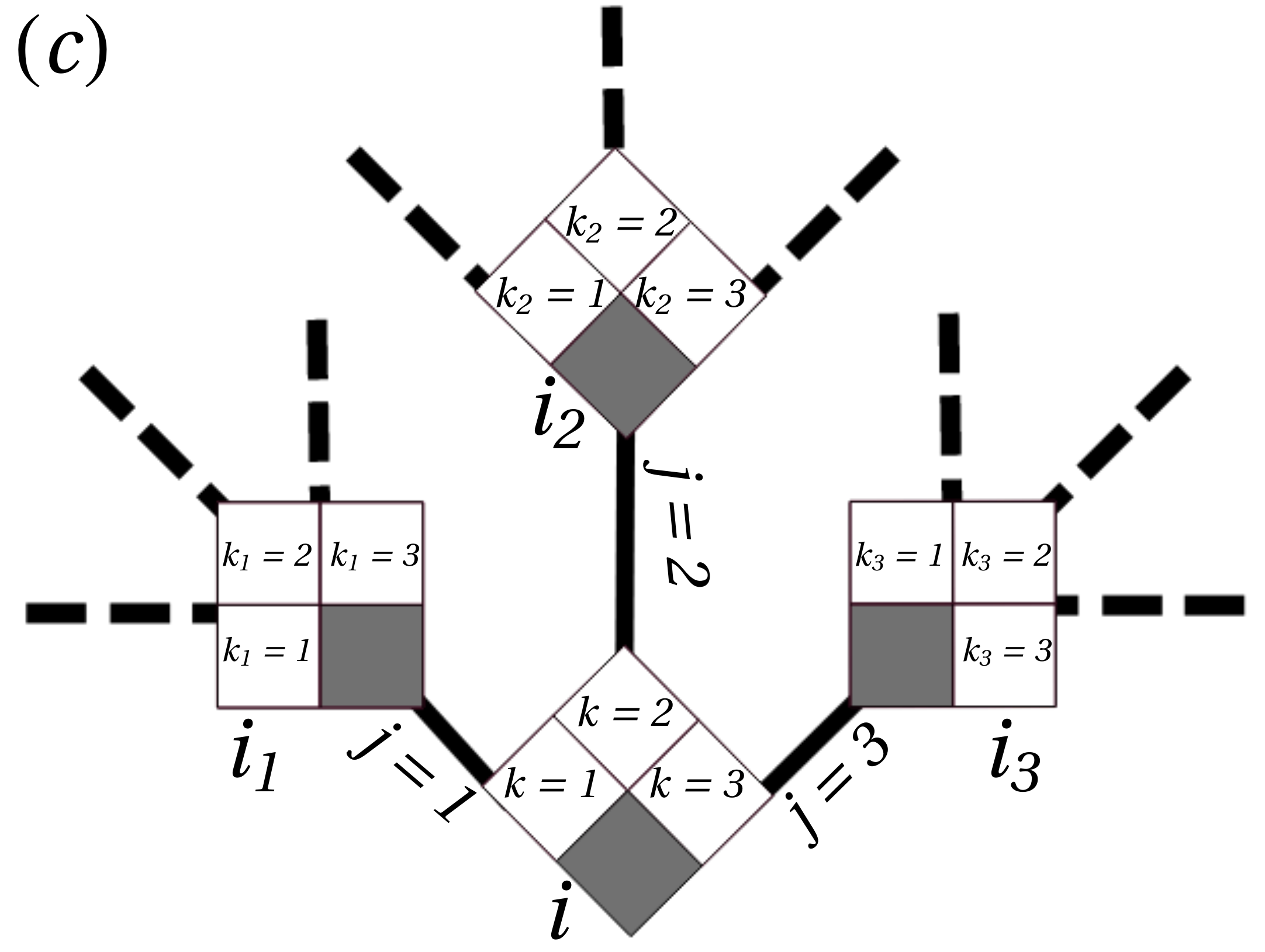}\\
\includegraphics[width=.98\columnwidth]{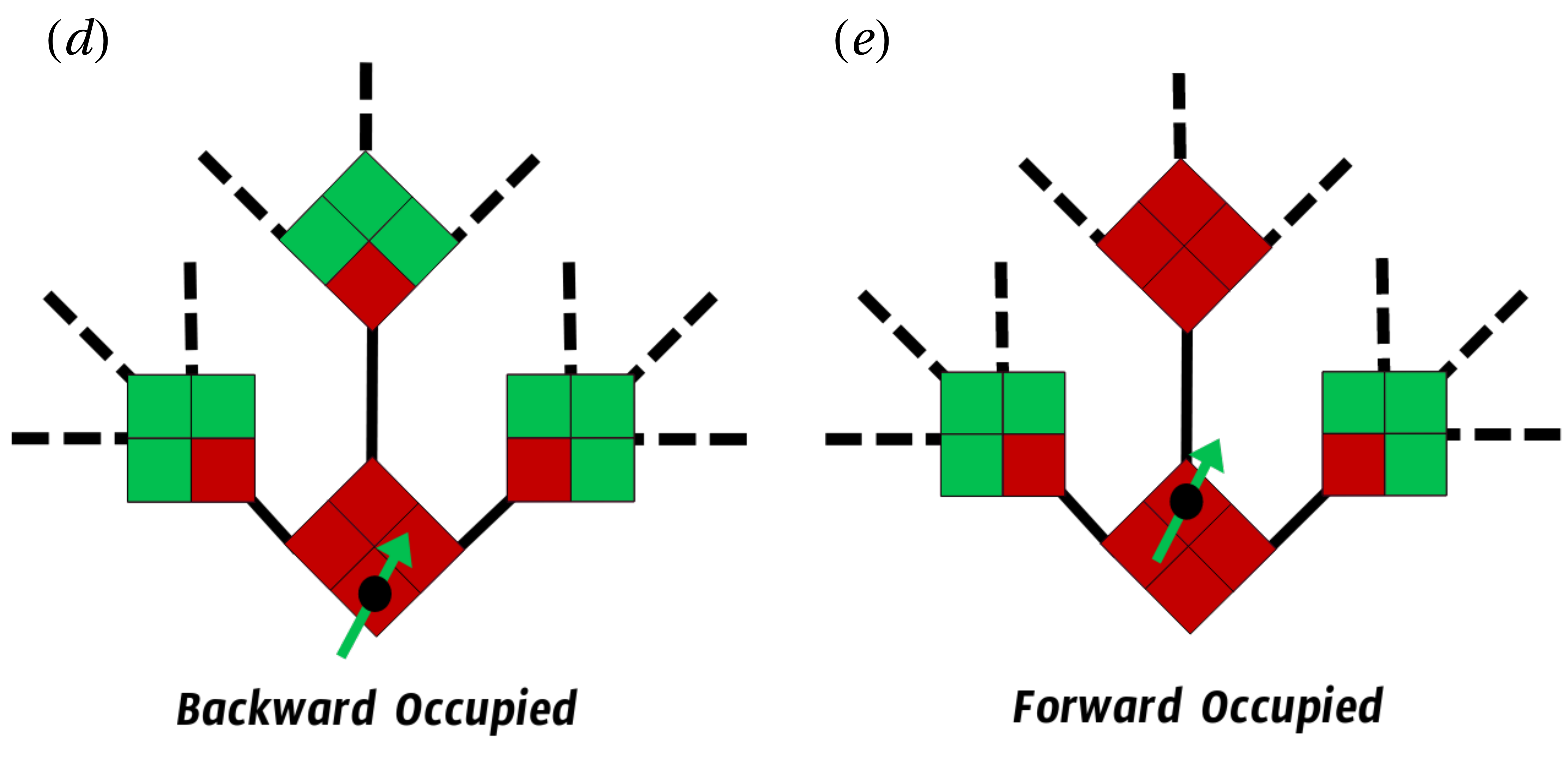}
\caption{{\bf{On-lattice interaction rules.}}
Exclusion and magnetic rules for $(a)$ an XY-spin Blume-Capel model, and $(b)$ the lattice model we use instead. 
When a square box contains a particle (black bullet), it forbids any other particle from occupying the red area.
In case $(a)$, it amounts to the usual exclusion rule. 
In case $(b)$, the exclusion rule depends on the location of the particle within the square box.
All particles carry an XY spin and interact ferromagnetically with their nearest neighbours, as defined by the square lattice (green area).
$(c)$ Rooted tree as obtained by digging a cavity in the Bethe lattice which approximates the original lattice.
$(d)$ and $(e)$ display the two types of occupied states of a given site that arise due to the anisotropy of the exclusion rules. 
\label{fig:ExclusionRules}}
\end{figure}
For $U \to \infty$, this model has been shown to feature a tricritical point atop a phase separation region, even though it then only contains hardcore exclusion and nearest-neighbour ferromagnetic interactions.~\cite{Blume1971}
This model has been shown to be closely related to $2d$ Ising-spin fluids,~\cite{Wilding1996} which suggests that an XY-spin extension could be relevant here.
However, simply replacing the Ising spins by XY spins would lead to an over-simplistic phase diagram: the full particle-hole symmetry in the $U\to\infty$ limit enforces a symmetric coexistence region around a tricritical point at half-filling.~\cite{Blume1971} 
To describe richer physics, we define a slightly different Hamiltonian, 
\begin{equation}
    H = - J \sum\limits_{<i,j>} n_i n_j \bm{S_i}\cdot\bm{S_j}  + U \sum\limits_{<i,j>} f\left(\sigma_i,\sigma_j\right) n_i n_j,
\end{equation}
with XY spins, and where we introduce a repulsion that depends on an internal degree of freedom $\sigma_i$, here used to make interactions slightly anisotropic in a way inspired by previous works on colloidal gels and glasses.~\cite{Tarzia2003,Tarzia2007,Krzakala2008}
We restrict ourselves to the case of hardcore exclusion, $U \to \infty$, as it simplifies calculations in the following.
To be more specific, for the direct XY-spin extension of the BC model (Fig.~\ref{fig:ExclusionRules}$(a)$), one uses standard exclusion rules, namely one particle per site at most (red area), and nearest-neighbour magnetic interactions (green area).
Here, the particle has 4 possible locations within each site of the square lattice (Fig.~\ref{fig:ExclusionRules}$(b)$), thus providing an anisotropy to the exclusion pattern (red area).
The magnetic interactions are unchanged.

We obtain a mean-field description of the lattice fluid by approximating the square lattice by a Bethe lattice, which only contains nearest-neighbour couplings.
We then compute the free-energy density resorting to the so-called cavity method.\cite{Mezard2009,Zdeborova2007} 
Any site $c$ in the Bethe lattice is connected to $4$ neighbouring sites.
Removing site $c$, each of its neighbours becomes the root of a rooted tree-like graph (Fig.~\ref{fig:ExclusionRules}$(c)$).
This is called ``digging a cavity".
On each of these rooted tree-like graphs, it is then possible to obtain recursively the (cavity) probability of the possible states of the root in terms of the (cavity) probabilities of the states of its nearest neighbours in absence of the root itself.
With our choice of interactions, the possible states of the root can be split into three kinds: empty, backward occupied when the particle is located closer to the root, and forward occupied when it is located closer to any of the branches (Fig.~\ref{fig:ExclusionRules}$(d)$-$(e)$).

Let us call $Z_{i\to c}^{B}(\theta_i)$, $Z_{i\to c}^{F_k}(\theta_i)$, and $Z_{i\to c}^{E}$ the cavity  partition functions defined on site $i$ in absence of $c$ and restricted to the backward occupied, forward occupied towards the $k$-th branch, and empty configurations, respectively. 
In cases corresponding to an occupied site, there is a dependence on the orientation of the spin of the particle, here noted as $\theta_i$. 
Let us finally introduce $Z_{i\to c}^{F}(\theta_i) = \sum_{k=1}^3 Z_{i\to c}^{F_k}(\theta_i)$.
In the grand-canonical ensemble the recursion rules for these three quantities and a given cavity site $c$ read,
\begin{subequations}\label{eq:Cavityeqs}
\begin{align}
&Z_{i\to c}^{F_k}(\theta_i) = e^{\beta \mu}  Z_{i_k\to i}^{E} \nonumber \\
&\hphantom{Z_{i\to c}^{F_k}(\theta_i) =} \prod\limits_{j(\neq k)}\left[Z_{i_j\to i}^{E} \vphantom{ \int } + \int d\theta_j Z_{i_j\to i}^{F}(\theta_j)  e^{\beta J \cos\theta_{ij}}  \right] \\
 &Z_{i\to c}^{B}(\theta_i)   = e^{\beta \mu} \prod\limits_{j}\left[Z_{i_j\to i}^{E} + \int d\theta_{j} Z_{i_j\to i}^{F}(\theta_j)e^{\beta J \cos\theta_{i j}}\right] \\
 &Z_{i\to c}^{E}  = \prod\limits_j \left[ Z_{i_j\to i}^{E} +  \int d\theta_j \left(Z_{i_j\to i}^{F}(\theta_j) + Z_{i_j\to i}^{B}(\theta_j) \right)\right]
\end{align}
\end{subequations}
where $\mu$ is the chemical potential, and $i_k$ a neighbouring site in the $k$-th direction (see panel Fig.~\ref{fig:ExclusionRules}$(c)$ for notations).
These equations look cumbersome, but they are in fact quite simple to understand as enumerations of allowed states of the neighbours of the root, $i$, depending on its state.
The first equation, for instance, means that if the root is forward occupied in a given direction, it forces the nearest neighbour in that direction to be empty ($j\neq k$), but the other two neighbours can be either forward occupied or empty.

From these quantities, we can also define recursions on normalized probabilities by dividing each equation by $Z_{i\to c}^T \equiv Z^E_{i\to c} + \int d\theta_i \left(Z^F_{i\to c}(\theta_i) + Z^B_{i\to c}(\theta_i) \right)$, the sum of all partial partition functions on the removed link between $i$ and $c$.
We denote these probabilities $\psi_{i\to c}^{S}$, where $S$ is $E$, $B$, $F_k$ or $F$ just like for their partial partition function counterparts.
We furthermore introduce $\psi_{i\to c}^{O}(\theta_i) = \psi_{i\to c}^{B}(\theta_i) + \psi_{i\to c}^{F}(\theta_i)$, the probability that site $i$ is occupied (in any way) in the absence of the cavity and carries a spin parametrized by $\theta_i$.
From these probabilities, the free energy $F$ and, therefore, the thermodynamic properties can be reconstructed by considering all the processes that allow to reconnect four cavity sites to obtain a well-defined Bethe lattice where all the sites have connectivity equal to four (see Refs.~\cite{Zdeborova2007,Krzakala2008,Tarzia2007,Mezard2009} for more details).
This free energy verifies
\begin{widetext}
\begin{equation}
\begin{aligned}
\beta F =& - \sum_{c=1}^N \ln \left[ \prod_{i=1}^4 \left ( \psi_{i\to c}^{E} + \int d\theta_i \psi_{i\to c}^{O} (\theta_i) \right)
+ e^{\beta \mu} \sum_{p=1}^4 \psi_{p \to c}^{E} \int d\theta_j \prod_{i \neq p} \left( \psi_{i \to c}^{E}  + \int d\theta_i  e^{\beta J \cos (\theta_i - \theta_j)} \psi_{i \to c}^{E} (\theta_i) \right) \right]\\
                 +& \sum_{\langle i,j \rangle} \ln\left[ \psi_{i\to j}^{E} \psi_{j\to i}^{E} +  
                 \psi_{i\to j}^{E} \left( \int d\theta_j \psi_{j\to i}^{O} (\theta_j) \right)
                 + \psi_{j\to i}^{E} \left( \int d\theta_i \psi_{i\to j}^{O} (\theta_i) \right)
                 + \int d\theta_i d\theta_i e^{\beta J \cos (\theta_i - \theta_j)} \psi_{i\to j}^{F} (\theta_i) \psi_{i\to j}^{F} (\theta_i) 
                 \right].
                 \end{aligned}
\end{equation}
\end{widetext}
Looking for homogeneous solutions, the recursive equations~(\ref{eq:Cavityeqs}) become a system of self-consistent algebraic equations, and they can be easily solved numerically once the spin orientation has been discretized (here, we use $16$ values).
Phase separations and phase transitions are then visible as singularities of the free energy or, equivalently, as density jumps or the onset of magnetization.
For completeness, we also seek the liquid-crystal coexistence curves using a similar technique.~\cite{Tarzia2007}
Note that the liquid-crystal transition is absent in the case of the usual BC model,~\cite{Blume1971} and is here a result of the anisotropic exclusion rules.

\begin{figure}
\includegraphics[width=\columnwidth]{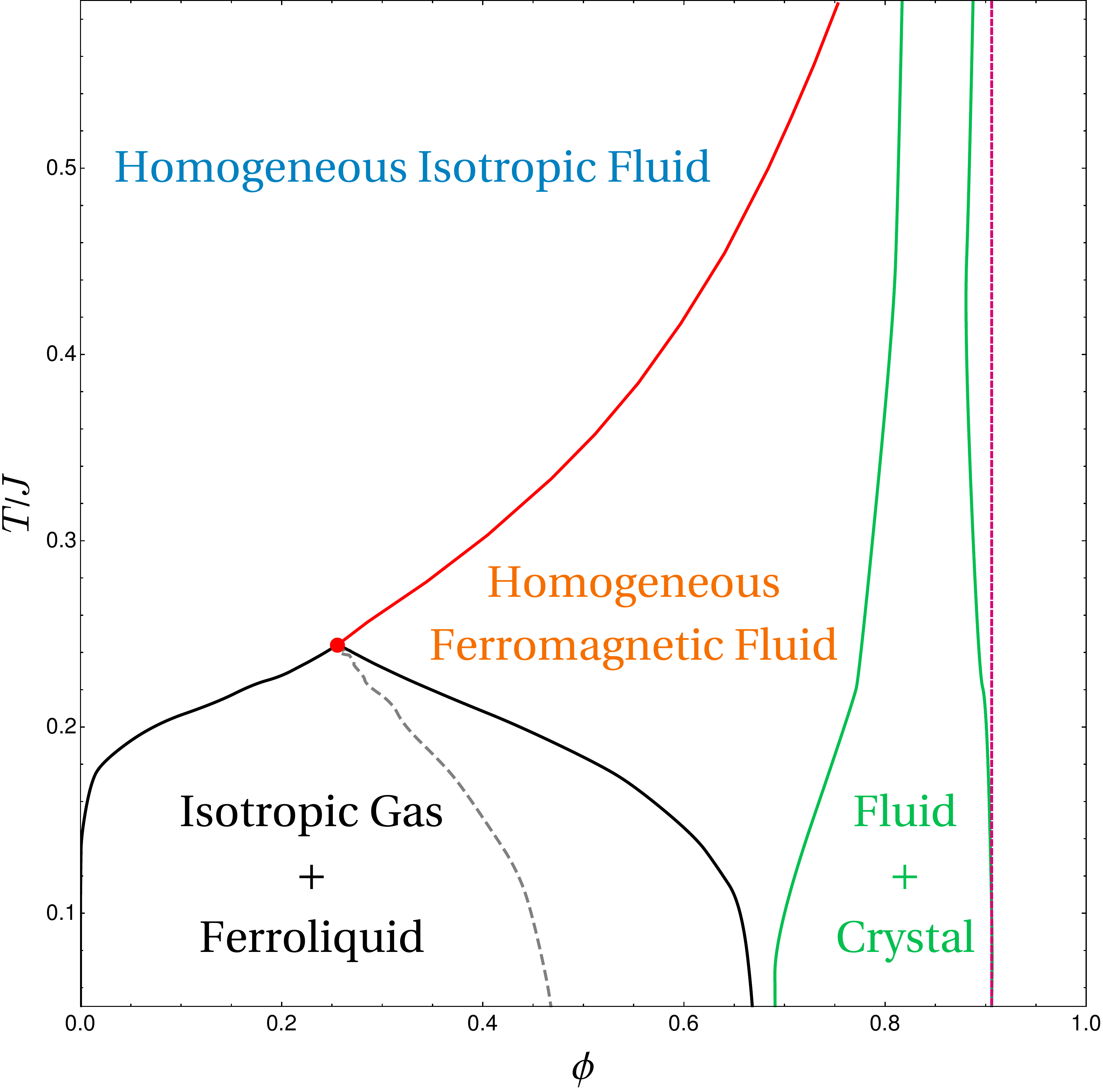}
\caption{{\bf{Bethe-lattice phase diagram.}}
Phase diagram corresponding to a numerical resolution of the coupled equations $(\ref{eq:Cavityeqs})$ for homogeneous and crystalline solutions, in the $(\phi,T/J)$ plane, using 16 discrete values for spin orientations.
We plot the coexistence line between the paramagnetic gas and the ferromagnetic liquid (black), the Curie line (red), the spinodal line associated to the liquid-gas separation (dashed gray), and the fluid-crystal coexistence region (green lines).
We also spot the close-packing density (magenta line).
The left-most branch of the spinodal is practically superimposed with the coexistence curve. 
The tricritical point is found at $\phi_c \approx 0.26$ and $T_c/J \approx 0.24$.  \label{fig:LatticePD}}
\end{figure}
Figure~\ref{fig:LatticePD} displays the so obtained phase diagram in the $(\phi,T/J)$ plane.
The packing fraction is calculated from the ratio of filled sites on the lattice (that can go up to $1$) by rescaling it so that for a completely filled lattice it takes the value of the $2d$ close-packing packing fraction, $\phi_{CP} = \pi \sqrt{3}/6 \simeq 0.9069$. 

The mean-field phase diagram in Fig.~\ref{fig:LatticePD} captures the phenomenology observed in the simulations: liquid-gas phase coexistence takes place, even for hard-core exclusion; it terminates at a tricritical point, located within this approximation at $\phi_c \approx 0.26$ and $T_c/J \approx 0.24$. 
These values, especially the packing fraction, are in good agreement with the simulation results considering that it is just a mean-field approximation and that, moreover, it approximates the soft potential of the simulations by a hard one.
Note that within mean field, the Curie line is a true critical line, which becomes a crossover in $2d$, as already stressed in Sec.~\ref{sec:MD}.
Our combined analysis of magnetic and density degrees of freedom give us access to the spinodal lines.
On the gas-side, we observe that the spinodal is essentially superimposed with the coexistence curve.
This corroborates the numerical observations.
Finally, we reckon that within the range of explored temperatures, the fluid-crystal coexistence region and the liquid-gas coexistence one are well-separated.
This ensures that not taking into account the solid phases when discussing the fluid ones is justified.

\section{Curie-Weiss-like approximation \label{sec:CurieWeiss}}

We now discuss an alternative off-lattice approach to describe a $2d$ spin fluid at the mean-field level while keeping some information about the liquid structure.
The idea is to introduce an equivalent of the Weiss molecular field~\cite{Weiss1948,Kittel1949} in continuous space, in order to get a self-consistent Curie-Weiss-like mean-field resolution.
To do so, we first consider, much like in Langevin's theory of paramagnetism, a single spin placed in a thermal bath at inverse temperature $\beta$, and in a magnetic field with amplitude $h$.
Let $\theta$ be the angle between the spin and the magnetic field.
The thermodynamic average magnetization of this spin, $m = \langle \cos\theta\rangle$, can be written as
\begin{equation}
    m = \frac{1}{Z_0}\int\limits_0^{2\pi} d\theta \cos\theta \, e^{\beta h \cos\theta},
\end{equation}
with $Z_0$ the partition function. 
The integral in the right-hand side can be rewritten in a more compact form by introducing $I_n$, the modified Bessel function of the first kind with parameter $n$,~\cite{Abramowitz1972}
\begin{equation}
    m = \frac{I_1 (\beta h)}{I_0(\beta h)}.
\end{equation}
Like in the usual Curie-Weiss approximation for lattice models, we now want to describe the spin fluid by a collection of spins that are only coupled through an effective field, defined through
\begin{equation}
    h_{\rm eff}(\beta,\phi,m) = \frac{1}{2} N_n(\langle\cos\theta\rangle = m) J m, \label{eq:heff}
\end{equation}
where $N_n$ is the number of magnetic neighbours of a particle in the spin fluid, $J$ is an effective amplitude of the ferromagnetic coupling over the interaction range, and $m$ is the magnetization.
The self-consistent equation for the magnetization in this Ansatz is then
\begin{equation}
    m = \frac{I_1 (\beta h_{\rm eff}(\beta,\phi,m))}{I_0(\beta h_{\rm eff}(\beta,\phi,m))}. \label{eq:CW1}
\end{equation}

Computing the magnetization from this equation requires the knowledge of the number of neighbours $N_n$ at a given temperature and packing fraction in the spin fluid.
The latter is closely related to the radial distribution function $g(r)$ introduced in Sec.~\ref{sec:MD}.
Indeed, recalling that the range of magnetic interaction is denoted $\sigma$, the number of magnetic neighbours reads
\begin{equation}
    N_n = 2\pi \rho \int\limits_{r = 0}^{\sigma}\,dr\,r\,g(r). \label{eq:CW2}
\end{equation}

We therefore need to estimate $g(r)$.
Computing $g(r)$ from the microscopic Hamiltonian is a central task of the theory of simple liquids, for which various methods have been devised.~\cite{Hansen2006}
In these methods, one generally first computes or chooses an approximate form for the direct correlation function, $c(r)$, the part of $g(r)$ that only contains the correlation between the position of two particles due to their direct interactions, as opposed to longer-range correlations that are mediated by other particles.~$c(r)$ and $g(r)$ are related by the Ornstein-Zernike equation,~\cite{Hansen2006}
\begin{equation}
    h(r_{12}) = c(r_{12}) + \rho \int d^2\bm{r}_3 c(r_{13}) h(r_{32}), 
\end{equation}
where $h(r) = g(r) - 1$ and $r_{ij} = \left|\bm{r}_i - \bm{r}_j\right|$.

The next two subsections correspond to two different methods to approximate $g(r)$ in a spin fluid.
In both cases, we consider square-shaped interaction potentials,
\begin{subequations}
\begin{align}
 U(r) &= u \Theta(\sigma_{\rm rep} - r), \\
 J(r) &= j \Theta(\sigma - r),
\end{align}
\end{subequations}
where $\Theta$ is a step function, $u>j$, and $\sigma_{\rm rep} \leq \sigma$.
We use $\sigma_{\rm rep}$ as the unit length scale, and $f = (\sigma / \sigma_{\rm rep})^{-1}$ parametrizes range ratio between repulsion and magnetic interactions. 
As before, $j$ is used as a unit energy scale.
The first one is inspired by the standard Percus-Yevick approach for hard spheres. 
It applies in the hard-disk limit, corresponding to $u/j \to \infty$, with finite $j$ and $f$ fixed. 
In this context, $f\in \left[0;1\right]$ with $f = 0$ corresponding to infinite-range magnetic coupling, and $f = 1$ corresponding to vanishing-range magnetic interactions. 
The second one, inspired by the virial expansion methods.
It provides results which are more analytically tractable.
In the third subsection, we discuss the respective advantages of these methods in the context of hard disk exclusion.
In a last subsection, we finally discuss the role played by $f$ and $u/j$ in the virial method, taking advantage of the tractability of this scheme.

\subsection{Percus-Yevick-Random-Phase-Approximation-Curie-Weiss (PY-RPA-CW)}

An approach that is rather usual in studies focussing on hard disks with an attractive tail consists in using a Percus-Yevick Ansatz for the core exclusion part of the potential, and a simple treatment of the attractive part, like for instance a Random-Phase Approximation (RPA), also called Mean-Spherical Approximation (MSA).~\cite{Hansen2006}
The Percus-Yevick Ansatz, applied to a liquid of hard disks, consists in approximating $c(r)$ by $g(r)(1 - \exp{\beta U(r)})$, which leads to a nice closure of the Ornstein-Zernike equation.
In so doing, the Percus-Yevick approach leads to an analytic form of the direct correlation function for a hard-sphere liquid in $3d$. 
This advantage is lost in even dimensions, but some efficient quasi-exact analytic form can be found by an extrapolation of their low density values. 
Here, we use a form proposed in the literature,~\cite{Ripoll1995} that reproduces the exact numerical values up to packing fractions that are very close to crystallization. 
The RPA, on the other hand, consists in a linearization of the attractive part of the potential.
Since the present particles carry spins, the attractive part of the potential not only depends on the distance between two particles, $r$, but also on their relative spin orientation $\theta$.
In this context, the direct correlation function reads
\begin{eqnarray}
 c(r,\theta) &=& c_{\rm PY}(r) + c_{\rm RPA}(r,\theta),
\end{eqnarray}
where
\begin{eqnarray}
 &c_{\rm RPA}(r,\theta) =& \beta \frac{j}{2} \cos\theta \, \Theta(\sigma - r)\Theta(r - \sigma_{\rm rep} ), \\
 &c_{\rm PY}(r) =& \Theta\left(1 - \frac{r}{\sigma_{\rm rep}}\right) c_0(\phi)  \nonumber\\
	      &\quad& \times \left(1 - 4 \phi + 4 \phi \omega_2(\frac{r}{2\sigma_{\rm rep}}) + s_2(\phi) \frac{r}{\sigma_{\rm rep}} \right),\nonumber
\end{eqnarray}
and
\begin{eqnarray*}
 c_0(\phi) &=& - \frac{1 + \phi + 3 p \phi^2 - p \phi^3}{(1 - \phi)^3}, \\
 s_2(\phi) &=& \frac{3}{8}\phi^2 \frac{8(1-2p) + (25-9p)p\phi - (7-3p)p\phi^2}{1 + \phi + 3 p \phi^2 - p \phi^3}, \\
 \omega_2(x)&=& \frac{2}{\pi}\left( \arccos x - x\sqrt{1 - x^2} \right), \\
 p &=& \frac{7}{3} - \frac{4 \sqrt{3}}{\pi}.
\end{eqnarray*}

\begin{figure}
\includegraphics[width=\columnwidth]{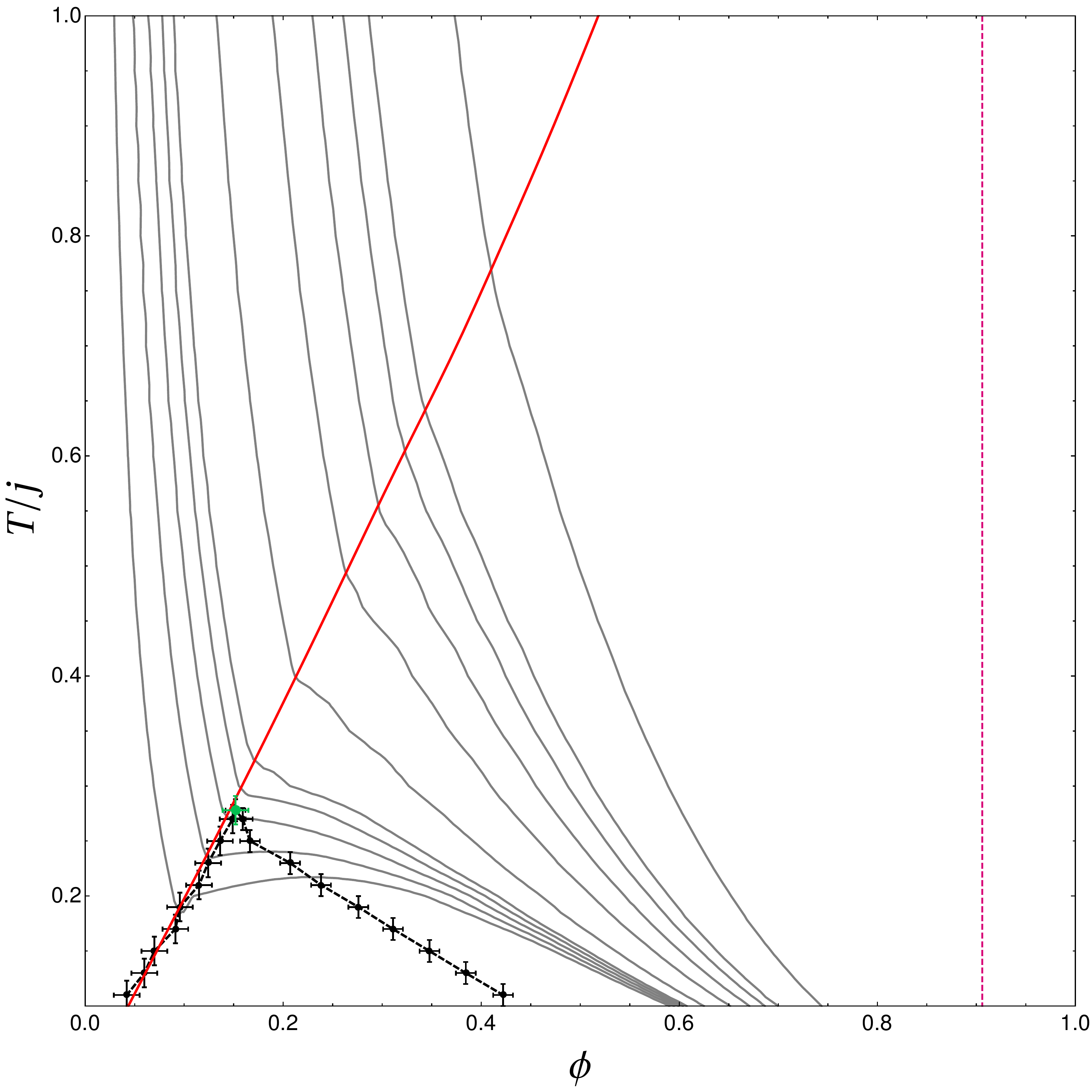}
\caption{{\bf{Isobaric curves and Curie line in the PY-RPA-CW description.}}
We show the Curie line (solid red), spinodal curves (solid black) and a few isobaric curves (solid gray), for the same magnetic range-core size ratio as in the on-lattice calculation ($f \approx 0.72$), close to the critical point. 
We added the close-packing line in magenta.  
Error bars are shown on the spinodal curve, corresponding to a rough evaluation of the stability of our algorithm searching for a vanishing of the inverse compressibility. 
We find a phase separation between an isotropic gas and a ferroliquid, with a tricritical point located at $\phi_c \approx 0.1518$, $T_c \approx 0.278$, plotted above as a green dot. 
Also notice that the Curie line seems to be playing the role of a left-most spinodal under the critical point, and isobaric curves display a cusp when they cross it rather than a smooth minimum.
\label{fig:PYRPAPD}}
\end{figure}

The Ornstein-Zernike equation is usually solved in Fourier space to deal with the otherwise cumbersome convolution.
Here, we resort to both a Fourier transform over space and a Fourier series over the angles.
Because of the linearization of attractive interactions in the context of the RPA, $c$ depends on $\theta$ only through $\cos n\theta$ with $n=0,1$, so that the series contains only two terms,
\begin{eqnarray*}
c(r,\theta) &=& \frac{1}{(2\pi)^2}\int d^2{\bm{k}} e^{- i \bm{k}\cdot \bm{r}} \sum\limits_{n = 0}^{1} \hat{c}_n (\bm{k}) \cos n\theta .
\end{eqnarray*}
The pair correlation function then reads
\begin{align}
 g(r,\cos\theta) &= \nonumber \\
 1 + &\frac{1}{(2\pi)^2}\int d^2\bm{k} e^{-i \bm{k}\cdot\bm{r}} \sum\limits_{n=0}^1 \frac{\hat{c}_n(k) \cos n\theta}{ 1 - \rho \hat{c}_n(k) }.
\end{align} 
Finally, the number of neighbours $N_n$, the spins of which have a given orientation $\theta$ relative to that of a central particle is
\begin{equation}
 N_n( \cos\theta) = 2 \pi \rho\int\limits_{\sigma_{\rm rep}}^{\sigma}dr\,r\,g(r,\cos\theta). \label{eq:Nn}
\end{equation}
Within the context of the present Curie-Weiss approximation, $\cos\theta = m$, so that Eq.~(\ref{eq:heff}), Eq.~(\ref{eq:CW1}) and Eq.~(\ref{eq:Nn}) form a self-consistent set of equations from which the magnetization $m_{\rm SC}(\phi,T)$ and, therefore, the Curie line can be computed.
Correspondingly, once the self-consistent magnetization is known, we obtain the liquid properties, and in particular the compressibility $\chi_c$ through the thermodynamic relation:~\cite{Hansen2006}
\begin{equation}
 \frac{1}{\rho k_B T \chi_c} = 1 - \rho \hat{c}(k = 0, m_{\rm SC}).
\end{equation}
The zeros of the inverse compressibility define the spinodal curve.
Finally, integrating this equation over the density yields the compressibility state equation for the pressure~\cite{Hansen2006}
\begin{equation}
 P = \rho k_B T - k_B T \int\limits_{\rho' = 0}^{\rho} d\rho'\,\rho'\,\hat{c}(k = 0, m_{\rm SC}).
\end{equation}

Figure~\ref{fig:PYRPAPD} displays the phase diagram obtained following the method explained above, together with a few isobaric curves.
The method predicts a liquid-gas phase separation at low temperature, which ends at a tricritical point on the Curie line.
Just like in the Bethe-lattice approach, the Curie line is here a true critical line as the Curie-Weiss self-consistent approach is mean-field-like, but it becomes a crossover in $2d$.
We also recover that the left-most spinodal curve seems to be exactly located on the Curie line.
Interestingly enough, the isobaric curves, when crossing the spinodal line, show a cusp instead of a flat minimum, indicating a sudden jump of compressibility associated to the nucleation of the liquid. 

\subsection{Virial Expansion - Curie-Weiss (VE-CW)\label{sec:Virial}}

In order to complement the PY-RPA-CW description and to discuss the role played by the amplitude of the repulsive potential and interaction ranges, we analyze the system in the low-density limit.
In this context, we write the direct correlation function up to first order in density using a standard Mayer expansion of the partition function~\cite{Hansen2006,Mayer1940}
\begin{align}
    c(r,\theta) &\simeq f(r,\theta) + \rho f(r,\theta) \left( f \star f \right)(r,\theta),
\end{align}
where $f$ is the Mayer function associated to the pairwise interaction potential
\begin{eqnarray}
 f(r,\theta) &=& \Theta(\sigma - r)\left(e^{\beta j \cos \theta} - 1 \right)  \\
	     &+&  \Theta(\sigma - r)  \Theta(\sigma_{\rm rep} - r) e^{\beta j \cos\theta} \left( e^{-\beta u} - 1\right), \nonumber
\end{eqnarray}
and $\star$ is a convolution product computed over both positions and angles,
\begin{eqnarray*}
f\star g(r_{12},\theta_{12}) &\equiv& \int d^2\bm{r_3} d\theta_3 f(r_{13},\theta_{13}) g(r_{32},\theta_{32}).
\end{eqnarray*}
We then use the Ornstein-Zernike equation in real space, cut it off at order $\rho$, and thence get the angle-dependent version of the pair correlation function,
\begin{eqnarray}
 g_{\theta}(r,\theta)		 &=& g_{\theta}^{(0)}(r,\theta) + \rho g_{\theta}^{(1)}(r,\theta),
 \end{eqnarray}
 with
 \begin{eqnarray}
 g_{\theta}^{(0)}(r,\theta)	 &=& 1 +  f(r,\theta), \\
 g_{\theta}^{(1)}(r,\theta) 	 &=& (1 + f(r,\theta)) \left( f \star f \right)(r,\theta).
\end{eqnarray}
The usual pair correlation function $g$ is obtained by averaging $g_\theta$ over the spin angles,
\begin{equation}
    g(r) = \int d\theta p (\theta) g_\theta(r,\theta),
\end{equation}
where $p(\theta)$ is the angle distribution of spins in the spin fluid at the considered density and temperature.
The actual angle distribution is a priori a complicated function to compute.
In the context of our self-consistent scheme, we choose a Von Mises distribution parametrized by temperature and an effective field $h$ that mimics the distribution observed for an isolated spin coupled to $h$:
\begin{equation}
 \Upsilon(\theta;\beta h) \equiv \frac{e^{\beta h \cos\theta}}{2 \pi I_0(\beta h)}. \label{eq:VM}
\end{equation}
This distribution provides a natural way to define the analogue of a Gaussian law wrapped on a circle,~\cite{Evans2000} whose two notable limits are the low- and high-variance regimes
\begin{eqnarray*}
 \Upsilon(\theta;\beta h) &\underset{\beta h \to 0}{\sim}& \frac{1}{2\pi} + O\left(\beta h\right), \\
  \Upsilon(\theta;\beta h) &\underset{\beta h \gg 1}{\sim}& \sqrt{\frac{\beta h }{2 \pi} } e^{- \frac{\beta h}{2} \theta^2} \underset{{\beta h \to \infty}}{\to} \delta(\theta).
\end{eqnarray*}
The number of magnetic neighbours reads
\begin{equation}
    N_n = 2 \pi \rho \int\limits_{r=0}^{\sigma}\,dr\,r\int d\theta \Upsilon (\theta; \beta h) g_\theta(r,\theta). \label{eq:NnVir}
\end{equation}
By setting $h = h_{\rm eff}$, the effective field introduced in Eq.~(\ref{eq:heff}), Eq.~(\ref{eq:CW1}) and Eq.~(\ref{eq:NnVir}) form a self-consistent set of equations from which we compute the magnetization and Curie line in the low-density limit. 

Regarding the liquid properties of the fluid, notice that the low-density expansion of the correlation functions is equivalent to a cut-off virial expansion, which is usually written as an equation of state for the pressure,~\cite{Hansen2006}
\begin{equation}
 \frac{\beta P}{\rho} = 1 + B_2 (T) \rho + B_3 (T) \rho^2 + \dots \label{eq:Virial}
\end{equation}
where $B_2$ and $B_3$ are the first two virial coefficients,
\begin{eqnarray}
 B_2 (T)  &=& - \frac{1}{2} \int d\bm{r} d\theta_1 d\theta_2 p(\theta_1)p(\theta_2)f(r,\theta_{12}), \\
 B_3 (T)  &=& - \frac{1}{3} \int d\bm{r_1} d\bm{r_2} d\theta_1 d\theta_2 d\theta_3 p(\theta_1)p(\theta_2)p(\theta_3) \nonumber \\
	     &\quad& \qquad \times f(r_{12},\theta_{12})f(r_{2},\theta_{23})f(r_{1},\theta_{31}).
\end{eqnarray}
These coefficients are averages over spin orientations and particle positions of the zero-th and first terms of the density expansion of $c(r)$, assuming an homogeneous spatial density $\rho$ and a spin orientation distribution $p(\theta)$.

Inserting the Von Mises distribution~(\ref{eq:VM}) into the definitions of $B_2$ and $B_3$, we then obtain exact expressions for these virial coefficients.
Details on the computation and their precise shapes are given in App.~\ref{app:VECW}.  
The first one, that is not too cumbersome, reads
\begin{align}
 B_2(h&,T) = \frac{\pi}{2}\sigma^2 \left\{\left[1 - I_0(\beta j) \right] - 2\sum\limits_{n=1}^\infty I_n(\beta j)\frac{I_n(\beta h)^2}{I_0(\beta h)^2}  \right\} \nonumber \\
	  &+\frac{\pi}{2}{\sigma_{\rm rep}}^2 \left[ I_0(\beta j) +   2\sum\limits_{n=1}^\infty I_n(\beta j)\frac{I_n(\beta h)^2}{I_0(\beta h)^2} - e^{-\beta u} \right].  \label{eq:B2}
\end{align}
$B_3(h,T)$ is given by a similar expression: both coefficients contain a series of Bessel functions $I_n(\beta j)$ and $I_n(\beta h)$ with $n\in\mathbb{N}$, which is the equivalent of a high-temperature expansion~\cite{Kardar2007a} of an XY model. 
We give in Eq. (\ref{eq:B2}) the full $n$ expansion, but the terms happen to be quickly decreasing with order $n$ at any value of $\beta j$ and $\beta h$.~\cite{Abramowitz1972} 
This allows us to cut off the series making negligible errors in the ensuing computations. 
We also check that these expressions, in the zero-field, standard hard-disks limit ($h \to 0, j\to 0, u\to \infty$) do yield the usual hard disks coefficients (see App.~\ref{app:VECW} for more details).~\cite{Hemmer1965}

\begin{figure}
\includegraphics[width=\columnwidth]{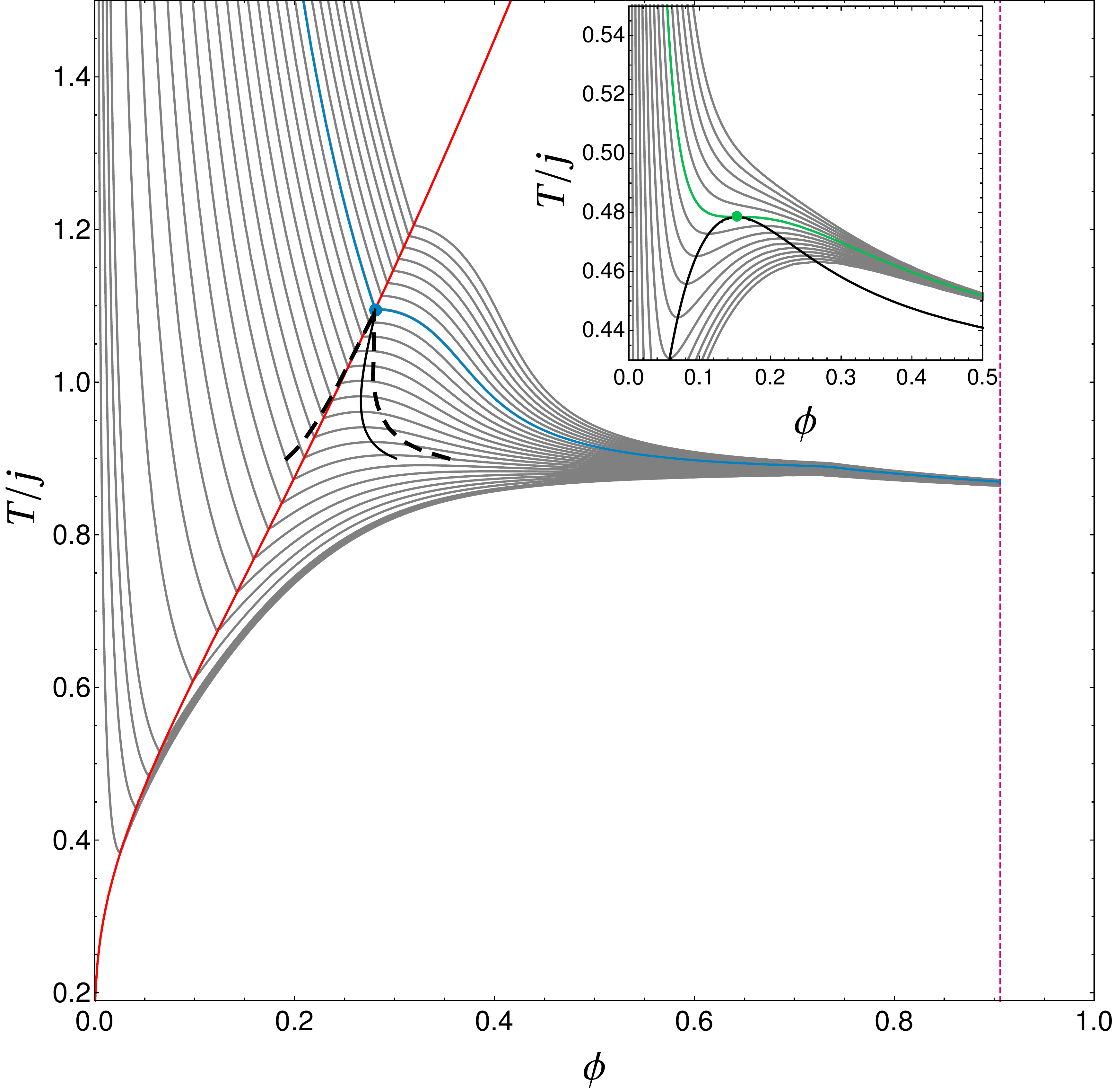}
\caption{{\bf{Isobaric curves and magnetization in the VE-CW description.}}
The isobaric curves (gray) and Curie line (red) were obtained with the VE-CW description of the spin fluid, in the $(\phi,T/j)$ plane, for $f = 1/2$. 
We find a tricritical point at $\phi_c \approx0.28$ , $T_c \approx 1.10 j$, here plotted in blue along with the critical isobaric curve.
The solid black line indicates the right-hand branch of the spinodal line, and the dashed black line is the coexistence curve found from a Maxwell construction.~\cite{Hansen2006}
In the inset, we plot the same quantities for a uniform distribution of spin angles (zero magnetization). 
A critical point is still observed, at $\phi_c \approx .15$ and $T_c/j \approx .48$, plotted in green along with its critical isobaric curve.
The solid black line is the spinodal line.
\label{fig:VECWPD}}
\end{figure}
We then write the virial equation of state for the pressure as given by Eq.~(\ref{eq:Virial}) using the first two virial coefficients and the effective field $h_{\rm eff}$ found from our self-consistent scheme.
We finally locate the spinodal curves as the points where the derivative of $P$ with respect to $\rho$ vanishes, which correspond to local extrema of isobaric curves in the $\left(\phi,T\right)$ plane.

The main panel of Fig.~\ref{fig:VECWPD} displays the phase diagram obtained using this method, together with a few isobaric curves in the hardcore exclusion limit ($u\to\infty )$ and for an interaction range ratio $f = (\sigma/\sigma_{\rm rep})^{-1}=1/2$.
We recover a liquid-gas phase separation between a paramagnetic gas and a ferromagnetic liquid, ending at a tricritical point. 
We find that, like in the on-lattice and PY-RPA-CW description, the Curie line plays the role of the left-most spinodal, where isobaric curves develop a cusp: this feature is thus a robust property of this system. 
The coexistence curve is also found close to the critical point using a standard Maxwell construction.
It starts off extremely close to the Curie line, as in the on-lattice approach, and has a similar shape as the one obtained in simulations.
The supercritical part of the obtained diagram is also interesting, as isobaric curves feature an inversion of curvature.
This feature is equivalent to the presence of a local maximum in the isobaric thermal dilatancy (or expansion coefficient),
\begin{equation}
 \alpha_P = - \frac{1}{\rho}\left( \frac{\partial\rho}{\partial T} \right)_P.
\end{equation}
Such maxima define the so-called Widom line, which separates gas-like and liquid-like regimes of the supercritical fluid.~\cite{Franzese2007}
While here this might very well be a high-density artifact of the low-order cut-off in the virial expansion, it is also observed in real polar liquids.~\cite{Franzese2007,Gallo2014} 
More in-depth studies of spin fluids would be required to confirm this feature.

Finally, we study the role played by the magnetization in this model by setting the effective field in the Von Mises angle distribution $\Upsilon(\theta; \beta h)$ to zero, so that $p (\theta) = 1/2\pi$.
We show the corresponding phase diagram in the inset of Fig.~\ref{fig:VECWPD}. 
Interestingly, we observe a liquid-gas transition caused solely by the spin-mediated attraction, although this system does not get magnetized as we assumed a flat angle distribution.
In this case, the isobaric curves feature smooth minima in the phase separation region, and do not change curvature in the supercritical fluid regime, indicating that both cusps on the isobars and local maxima of $\alpha_P$ are linked to the onset of the magnetization.

\subsection{Comparison between PY-RPA-CW and VE-CW}

\begin{figure}[b!]
    \centering
    \includegraphics[width=.31\columnwidth,height=.31\columnwidth]{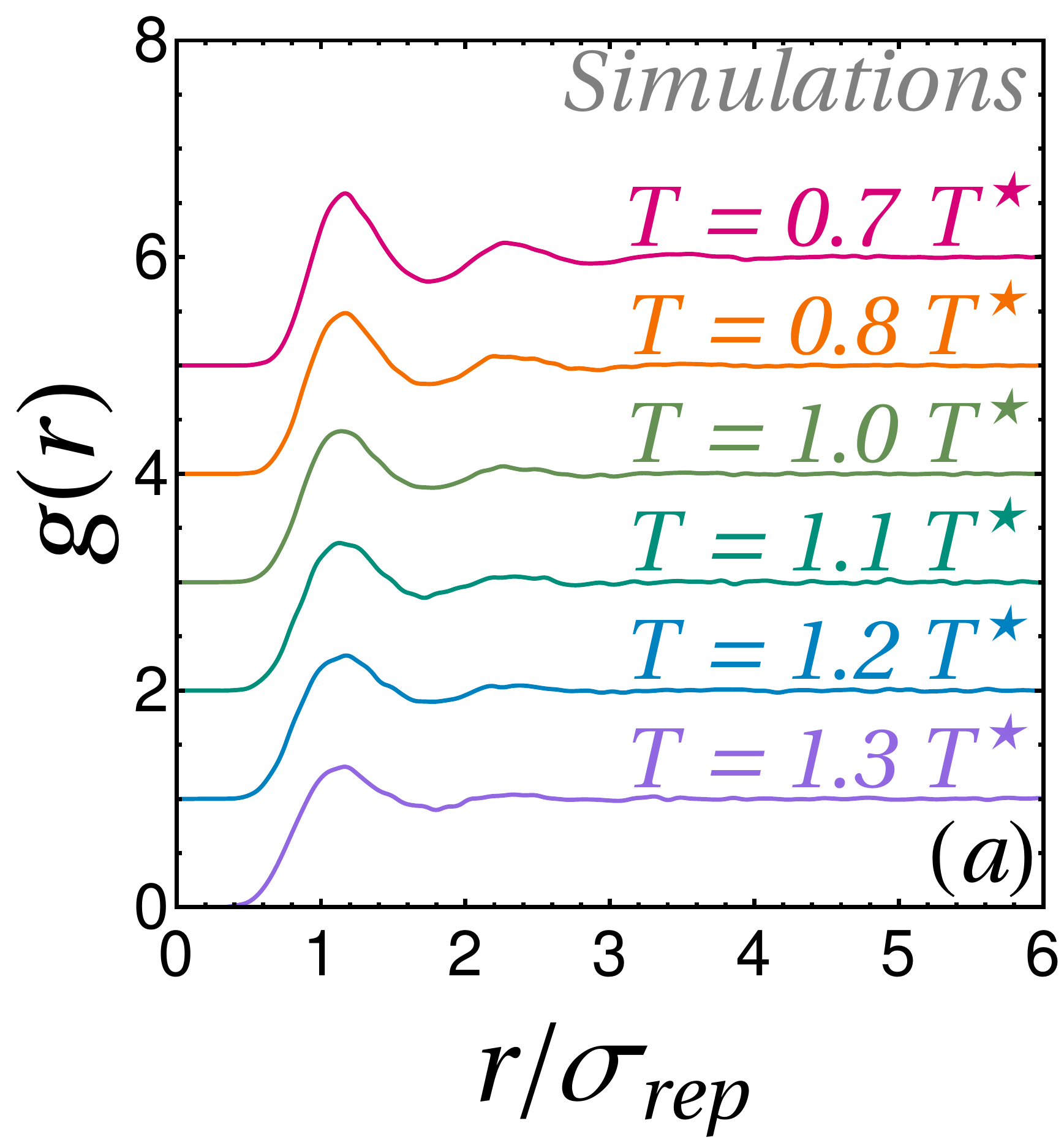}
    \includegraphics[width=.31\columnwidth,height=.31\columnwidth]{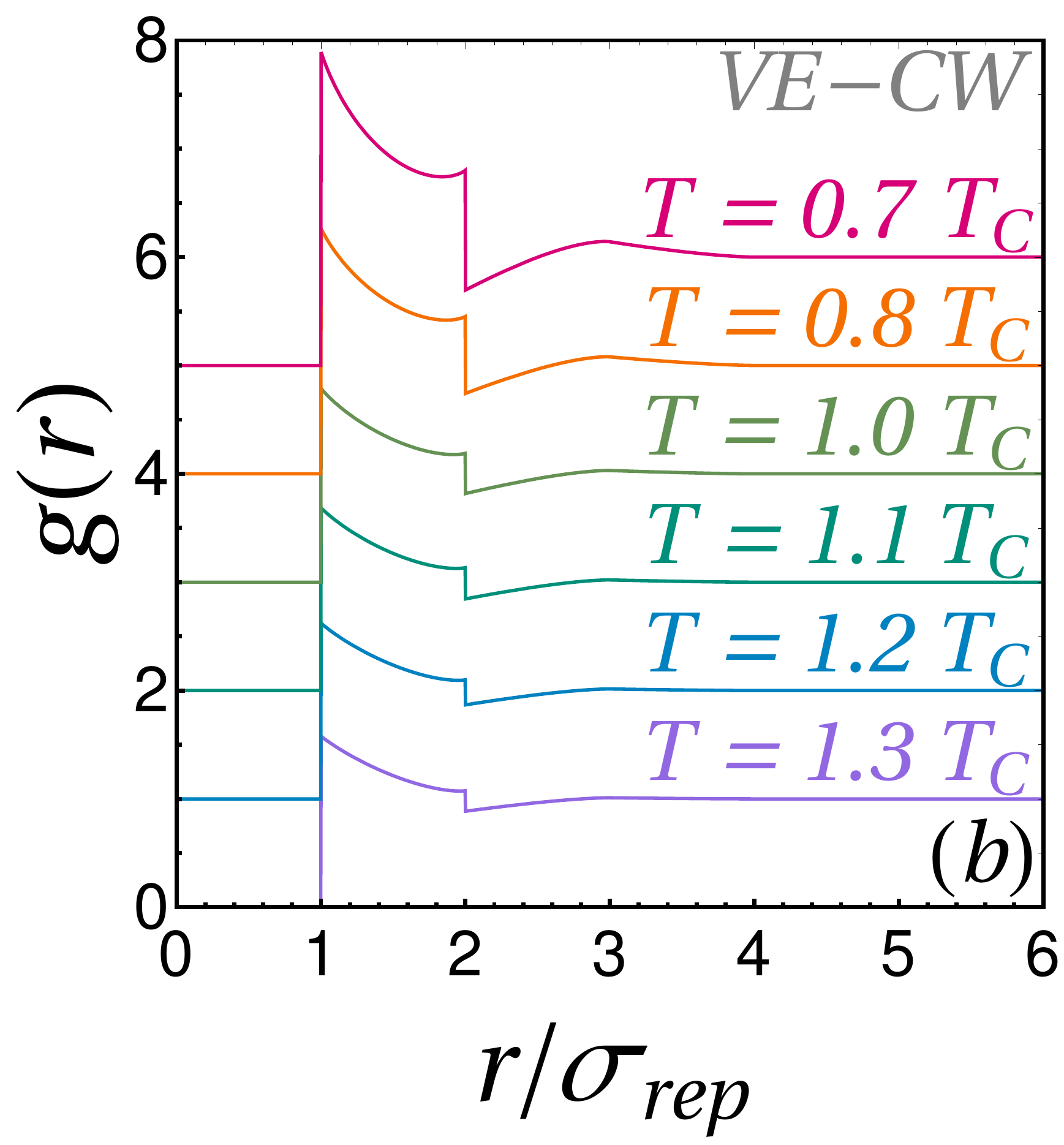}
    \includegraphics[width=.31\columnwidth,height=.31\columnwidth]{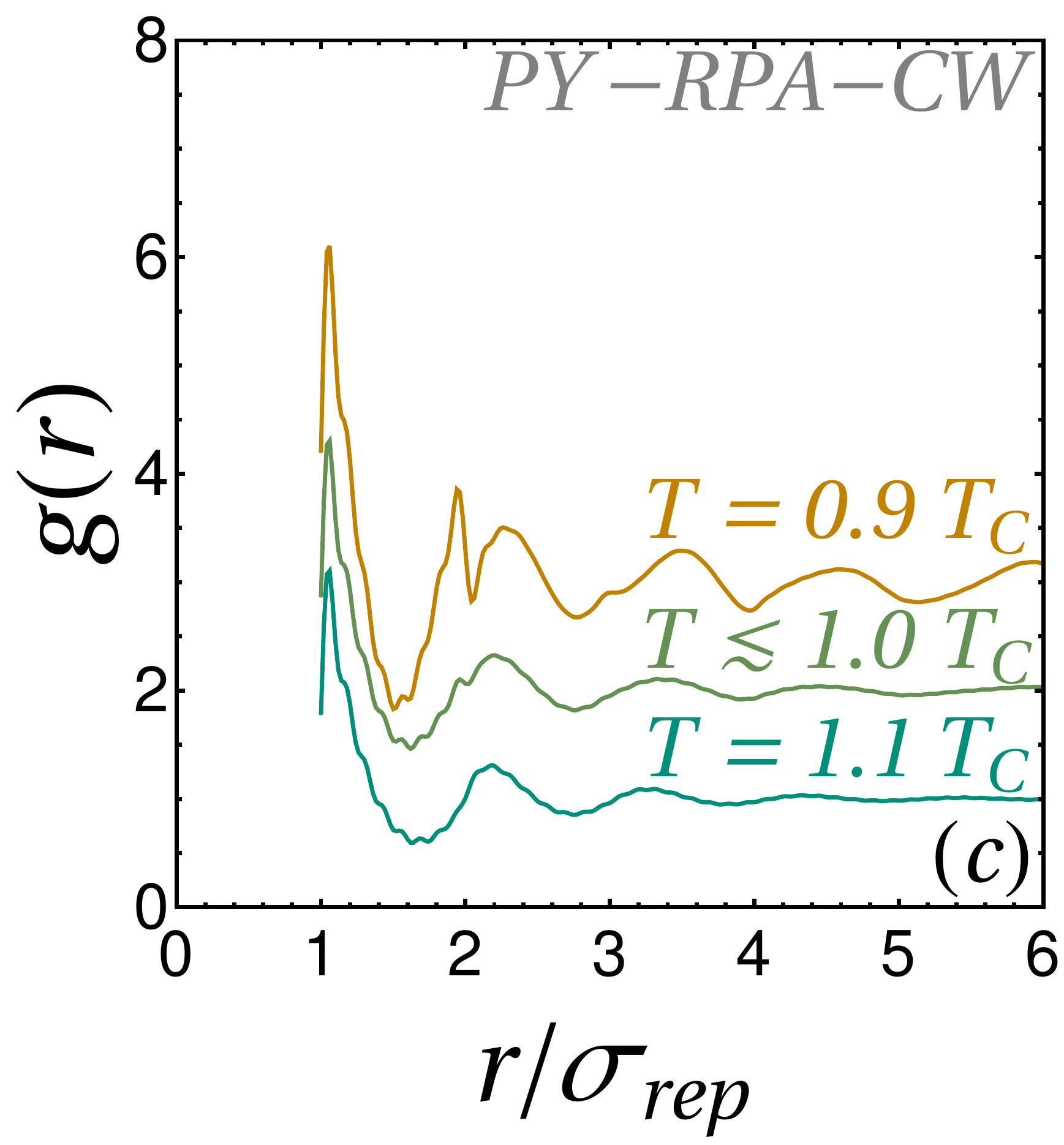}
    \caption{{\bf{Comparison of analytical and numerical radial distribution functions.}}
    We show the Radial Distribution Functions measured at $\phi = 0.55$ $(a)$  in equilibrated simulations with $N=8192$ particles, $(b)$ computed using the VE-CW approximation, and  $(c)$ computed using the PY-RPA-CW approximation. 
    The curves in $(b)$ and $(c)$ are computed in the hardcore limit $u\to\infty$ and for $f=1/2$.
    The simulation curves are measured around the finite-size crossover temperature $T^\star$ (see Sec.~\ref{sec:MD}).
    The theoretical curves are plotted around the Curie temperature $T_C$ found in each approach.
    In all three subplots, curves are shifted by a constant for better comparison.}
    \label{fig:gComp}
\end{figure}

In order to understand the qualitative difference between the outcome of the PY-RPA-CW and VE-CW approaches and how they compare to simulations, it is interesting to compare the pair correlation functions $g(r)$ and the magnetization they predict around the magnetization transition.
Figure~\ref{fig:gComp} displays the pair correlation functions at a packing fraction $\phi = 0.55$ around the Curie temperature in simulations of $N = 8192$ particles (panel $(a)$), in the VE-CW scheme (panel $(b)$) and in the PY-RPA-CW approach (panel $(c)$). 
The last two panels, are computed in the hardcore limit ($u\to\infty$) and for an interaction range ratio $f = 1/2$.
Recall that in simulations, the interaction potentials are instead those defined in Eq.~(\ref{eq:Int}).
The pair correlation functions obtained in the simulations show that the onset of magnetization is accompanied by an increase of structure, with the appearance of a second and third peak.
At the order we consider in this paper, the VE-CW predicts unrealistically sharp features in the pair correlation function, and no structure beyond twice the magnetic interaction range.
This is due to the very nature of the cut-off in the virial expansion: we neglected any event other than two- and three-body interactions, so that the sharpness of hardcore exclusions remains apparent in $g(r)$.
Still, the VE-CW approach captures the structuration of the liquid at the onset of the magnetization, despite the strength of the approximations.
By construction, the PY-RPA-CW route very well describes the hardcore repulsion and avoids the sharp discontinuities in $g(r)$ reported above.
However, long-ranged oscillations develop for temperatures as close as $0.9 T_C$.
This suggests that the effective attraction is very much overestimated.

\begin{figure}
    \centering
    \includegraphics[width=.48\columnwidth]{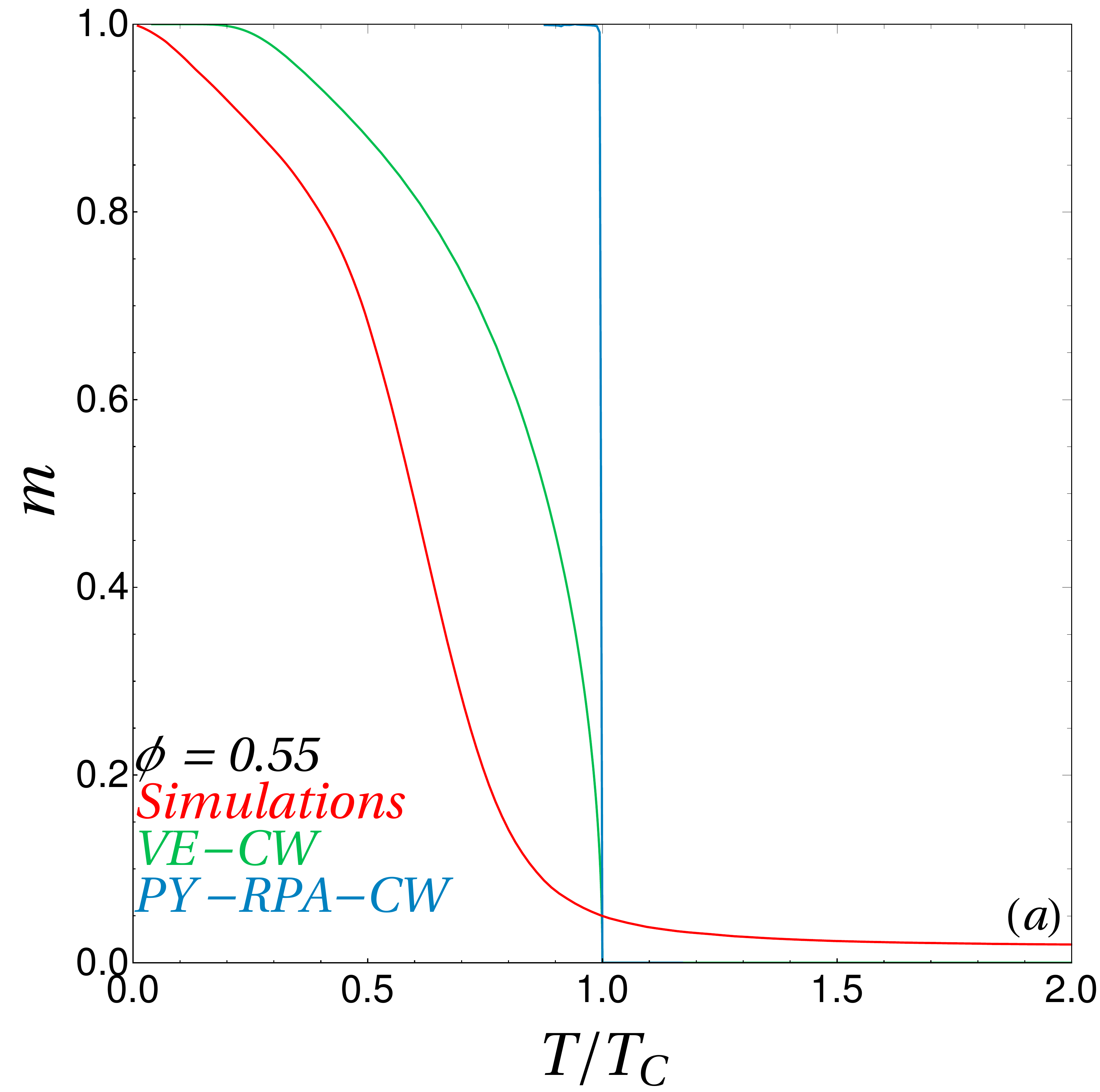}
    \includegraphics[width=.48\columnwidth]{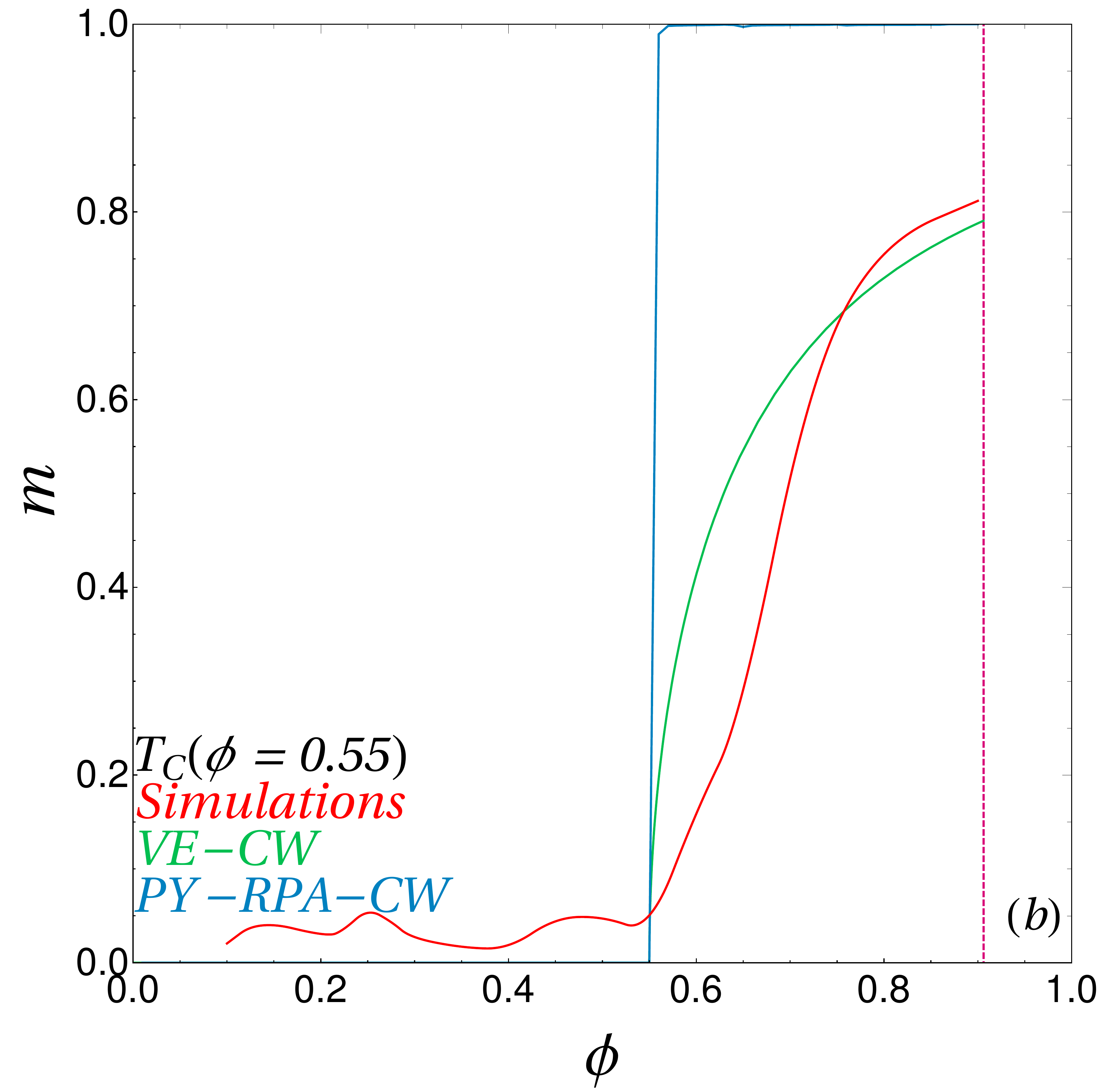}
    \caption{{\textbf{Comparison of the numerical and analytical predictions for the magnetization.}}
    $(a)$ Magnetization versus temperature rescaled by $T_C(\phi = 0.55)$, the corresponding Curie temperature at $\phi = 0.55$.
    $(b)$ Magnetization against the packing fraction at $T_C(\phi = 0.55)$.
    In both panels, simulations of $N = 8192$ particles are shown in red, VE-CW calculations are shown in green, and PY-RPA-CW calculations are shown in blue.
    In both theoretical approaches, $u\to\infty$ and $f = 1/2$.}
    \label{fig:VECWmag}
\end{figure}

This is further confirmed when looking at the magnetization curves.
Figure~\ref{fig:VECWmag} shows the magnetizations observed in simulations and predicted in both approaches against the temperature (panel $(a)$) and the packing fraction (panel $(b)$).
In panel $(a)$, we choose $\phi = 0.55$, and the temperature is normalized by the Curie temperature $T_C(\phi = 0.55)$ observed or predicted in each case.
Although the two analytical approaches are mean-field and therefore display mean-field critical exponents (e.g. $\beta = 1/2$), they predict very different behaviours away from the transition.
Indeed, the PY-RPA-CW approach predicts a very sharp increase to unit magnetization, while the VE-CW approach features a smoother variation, that compares better with simulation data.
In panel $(b)$, the temperature is set to $T_C(\phi = 0.55)$.
Again, the PY-RPA-CW approach predicts a sharp increase to unit magnetization, while the VE-CW grows smoothly and qualitatively reproduces numerical results.

These curves can be used to understand qualitatively the shapes of the isobaric curves in both cases.
In the PY-RPA-CW description, the sharp onset of the magnetization induces a transition across the Curie line from a purely repulsive fluid to an attractive fluid whose attraction is essentially fixed.
This explains the sudden change of slope at the Curie line, which connects the isobaric lines corresponding to these two fluids.
In the VE-CW description, the effective attraction grows smoothly, and the slope of the isobaric curves is governed by the rate of change of the magnetization with both the temperature and packing fraction.
In particular, close to the Curie line, the magnetization increases sharply, and favours a rapid compaction with very little cooling.

\subsection{Role of the Interaction Parameters using the VE-CW scheme}
\subsubsection{Magnetic Interaction Range}

\begin{figure}[b!]
\includegraphics[width=.47\columnwidth]{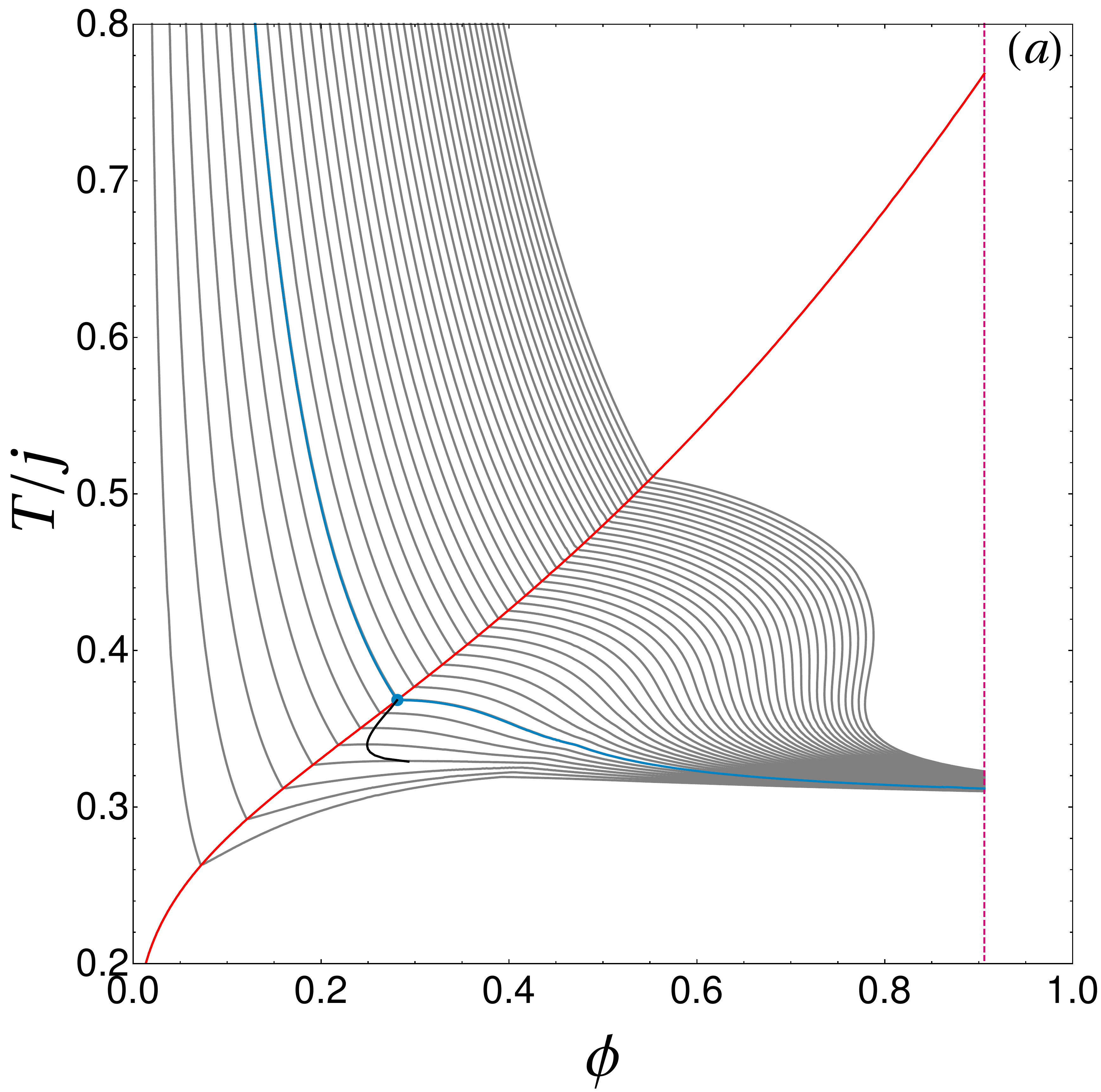}
\includegraphics[width=.50\columnwidth]{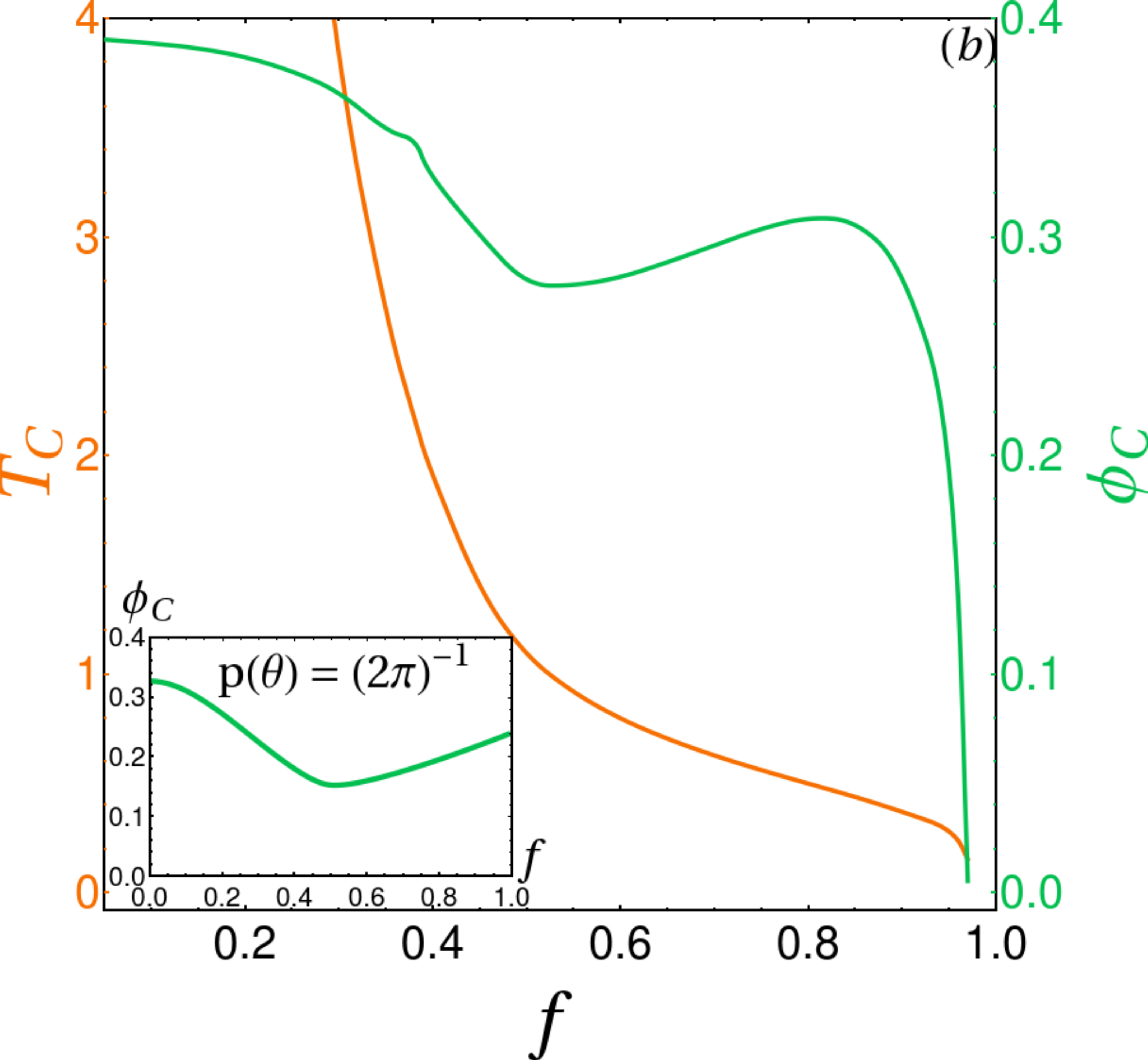}
\caption{{\bf{Effects of the interaction range.}}
$(a)$ Phase diagram in the $(\phi,T)$ plane found in the VE-CW approach diagram for hard disks and $f = 0.9$.
We plot the Curie line (red), the right-most spinodal line (black), some isobaric lines (gray) and we highlight the critical isobaric line as well as the tricritical point (blue).
We also spot the close-packing density (magenta).
$(b)$ Displacement of the tricritical temperature (orange) and packing fraction (green) in the VE-CW description when varying $f$ for hardcore exclusion.
In the inset, we plot the critical packing fraction found from the virial approach with a flat distribution of spins (no magnetization).
 \label{fig:VECWPDf}}
\end{figure}

Using the virial approach, that proved to yield a good qualitative representation of the system in fluid phases, it is also easy to tune the shape of the interaction potentials.
First, let us keep hard disks and change the interaction range ratio $0<f<1$. 
Regardless of this value, we always observe a liquid-gas phase separation and a Curie line that is also the left-most spinodal and the locus of cusps on the isobaric curves, as illustrated in the example of Fig.~\ref{fig:VECWPDf}$(a)$ ($f = 0.9$).

Figure~\ref{fig:VECWPDf}$(b)$ shows how the tricritical point moves in the $(\phi,T)$ plane as $f$ is varied.
As magnetic interactions are made increasingly short-ranged ($f$ increases), the Curie line is sent to lower temperature, and the critical point follows. 
The reason is that as each particle gets fewer magnetic neighbours, magnetic order becomes harder to establish.
The behaviour of the tricritical density is more intriguing. 
It essentially decreases as $f$ increases, suggesting that in the $f\to1$ limit, the coexistence region shrinks towards the origin of the $(\phi,T)$ plane. 
Also, a local minimum of the tricritical density sits close to $f = 1/2$.
As shown in the inset, it is also present in the absence of magnetization, as obtained for a uniform distribution of spins ($p(\theta) = 1/2\pi)$.
It is seemingly related to the fact that for $f=1/2$, one cell of the regular hexagonal packing of the hard disks perfectly matches the attraction range.

Coming back to Fig.~\ref{fig:VECWPDf}$(a)$, it is interesting to note that for high values of $f$ such as $f=0.9$, the isobaric curves present an s-shaped feature at high density. 
Notwithstanding that this feature can be an artifact of the low order cut-off of the virial expansion, if real, it would correspond to a region with a negative expansion coefficient.
Interestingly, this kind of behaviour has also been reported in realistic polar liquids.
Water, for instance, displays a similar anomaly of its expansion coefficient associated to a liquid-liquid transition caused by the ordering of molecules due to H bonds.~\cite{Gallo2016} 
Models of single-component systems with isotropic interactions have successfully reproduced such anomalies using specifically devised attractive potentials.~\cite{Rechtsman2007}  
Our observation that spin fluids present similar features suggests that they could be useful alternative models in which an effective attraction with the appropriate shape sets in spontaneously.
Following another line of thought, it could be interesting to make the connection with the gelation of hard disks with very short attractive interactions, where unusual thermodynamic features are associated to the formation of long ramified chains,~\cite{Zaccarelli2004} and unusual glassy behaviour has been reported for polar ``patchy'' colloids.~\cite{Yoshino2018}
Indeed, the here reported anomalous behaviour takes place close to the sticky disk limit.~\cite{Baxter1968}

\subsubsection{Softness of the Repulsion}

\begin{figure}
\includegraphics[width=.48\columnwidth]{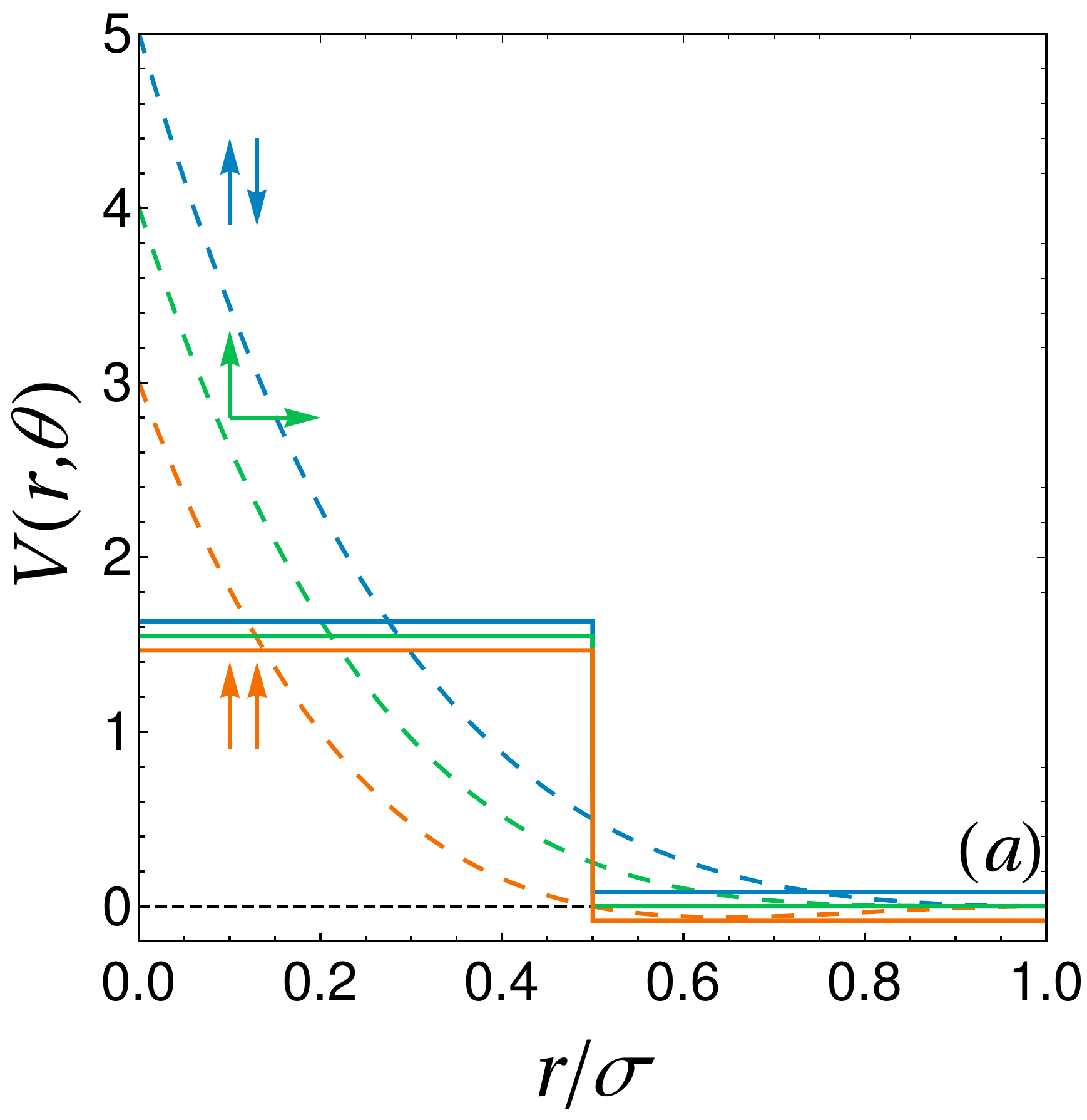}
\includegraphics[width=.48\columnwidth]{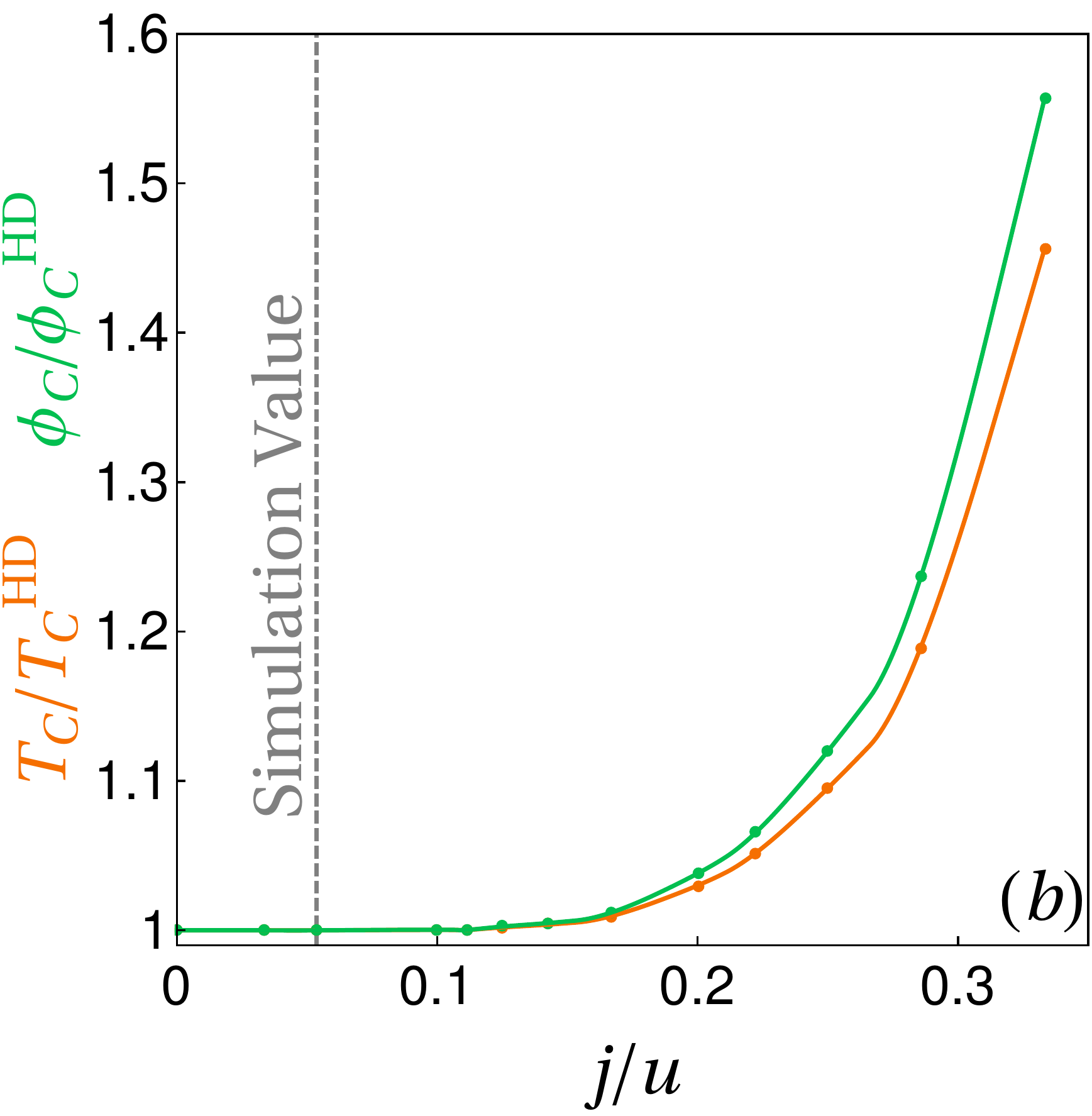}
\caption{{\bf{Effects of the repulsion amplitude.}}
$(a)$ Typical aspect of the square-shaped interaction potential used in the soft VE-CW approach (full lines), compared to the potentials used in simulations (dashed lines). 
The $V = 0$ line, corresponding to the separation between attractive and repulsive parts, is highlighted by a dashed black line.
$(b)$ Effect of $j/u$ on the coordinates of the tricritical point for $f = 1/2$. 
We plot the critical temperature (orange) and packing fraction (green), normalized by the hard-disk value found at the same $f$.
\label{fig:VECWSoft}}
\end{figure}

Finally, we briefly discuss the role of having a soft square-potential repulsion (finite $u$) instead of a hardcore exclusion. 
The corresponding shapes of the effective potential $V(r,\theta) = U(r) - J(r)\cos\theta$ are plotted in Fig.~\ref{fig:VECWSoft}$(a)$, and compared to the potentials used in simulations.
To set the comparison, the amplitude of the square repulsion potential $u$ corresponding to the simulations potential $U(r)$ is taken to be:
\begin{equation}
    u_{\rm sim} \equiv \frac{2}{\sigma}\int\limits_{0}^{\sigma/2}\,dr\,U(r) = \frac{31}{20}.
\end{equation}
When varying $j/u$, the softness parameter, with $j \neq 0$ and $u\neq 0$, we still observe the same features on the phase diagram, but with a shifted tricritical point.
Figure~\ref{fig:VECWSoft}$(b)$, displays the evolution of the coordinates of the tricritical point against $j/u$.
Starting from the hard disk limit ($j/u = 0$), when the particles become softer, the tricritical point goes up in temperature and in packing fraction.
Being softer allows for more neighbours and therefore makes both magnetization and compaction easier, hence the shift to higher temperatures.
The shift to higher packing fractions, follows from the fact that the tricritical point has to remain on the Curie line.
For values corresponding to our simulations ($j/u_{\rm sim} = 5/93$), the repulsion is hard enough that the displacement is barely noticeable.

\section{Conclusion \label{sec:Conclusion}}

In this paper, we discussed several equilibrium and non-equilibrium properties of XY spin fluids in $2d$, with a variety of theoretical approaches and simulations.

We numerically showed that these systems do not follow the  Berezinskii-Kosterlitz-Thouless scenario of vortex unbinding in homogeneous phases. 
The zero-temperature ferromagnetic order is instead destroyed solely by low-energy spin-wave excitations. 
Accordingly the magnetic correlation length diverges exponentially at $T = 0$. 
Hence, finite-size systems are in practice also smaller than the correlation length and appear to behave in a mean-field-like fashion.
It would be interesting to see if this feature breaks down in solid phases, in which the quasi-long-range positional order of the particles constrains them to be near the vertices of a lattice, and the magnetic behaviour should resemble the one of the conventional $2d$ XY model. 
In the same line of thought, it would be interesting to see whether BKT phenomenology would be recovered by tuning the interaction potentials we used here, for instance by making $J(r)$ deeper, thereby favouring crystalline phases.

We also showed the presence of a phase separation between a paramagnetic gas and a ferromagnetic liquid at low temperatures due to the effective spin-mediated attraction between the particles.
This last finding, which is quite similar to previous results in $3d$ for Heisenberg and planar spins, is recovered with various analytical mean-field approximations, and is found to be robust against the shape of interaction potentials.

The out-of-equilibrium relaxation after a quench into the phase-separated phase indicates that, even though the correlation lengths associated to magnetic and liquid properties grow in standard way, the order parameter associated to the liquid-vapour phase separation grows with an unusual exponent.
The liquid droplet growth thus escapes the locally non-conserved order parameter dynamic universality class exemplified by Model B, and is instead synchronized with the growth of the magnetization.
This synchronization suggests that the Curie line plays a special role in the phase-separating regime, the one of the left-most branch of the spinodal curve, as also shown with analytical mean-field-level calculations.
Physically, the coincidence of the Curie line and gas-side spinodal means that the development of magnetization destabilizes the gas and stabilizes the liquid.
This could have interesting implications on the precise clustering dynamics going beyond our study of domain growth, including in the long time regime and after quenches of varying speed that could lead to nucleation-dominated dynamics.~\cite{Cahn1961}

The study of spin fluids of the kind we used here could be relevant in a variety of fields, ranging from the study of ferromagnetic gases~\cite{Jo2009} and liquids~\cite{Boudalis2017} to more general occurrences of polar fluids in physics. 
Those occurrences include usual liquids, in which polarity seem to be responsible for still not-well-understood properties, as in the case of water,~\cite{Gallo2016} and systems of polar active matter, some properties of which could possibly be linked to ours through the introduction of spin-velocity couplings.~\cite{Bore2016} 

\begin{acknowledgments}
We acknowledge fruitful discussions and help from Michael Schindler on the numerical and theoretical aspects of this work, and we warmly thank Peter C. W. Holdsworth for his manifold and very interesting suggestions that have rendered our paper much more interesting. We would also like to thank Gilles Tarjus, Nicolas Sator, and Pascal Viot for useful suggestions. Leticia F. Cugliandolo and Marco Tarzia are members of the {\it{Institut Universitaire de France}}.
\end{acknowledgments}

\appendix
\section{Virial Coefficients Computation  \label{app:VECW}}

In this Appendix, we give more details on the computation of the virial coefficients used in the VE-CW approach, for arbitrary values of $\beta$ and $h$ in our choice of parametrization of the angle distribution of spins. 
As mentioned in the main text, the integrals we need to compute are
\begin{eqnarray*}
 B_2 (\beta, h)  &=& - \frac{1}{2} \int d\bm{r} d\theta_1 d\theta_2 p(\theta_1, \beta h)p(\theta_2, \beta h)f(r,\theta_{12}), \\
 B_3 (\beta, h)  &=& - \frac{1}{3} \int d\bm{r_1} d\bm{r_2} d\theta_1 d\theta_2 d\theta_3 \\
	     &\quad& \qquad \times p(\theta_1, \beta h)p(\theta_2, \beta h)p(\theta_3, \beta h) \nonumber \\
	     &\quad& \qquad \times f(r_{12},\theta_{12})f(r_{2},\theta_{23})f(r_{1},\theta_{31}),
\end{eqnarray*}
with
\begin{eqnarray*}
 p(\theta;\beta h) &=& \frac{e^{\beta h \cos\theta}}{2 \pi I_0(\beta h)}, \\
 f(r,\theta) &=& \Theta(\sigma - r)\left(e^{\beta j \cos \theta} - 1 \right) \nonumber \\
	     &+&  \Theta(\sigma - r)  \Theta(\sigma_{\rm rep} - r) e^{\beta j \cos\theta} \left( e^{-\beta u} - 1\right).
\end{eqnarray*}

A nice feature of these expressions is that in both cases, integrals over space and integrals over angles can be decoupled. 
Furthermore, as we only used square-shaped potentials for both repulsion and magnetic alignment, the space integrals only amount to disk overlap computations. 
Finally, we can comment on several limits of the amplitudes of the interactions that can still be taken smoothly at this level. 
For $j\to0$, we recover a square-potential liquid, with either hardcore exclusion ($u\to\infty$) or soft square-potential repulsion.
Likewise, for $0<j<\infty$, the hard-disk limit can be taken smoothly, as the $u \to \infty$ limit simply amounts, for a finite value of $j$, to taking $e^{- \beta u} \to 0$.

We will treat each virial coefficient separately, starting by the lowest-order one as it happens to be easier to compute.

\subsection{Computation of $B_2$}

The space integral in $B_2$ is the one of the surface of a disk, and the only non-trivial integral to compute is the one over the angles. 
In order to compute it, we write the exponential of the difference between angles as~\cite{Abramowitz1972}
 \begin{eqnarray*}
  e^{\beta j \cos\theta_{12}} &=& I_0(\beta j) + 2 \sum\limits_{n\geq 1}I_n(\beta j) \cos(n\theta_{12}) \\
			      &=& I_0(\beta j) + 2 \sum\limits_{n\geq 1}I_n(\beta j) \left[\cos(n\theta_{1})\cos(n\theta_{2}) \right. \\
			      &\quad& \qquad \qquad \qquad \qquad \left. + \sin(n\theta_{1})\sin(n\theta_{2})\right].
 \end{eqnarray*}

It is then rather simple to write the full form of $B_2$ in terms of Bessel functions
\begin{align*}
 B_2(&h,T) = \frac{\pi}{2}\sigma^2 \left\{\left[1 - I_0(\beta j) \right] - 2\sum\limits_{n=1}^\infty I_n(\beta j)\frac{I_n(\beta h)^2}{I_0(\beta h)^2}  \right\} \nonumber \\
	&+ \frac{\pi}{2}f^2 {\sigma}^2 \left\{ I_0(\beta j) +   2\sum\limits_{n=1}^\infty I_n(\beta j)\frac{I_n(\beta h)^2}{I_0(\beta h)^2}  - e^{-\beta u} \right\}.
\end{align*}
 
In this work, we also used the values of $B_2$ in the hard-disk limit $u \to \infty$, which can be taken smoothly,
\begin{eqnarray*}
 B_2^{HS}(h,T) &=& \frac{\pi}{2}\sigma^2 \left\{\left[ \vphantom{e^{x^x}}1 + (f^2-1)I_0(\beta j) \right]\vphantom{\sum\limits_{n=1}^\infty}\right. \\
	       &\quad& \left. \quad + 2(f^2-1) \sum\limits_{n=1}^\infty I_n(\beta j)\frac{I_n(\beta h)^2}{I_0(\beta h)^2}  \right\}, \nonumber 
\end{eqnarray*}
and, in some cases, also in its $h \to 0$ limit, which reads:
 \begin{eqnarray}
   B_2^{HS}(0,T) &=& \frac{\pi}{2}\sigma^2 \left[ \vphantom{e^{x^x}} 1 + (f^2-1)I_0(\beta j) \right]. \nonumber 
 \end{eqnarray}
 
 In the limit $j \to 0$, this coefficient simply becomes the well-known hard-disk coefficient,~\cite{Hemmer1965}
  \begin{eqnarray}
   B_2^{HS} &=& \frac{\pi}{2}f^2 \sigma^2. \nonumber 
 \end{eqnarray}
 
 \subsection{Computation of $B_3$}
 
 When computing $B_3$ explicitly, the integrals over the angles are computed by using the same trick as the one used to calculate $B_2$. 
 The integrals over space, however, are a bit more complicated to calculate, as they now comprise overlap surface computations between disks of unequal radii
 \begin{eqnarray*}
  J_f    &\equiv& \int d^2\bm{r_1}d^2\bm{r_2} \Theta(f\sigma - r_{12}) \Theta(\sigma - r_2) \Theta(\sigma - r_1), \\
  J_{ff} &\equiv& \int d^2\bm{r_1}d^2\bm{r_2} \Theta(\sigma - r_{12}) \Theta(f\sigma - r_2) \Theta(f\sigma - r_1). 
 \end{eqnarray*}
 
 It is useful for the low-density radial distribution function calculation explained in Sec. \ref{sec:Virial} to define the integrals over, say, $\bm{r}_1$ separately. 
 They can, in fact, be grouped under the definition
 \begin{equation}
  A_{f_1,f_2}(r_2) \equiv \frac{1}{\sigma^2} \int d^2\bm{r_1} \Theta(f_1 \sigma - r_1) \Theta( f_2 \sigma - r_{12}).
 \end{equation}

 The trick is then to write this integral in polar coordinates and with the changes of variables $\bm{r}_{1,2} \to \bm{r}_{1,2} / \sigma$ (which should be taken into account in the integral over $\bm{r}_2$ afterwards), 
  \begin{eqnarray}
  A_{f_1,f_2}(r_2) &=& \int\limits_{r_1 = 0}^{f_1} r_1 dr_1 \int\limits_{\phi = - \pi}^{\pi} d\phi  \nonumber \\
		   &\quad&  \quad \Theta( {f_2}^2  - \left(r_2^2 + r_1^2 - 2 r_2 r_1 \cos\phi \right)) \nonumber
 \end{eqnarray}

This integration can be carried out by noticing that the geometric domain defined through the $\Theta$ function can be cut into two complementary parts that are simple to integrate over $\phi$.
Geometrically, the domain we are drawing is the overlap between a disk centered on the point located (in polar coodinates) at $\left(r_1,\phi\right)$, and a second disk centered on $\left(r_2,0\right)$, with disk radii $f_1$ and $f_2$, respectively.
Below a given value of the distance between the two disks and the origin, all values of $\phi$ belong to the domain defined by the step function, while over this distance there is a finite interval of $\phi$, symmetric around zero, that belong to this domain. 
Using this, the integral can be rewritten in the general form:
   \begin{eqnarray}
  A_{f_1,f_2}(r_2) &=&  2 \pi \left[\frac{r_1^2}{2} \right]_{0}^{\min\left(f_1, f_2 - r_2\right)} \Theta(f_2  - r_2) \nonumber \\
		   &+& \int_{0}^{f_1 \sigma} d{r_1} r_1 \left[ 2 \arccos\left(\frac{r_1^2+r_2^2 - f_2^2 \sigma^2}{2 r_1r_2} \right) \nonumber \right. \\
		   &\quad& \left. \times \Theta(r_1 + r_2  - f_2 )\Theta (|r_1 - r_2| - f_2 ) \vphantom{\frac{r_1^2+r_2^2 - f_2^2 \sigma^2}{2 r_1r_2}}\right],
 \end{eqnarray}
 where we used the notation $\left[ f(x) \right]_a^b = f(b) - f(a)$.
 
 The last remaining integral can then be computed as follows.
 First, we use an integration by parts to get a derivative of the arccos out.
 Then, we change variables first switching to $u = r_1^2$, and then to an angle $\varphi$ such that: $u = u_- \cos^2\varphi + u_+ \sin^2 \varphi$, where $u_\pm \equiv (f_2 \pm r_2)^2$. 
 This allows us to derive an exact expression, that is a bit cumbersome to write here. The integrals $J_f$ and $J_{ff}$ finally read
  \begin{align*}
  J_f &=\pi \sigma^4\left[ \vphantom{\arctan\left(\frac{f}{\sqrt{4-f^2}}\right)} \pi f^2 -  \sqrt{4 - f^2} \left( \frac{f}{2} + \frac{f^3}{4}\right) \right.\\
  &+ \arccos\left(1 - \frac{f^2}{2} \right)  - f^2 \arcsin\left(\frac{f}{2} \right) \\
      &\left.  - f^2 \arctan\left(\frac{f}{\sqrt{4-f^2}}\right) \right], \\
  J_{ff} &= 4 \pi^2 \sigma^4 J^L_{ff} + 4 \pi \sigma^4 J^{U}_{ff},
 \end{align*}
 where
 \begin{eqnarray*}
 J^L_{ff} &=& \frac{f^4}{4} \Theta\left(\frac{1}{2} - f\right) \nonumber \\
	  &\quad& + \frac{1}{24}\left(6f^4 - 16f^3 +12f^2 - 1 \right) \Theta\left(f - \frac{1}{2}\right), \\
 J^U_{ff} &=& \Theta\left(f - \frac{1}{2}\right) \left[\frac{\pi }{24} -\frac{\pi }{4} f^2 +\frac{2 \pi }{3}  f^3 -\frac{\pi }{4} f^4 \right. \nonumber \\
	  &\quad& -\frac{ f^2}{8} \sqrt{4 f^2-1}  \nonumber \\
	  &\quad& -\frac{1}{16} \sqrt{4 f^2-1} -\frac{f^2}{2}  \arctan\left(\frac{1}{\sqrt{4 f^2-1}}\right)  \nonumber\\
	  &\quad& -\frac{1}{2} f^2 \arccos\left(\frac{1}{2}\sqrt{\frac{1}{f}+2}\right)+\frac{1}{4} f^2 \arccos\left(\frac{1}{2   f}\right) \nonumber \\
	  &\quad& \left. +\frac{1}{4} f^4 \arccos\left(1-\frac{1}{2 f^2}\right)\right].
\end{eqnarray*}
 
 Putting all these results together, we find that $B_3$ reads
  \begin{eqnarray}
   B_3(h,T)  &=& - \frac{1}{24 \pi^3 I_0(\beta h)^3} \left( I_2^{(3)}  J_\triangle^{(3)} - 3 I_2^{(2)}  J_\triangle^{(2)} \nonumber \right. \\
	     &\qquad& \qquad  \quad + \left.  3 I_2^{(1)}  J_\triangle^{(1)} - I_2^{(0)}  J_\triangle^{(0)} \right), \nonumber\\
  \end{eqnarray}
  where we defined quite a few intermediary integrals,
  \begin{eqnarray}
  I_2^{(3)}&=&  J_1 + 3 (e^{-\beta u} - 1) J_{f}, \nonumber \\
	   &\quad& + 3 (e^{-\beta u} - 1)^2  J_{ff} + (e^{-\beta u} - 1)^3 f^4 J_1, 	 \nonumber  \\
  I_2^{(2)}&=& J_1 + 2 (e^{-\beta u} - 1)J_{f} + (e^{-\beta u} - 1)^2  J_{ff},\nonumber \\
  I_2^{(1)}&=& J_1  +  (e^{-\beta u} - 1) J_{f}, \nonumber\\
  I_2^{(0)}&=&  J_1, \nonumber
 \end{eqnarray}
 with $J_1$ the value of $J_f$ for $f = 1$, and
 \begin{widetext}
 \begin{eqnarray*}
   J_\triangle^{(0)} &=& 8 \pi^3 I_0(\beta h)^3, \\ 
   J_\triangle^{(1)} &=& 8 \pi^3 I_0(\beta h) \left[ I_0(\beta j)I_0(\beta h)^2 + 2 \sum\limits_{n\geq 1}I_n(\beta j) I_n(\beta h)^2 \right], \\
   J_\triangle^{(2)} &=& (2 \pi  I_0 (\beta h))^3 I_0(\beta j)^2 + 32 \pi^3 I_0 (\beta j)I_0(\beta h) \sum\limits_{n\geq 1} I_n(\beta j) I_n(\beta h)^2 \\
		     &+& 16 \pi^3 \sum\limits_{m,n\geq 1} I_m(\beta j) I_n(\beta j) I_m(\beta h) I_n(\beta h) (I_{m+n}(\beta h) + I_{m-n}(\beta h)), \nonumber \\
J_\triangle^{(3)} &=& 8 \pi^3  (I_0(\beta h) I_0(\beta j))^3 + 48 \pi^3 I_0(\beta h) I_0(\beta j)^2 \sum\limits_{n = 1}^\infty  (I_{n}(\beta h))^2  I_{n}(\beta j)  \nonumber \\
                     &+& 48 \pi^3 I_0(\beta j) \sum\limits_{\epsilon = \pm 1} \sum\limits_{m,n = 1}^\infty I_{m}(\beta h) I_{n}(\beta h) I_{m+ \epsilon n }(\beta h) I_{m}(\beta j) I_{n}(\beta j)   \nonumber \\
		     &+&  8 \pi^3 \sum\limits_{\epsilon,\eta, \nu = \pm 1}\sum\limits_{m,n,p = 1}^\infty I_{m +\epsilon n}(\beta h) I_{n +\eta p}(\beta h) I_{p +\nu m}(\beta h)  I_{m}(\beta j)I_{n}(\beta j)I_{p}(\beta j)\nonumber \\
		     &-& 8 \pi^3 \sum\limits_{\epsilon,\eta,\nu = \pm 1} \epsilon \eta \nu\sum\limits_{m,n,p = 1}^\infty I_{m +\epsilon n}(\beta h)  I_{n +\eta p}(\beta h) I_{p +\nu m}(\beta h)  I_{m}(\beta j)I_{n}(\beta j)I_{p}(\beta j).\nonumber \\
\end{eqnarray*}
 \end{widetext}
Just like for $B_2$, we can check that we recover the hard-disk limit given in the literature~\cite{Hemmer1965} for $u\to\infty, h\to0,$ and $j\to 0$,
 \begin{eqnarray*}
   B_3^{HS}   &=& \pi f^4 {\sigma}^4\left(\frac{\pi}{3} - \frac{\sqrt{3}}{4} \right).
 \end{eqnarray*}

\bibliography{Bibtex-SpinFluid}%/users/invites/casiulis/Documents/BibTeX/Bibtex-SpinFluid}

\end{document}